\documentclass[twoside,11pt]{article}

\PassOptionsToPackage{sort}{natbib}

\usepackage{amsmath}
\usepackage{mathtools}
\usepackage{bbold}
\usepackage{tikz}
\usepackage{xcolor}
\usepackage{subcaption}
\usepackage[ruled,algo2e,linesnumbered]{algorithm2e}

\usepackage[toc,page]{appendix}

\usepackage{etoolbox}
\appto\appendix{\addtocontents{toc}{\protect\setcounter{tocdepth}{2}}}

\usepackage[preprint]{jmlr2e}

\graphicspath{{figs/}}

\makeatletter
    \newtheorem{case@}{Case}
    \newenvironment{case}[1][]
    {%
      \if\relax\detokenize{#1}\relax\begin{case@}%
      \else\begin{case@}[#1]\fi\upshape%
    }
    {\end{case@}}
\makeatother
\renewenvironment{proof}[1][]
  {\medskip\par\noindent{%
    \bf\if\relax\detokenize{#1}\relax Proof\ \else #1\ \fi}%
  }
  {\hfill\BlackBox\\[2mm]}

\DeclareFontFamily{U}{matha}{\hyphenchar\font45}
\DeclareFontShape{U}{matha}{m}{n}{
  <-6> matha5 <6-7> matha6 <7-8> matha7
  <8-9> matha8 <9-10> matha9
  <10-12> matha10 <12-> matha12
  }{}

\DeclareSymbolFont{matha}{U}{matha}{m}{n}
\DeclareMathSymbol{\Lt}{3}{matha}{"CE}

\newcommand{\B}[1]{\mathbb{#1}}
\newcommand{\C}[1]{\mathcal{#1}}
\newcommand{\D}[1]{\operatorname{\textsc{#1}}}
\newcommand{\T}[1]{\textrm{#1}}
\newcommand{\EE}{\mathbb{E}}
\newcommand{\bb}[1]{\mathbf{#1}}
\newcommand{\indep}{\protect\mathpalette{\protect\independenT}{\perp}}
\newcommand{\dx}[1]{\mathrm{d}#1}
\newcommand{\sgn}{\mathop{\mathrm{sign}}}
\def\independenT#1#2{\mathrel{\rlap{$#1#2$}\mkern2mu{#1#2}}}

\DeclareMathOperator{\unif}{\textsc{Unif}}
\DeclareMathOperator{\bern}{\textsc{Bern}}
\DeclareMathOperator{\bbeta}{\textsc{Beta}}

\DeclareMathOperator{\logit}{\text{logit}}
\DeclareMathOperator{\rest}{\upharpoonright}
\DeclareMathOperator{\Span}{\textsc{Span}}
\DeclareMathOperator{\supp}{\textsc{Supp}}

\DeclareMathOperator{\pPr}{Pr}
\DeclareMathOperator{\up}{{Up}}
\DeclareMathOperator{\low}{{Lo}}

\DeclareMathOperator{\img}{\textsc{Img}}
\DeclareMathOperator{\abs}{\textsc{Abs}}

\DeclareMathOperator{\id}{id}
\DeclareMathOperator{\st}{st}

\newcommand*\diff{\mathop{}\!\mathrm{d}}

\renewcommand{\Pr}{\pPr}

\newlength{\myl}
\usepackage{lastpage}
\jmlrheading{XX}{2023}{1-\pageref{LastPage}}
  {12/22; Revised 08/23}
  {XX}
  {XX}
  {Corbett-Davies, Gaebler, Nilforoshan, Shroff, and Goel}

\ShortHeadings{The Measure and Mismeasure of Fairness}
  {Corbett-Davies, Gaebler, Nilforoshan, Shroff, and Goel}
\firstpageno{1}

\begin{document}

\title{The Measure and Mismeasure of Fairness}

\author{\name Sam Corbett-Davies$^*$ \email samcorbettdavies@gmail.com \\
       \AND
       \name Johann D. Gaebler$^*$ \email jgaebler@fas.harvard.edu \\
       \addr Department of Statistics\\
       Harvard University\\
       Cambridge, MA 02138, USA
       \AND
       \name Hamed Nilforoshan$^*$ \email hamedn@cs.stanford.edu \\
       \addr Department of Computer Science\\
       Stanford University\\
       Stanford, CA 94305, USA
       \AND
       \name Ravi Shroff \email ravi.shroff@nyu.edu \\
       \addr Department of Applied Statistics, Social Science, and Humanities\\
       New York University\\
       New York, NY 10003, USA
       \AND
       \name Sharad Goel \email sgoel@hks.harvard.edu \\
       \addr Harvard Kennedy School\\
       Harvard University\\
       Cambridge, MA 02138, USA\\
       \\
       $^*$Authors contributed equally.
       }

\editor{Kilian Weinberger}

\maketitle

\begin{abstract}%
The field of fair machine learning aims to ensure that decisions guided
by algorithms are equitable. Over the last decade, several formal,
mathematical definitions of fairness have gained prominence. Here we first
assemble and categorize these definitions into two broad families: (1) those
that constrain the effects of decisions on disparities; and (2) those that
constrain the effects of legally protected characteristics, like race and
gender, on decisions. We then show, analytically and empirically, that both
families of definitions typically result in strongly Pareto dominated decision
policies. For example, in the case of college admissions, adhering to popular
formal conceptions of fairness would simultaneously result in lower student-body
diversity and a less academically prepared class, relative to what one could
achieve by explicitly tailoring admissions policies to achieve desired outcomes.
In this sense, requiring that these fairness definitions hold can, perversely,
harm the very groups they were designed to protect. In contrast to axiomatic
notions of fairness, we argue that the equitable design of algorithms requires
grappling with their context-specific consequences, akin to the equitable design
of policy. We conclude by listing several open challenges in fair machine
learning and offering strategies to ensure algorithms are better aligned with
policy goals.
\end{abstract}

\vspace{5mm}

\clearpage
\tableofcontents
\clearpage

\section{Introduction}

In banking, criminal justice, medicine, and beyond, consequential decisions are
often informed by machine learning algorithms~\citep{chouldechova2018case,
barocas2016big, berk2012criminal, shroff2017predictive}. As the influence and
scope of algorithms increase, academics, policymakers, and journalists have
raised concerns that these tools might inadvertently encode and entrench human
biases. Such concerns have sparked tremendous interest in developing \emph{fair}
machine-learning algorithms, and, accordingly, a plethora of formal fairness
criteria have been proposed in the computer science community~\citep{
  darlington1971another, cleary1968test, zafar2017parity, dwork2012fairness,
  chouldechova2017fair, hardt2016equality, kleinberg2016inherent,
  woodworth2017learning, zafar2017fairness, corbett2017algorithmic,
  chouldechova2020snapshot, berk2021fairness, coston2020counterfactual,
  imai2020principal, imai2020experimental, wang2019equal,
  kusner2017counterfactual, nabi2018fair, wu2019pc, mhasawade2021causal,
  kilbertus2017avoiding, zhang2018fairness, zhang2016causal, chiappa2019path,
  loftus2018causal, galhotra2022causal, carey2022causal%
}.
Here we synthesize and critically examine the statistical properties of popular
formal fairness approaches as well as the consequences of enforcing them. Using
both theory and empirical evidence, we argue that these approaches, when used as
algorithmic design principles, can often cause more harm than good. In contrast
to popular axiomatic approaches to algorithmic fairness, we advocate for a
consequentialist perspective that directly grapples with the difficult policy
trade-offs inherent to many algorithmically guided decisions.

We begin, in Section~\ref{sec:defn}, by proposing a two-part taxonomy of formal
fairness definitions. Our first category of definitions encompasses those that
consider the effects of decisions on disparities. Imagine, for example,
designing an algorithm to guide decisions for college admissions. Under the
principle that fair algorithms should have comparable performance across
demographic groups~\citep{hardt2016equality}, one might check that among
applicants who were ultimately academically ``successful'' (e.g., who eventually
earned a college degree, either at the institution in question or elsewhere),
the algorithm would recommend admission for an equal proportion of candidates
across race groups. Our second category of definitions encompasses those that
seek to limit both the direct and indirect effects of one's group membership on
decisions. Following the principle that decisions should be agnostic to legally
protected attributes like race and gender~\citep[cf.{}][]{dwork2012fairness},
one might mandate that these features not be provided to the algorithm. Further,
because one's race might impact earlier educational opportunities, and hence
test scores, one might require that admissions decisions are robust to the
effect of race along such causal paths.

These formalizations of fairness have considerable intuitive appeal. It can feel
natural to exclude protected characteristics in a drive for equity; and one
might understandably interpret disparities in error rates as indicating problems
with an algorithm's design or with the data on which it was trained. However, in
Sections~\ref{sec:absence} and \ref{sec:presence}, we show that both classes of
algorithmic fairness definitions suffer from deep statistical limitations. For
example, for natural families of utility functions---like those that prefer both
higher academic preparedness and more student-body diversity---we prove that
common fairness criteria almost always, in a measure theoretic sense, lead to
strongly Pareto dominated decision policies.\footnote{%
  A policy is strongly Pareto dominated if there is an alternative feasible
  policy that is preferred under every utility function in the family (cf.\
  Section~\ref{sec:theory}).
}
In particular, in our running college admissions example, adhering to several of
the popular conceptions of fairness we consider would simultaneously result in
lower student-body diversity and a less academically prepared class, relative to
what one could attain by explicitly tailoring admissions policies to achieve
desired outcomes. In fact, under one prominent definition of fairness, we prove
that the induced policies require simply admitting all applicants with equal
probability, irrespective of one's academic qualifications or group membership.
These formal fairness criteria are thus often at odds with policy goals, and,
perversely, can harm the very same groups one ostensibly sought to protect by
developing and adopting axiomatic notions of fairness.

How, then, can we ensure algorithms are fair? There are no easy solutions, but
we conclude in Section~\ref{sec:forward} by offering several observations and
suggestions for designing more equitable algorithms. Most importantly, we
believe it is critical to acknowledge and tackle head-on the substantive
trade-offs at the heart of many decision problems. For example, when creating a
college admissions policy, one must necessarily make difficult choices that
balance competing priorities. Formal fairness axioms are poor tools for engaging
with these challenging issues. Our overarching exhortation is thus to recognize
algorithms as encoding policy choices, and to accordingly tailor their design.

\medskip\noindent\emph{Contributions.}\qquad
To summarize, we offer three main contributions. First, we survey the fairness
literature, describing existing fairness definitions and organizing them into a
two-part taxonomy. Our categorization of formal fairness definitions proposed in
the computer science literature highlights their connections to influential
legal and economic notions of discrimination. Second, we lay out a
consequentialist framework for designing equitable algorithms. Our framework is
motivated by viewing algorithmic fairness as a policy objective rather than as a
technical problem. This approach exposes the statistical and normative
limitations of many popular formal fairness definitions. Finally, we apply our
consequentialist framework to develop a positive vision for addressing problems
of fairness and equity in algorithm design.

Much of the content we present synthesizes and builds on research that we and
our collaborators have conducted over the last several years
\citep{corbett2017algorithmic, cai2020fair, L2BF, chohlas2023designing,
koenecke2023}. In particular, we draw heavily on two papers by
\citet{corbett2018measure} and \citet{nilforoshan2022causal}. In addition to
synthesis, we broaden the formal theoretical results presented in this line of
work and offer new, concrete illustrations of our theoretical arguments. Some of
the results and arguments we present date back five years, and the field of
algorithmic fairness has since moved forward in many ways. For example, in the
intervening time, there has been increasing recognition of the shortcomings of
popular formal fairness definitions \citep{barocas-hardt-narayanan}.
Nevertheless, we believe our message is as relevant as ever. For instance,
within the research community, new algorithmic fairness definitions are
regularly introduced that, while different in some respects, frequently suffer
from the same statistical and conceptual limitations as the notions we survey
here. In the broader world, policymakers, algorithm designers, journalists, and
advocates often still evaluate algorithms---and accordingly influence
decisions---by turning to these formal fairness definitions without necessarily
appreciating their shortcomings. For example, proposed legislation in Idaho
sought to require that pretrial risk assessment algorithms have equal error
rates across groups~\citep{idaho}. Although the proposed bill was never passed,
it the illustrates the ways in which these formal measures have garnered
significant attention beyond the academic community.

The call to build equitable algorithms will only grow over time as automated
decisions become even more widespread. As such, it is imperative to address
limitations in past formulations of fairness, to identify best practices moving
forward, and to outline important open research questions. By synthesizing and
critically examining recent developments in fair machine learning, we hope to
help both researchers and practitioners advance this increasingly
influential field.

\section{Mathematical Definitions of Fairness }
\label{sec:defn}

We start by assembling and categorizing definitions of algorithmic fairness into
a two-part taxonomy: those that seek to limit the effect of decisions on
disparities, and those that seek to limit the effect of protected attributes
like race or gender on the decisions themselves. We first introduce formal
notation and concrete examples of decision problems in which one might seek to
apply these fairness definitions, before reviewing prominent examples of both
approaches in turn.

\subsection{Formal setting}

Consider a population of individuals with observed covariates \(X\), drawn
i.i.d.\ from a set \(\mathcal{X} \subseteq \mathbb{R}^n\) with distribution
\(\mathcal{D}_{X}\). Further suppose that \(A \in \mathcal{A}\) describes one or
more discrete protected attributes, such as race or gender, which can be derived
from \(X\) (i.e., \(A = \alpha(X)\) for some function \(\alpha\)). Each
individual is subject to a binary decision \(D \in \{0, 1\}\), determined by a
(randomized) rule \(d(x) \in [0, 1]\), where \(d(x) = \Pr(D = 1 \mid X = x)\) is
the probability of receiving a positive decision, \(D=1\).\footnote{%
  That is, \(D = \B 1_{U_D \leq d(X)}\), where \(U_D\) is an independent uniform
  random variable on $[0, 1]$.
}\textsuperscript{,}\footnote{%
    By ``positive,'' we simply mean the decision D is greater than zero, without
    ascribing any normative position to the decision. Individuals may or may not
    have a preferences for ``positive'' decisions in this sense.
}
Given a budget \(b\) with \(0 < b \leq 1\), we require the decision rule to
satisfy \(\EE[D] \leq b\). Finally, we suppose that each individual has some
associated binary outcome \(Y\). In some cases, we will be concerned with the
causal effect of the decision \(D\) on \(Y\), in which case we imagine that
there exist two potential outcomes, \(Y(0)\) and \(Y(1)\), corresponding to what
happens to the individual depending on whether they receive a negative or
positive decision.\footnote{%
  As is implicit in our notation, we assume that there are no spillover effects
  between units \citep{imbens2015causal}.
}

To make our discussion concrete, we imagine two running examples corresponding
to this formal setting: diabetes screening and college admissions. As we discuss
in detail below, these two examples differ in the extent to which there is
agreement about the ultimate value of different decision policies, which in turn
impacts our mathematical analysis. Diabetes is a common and serious health
condition that afflicts many American adults. If caught early, it is often
possible to avoid some of the most significant consequences of the disease
through, for example, changes to one's diet and physical routine. A blood test
can be used to determine whether an individual has diabetes, but as with many
screening tools, there are risks and inconveniences associated with screening
(e.g., a patient may need to take time off from work). In particular, if an
individual were \emph{certain} that they did not have diabetes, then they would
prefer not to undergo screening. Our goal is to design an equitable screening
policy \(d(x)\) to determine which patients have (\(Y = 1\)) or do \emph{not}
have (\(Y = 0\)) diabetes, based on a set of covariates \(X\). For example,
following \citet{agarwal2022diabetes}, the screening decision may be based on a
patient's age, body mass index (BMI) and race. (Those authors argue that
consideration of race, while controversial, leads to more precise and equitable
estimates of diabetes risk, a point we return to in
Section~\ref{sec:anti-classification}.) We further imagine the budget \(b\)
equals \(1\), corresponding to the fact that everyone could be screened in
principle.

Our second example concerns college admissions. Here, the population of interest
is applicants to a particular college, and the decision \(D\) is the admissions
committee's binary admissions decision. To simplify our exposition, we assume
all admitted students attend the school. In this setting, the covariates \(X\)
may, for example, consist of an applicant's test score and race \(A \in \{a_0,
a_1\}\), and \(Y\) is a binary variable that indicates college graduation (i.e.,
degree attainment). In contrast to our diabetes example, here we imagine that
the decision itself may affect the outcomes. Specifically, \(Y(1)\) and \(Y(0)\)
describe whether an applicant \emph{would} attain a college degree if admitted
to or if rejected from the school we consider, respectively. Note that \(Y(0)\)
is not necessarily zero, as a rejected applicant may attend---and graduate
from---a different university. Further, in this case we set the budget \(b\) to
be less than one to reflect the fact that the admissions committee has limited
resources and is unable to admit every candidate.

As mentioned above, a key distinction between these two examples is the extent
to which stakeholders may agree on the value of different potential decision
policies. For example, in college admissions, there may be significant
disagreement on how to balance competing priorities, such as academic
preparedness and class diversity.\footnote{%
  In some jurisdictions, explicit considerations of racial diversity may be
  prohibited. For instance, a recent U.S. Supreme Court case bars colleges from
  explicitly considering race in admissions; however, colleges may consider ``an
  applicant's discussion of how race affected the applicant's life''
  \citep{2023students}. U.S. colleges may also consider other forms of
  diversity, such as economic or geographic diversity.
}
Admissions committees may seek to increase both dimensions, but there is often
an inherent trade-off, particularly since there is a limit on the number of
students that can be admitted by the college (i.e., \(b < 1\)). Our diabetes
example, in contrast, reflects a setting where there is ostensibly broader
agreement on the value of different decision policies. Indeed, since there is
effectively no limit on the number of diabetes tests that can be administered
(i.e., \(b = 1\)), we can model the value of a decision policy as the sum of
each individual's value for being screened.\footnote{%
  In the case of infectious diseases---which involve greater
  externalities---there is again often disagreement about the value of different
  screening and vaccination policies. \citet{paulus2020predictably} similarly
  draw a distinction between \emph{polar} settings (in which parties have
  competing interests, like our admissions example) and \emph{non-polar}
  settings (where there is broad alignment, as in our diabetes example).
}
In Sections~\ref{sec:absence} and \ref{sec:presence}, we in turn examine the
structure of equitable decision making in the absence and presence of such
trade-offs. First, though, we introduce several formal fairness criteria.

\subsection{Limiting the Effect of Decisions on Disparities}

A popular class of fairness definitions requires that error rates (e.g., false
positive and false negative rates) are equal across protected
groups~\citep{hardt2016equality}.\footnote{%
  Some work relaxes strict equality of error rates or other metrics to requiring
  only that the difference be at most some fixed \(\epsilon\)
  \citep[e.g.,][]{nabi2018fair}. For ease of exposition, we consider strict
  equality throughout, though we emphasize that the spirit of the critique we
  develop applies also in cases where fairness constraints are approximate,
  rather than exact.
}
We refer to these definitions as examples of ``classification parity,'' meaning
that some given measure of classification error is equal across groups defined
by attributes such as race and gender. In particular, we include in this
definition any measure that can be computed from the two-by-two confusion matrix
tabulating the joint distribution of decisions \(D\) and outcomes \(Y\) for a
group. \citet{berk2021fairness} enumerate seven such statistics, including
false positive rate, false negative rate, precision, recall, and the proportion
of decisions that are positive. The proportion of positive decisions is not,
strictly speaking, a measure of ``error'', but we nonetheless include it under
classification parity since it can be computed from a confusion matrix. We also
include the area under the ROC curve (AUC), a popular measure among
practitioners examining the fairness of algorithms~\citep{skeem2015risk}.

Two of the above measures---the proportion of decisions that are positive, and
the false positive rate---have received considerable attention in the machine
learning community~\citep{
  feldman2015, hardt2016equality, calders2010three,
  pedreshi2008discrimination, zemel2013learning, kamiran2013,
  edwards2015censoring, agarwal2018reductions, zafar2017fairness,
  zafar2015fairness, chouldechova2017fair, jung2020fair, blum2019recovering%
}.

\begin{definition}
\label{defn:demographic_parity}
  We say that \emph{demographic parity} holds when\footnote{%
    We use the notation \(X \indep Y\) throughout to mean that the random
    variables \(X\) and \(Y\) are independent.
  }
  \begin{equation}
  \label{eq:demographic_parity}
    D \indep A.
  \end{equation}
\end{definition}

\begin{definition}
\label{defn:equalized_false_positive_rates}
  We say that \emph{equalized false positive rates} holds when
  \begin{equation}
  \label{eq:equalized_false_positive_rates}
    D \indep A \mid Y = 0.
  \end{equation}
\end{definition}

In our running diabetes example, demographic parity means that the proportion of
patients who are screened for the disease is equal across race groups.
Similarly, in our college admissions example, demographic parity means an equal
proportion of students is admitted across race groups. Equalized false positive
rates, in our diabetes example, means that among individuals who in reality do
not have diabetes---and thus for whom screening, \emph{ex post}, would not have
been beneficial---screening rates are equal across race groups.\footnote{%
  In our college admissions example, the decision \(D\) impacts the outcome
  \(Y\). One could, in theory, apply the definition of error rate parity above
  to that case by recognizing that \(Y = Y(D)\). However, that interpretation
  does not seem aligned with the original intent of the definition. We instead
  discuss the admissions example in the context of the explicitly causal
  definitions of fairness below.
}

Causal analogues of these definitions have also recently been proposed~\citep{%
  coston2020counterfactual, imai2020principal, imai2020experimental,
  mishler2021fairness%
},
which require various conditional independence conditions to hold between the
potential outcomes, protected attributes, and decisions.\footnote{%
  In the literature on causal fairness, there is at times ambiguity between
  ``predictions'' \(\hat{Y} \in \{0,1\}\) of \(Y\) and ``decisions'' \(D \in
  \{0,1\}\). Following past work~\citep[e.g.,][]{%
    corbett2017algorithmic, kusner2017counterfactual, wang2019equal%
  },
  here we focus exclusively on decisions, with predictions implicitly impacting
  decisions but not explicitly appearing in our definitions.
}
Below we list three representative examples of this class of fairness
definitions: counterfactual predictive parity~\citep{coston2020counterfactual},
counterfactual equalized odds~\citep{mishler2021fairness,
coston2020counterfactual}, and conditional principal
fairness~\citep{imai2020principal}.\footnote{%
  Our subsequent analytical results extend in a straightforward manner to
  structurally similar variants of these definitions (e.g., requiring \(Y(0)
  \indep A \mid D=1\) or \(D \indep A \mid Y(0)\), variants of counterfactual
  predictive parity and counterfactual equalized odds, respectively).
}

\begin{definition}
\label{defn:predictive-parity}
  We say that \emph{counterfactual predictive parity} holds when
  \begin{equation}
  \label{eq:counterfactual_predictive_parity}
    Y(1) \indep A \mid D = 0.
  \end{equation}
\end{definition}

In our college admissions example, counterfactual predictive parity means that
among rejected applicants, the proportion who would have attained a college
degree, had they been accepted, is equal across race groups. (For our diabetes
example, because the screening decision does not affect whether a patient
actually has diabetes, \(Y(0) = Y(1) = Y\), and so counterfactual predictive
parity, as well as the causal definitions below, reduce to their non-causal
analogues).

\begin{definition}
\label{defn:cf-eo}
  We say that \emph{counterfactual equalized odds} holds when
  \begin{equation}
  \label{eq:counterfactual_equalized_odds}
    D \indep A \mid Y(1).
  \end{equation}
\end{definition}

In our running college admissions example, counterfactual equalized odds is
satisfied when two conditions hold: (1) among applicants who would graduate if
admitted (i.e., \(Y(1) = 1\)), students are admitted at the same rate across
race groups; and (2) among applicants who would not graduate if admitted (i.e.,
\(Y(1) = 0\)), students are again admitted at the same rate across race groups.

\begin{definition}
\label{defn:principal-fairness}
  We say that \emph{conditional principal fairness} holds when
  \begin{equation}
  \label{eq:conditional_principal_fairness}
    D \indep A \mid Y(0), Y(1), W,
  \end{equation}
  where, for some function \(\omega\) on \(\C X\), \(W = \omega(X)\)
  describes a reduced set of the covariates \(X\). When \(W\) is constant (or,
  equivalently, when we do not condition on \(W\)), this condition is called
  \emph{principal fairness}.
\end{definition}

In the college admissions example, conditional principal fairness means that
``similar'' applicants---where similarity is defined by the potential outcomes
and covariates \(W\)---are admitted at the same rate across race groups.

\subsection{Limiting the Effect of Attributes on Decisions}

An alternative framework for understanding fairness considers the effects of
protected attributes on decisions. This approach can be understood as codifying
the legal notion of disparate treatment~\citep{goel2017combatting,
zafar2017fairness}---which we discuss further in Section~\ref{sec:law}. Perhaps
the simplest way to limit the effects of protected attributes on decisions is to
require that the decisions do not explicitly depend on them, what some call
``fairness through unawareness''~\citep[cf.{}][]{dwork2012fairness}.

\begin{definition}
  Suppose that the covariates can be partitioned into the protected attributes
  and all other covariates, i.e., that \(\C X = \C X_u \times \C A\), where
  \(\C X_u\) consists of ``unprotected'' attributes. Then, we say that
  \emph{blinding} holds when, for all \(a, a' \in \C A\) and \(x_u \in \C
  X_u\),
  \begin{equation}
  \label{eq:anticlassification}
    d(x_u,a) = d(x_u,a').
  \end{equation}
\end{definition}
In our running diabetes example, blinding holds when the screening decision
depends solely on factors like age and BMI, and, in particular, does not depend
on the patient's race. We similarly say college admissions decisions satisfy
blinding when the decisions depend on factors like test scores and
extracurricular activities, but not race.

Blinding is closely tied to the notion of \emph{calibration}, the requirement
that, conditional on the estimated probability of some outcome (such as
graduation from college or having diabetes), the outcome is independent of group
membership. For example, among people with an estimated diabetes risk of 1\%,
calibration would require that the proportion of individuals who actually have
diabetes be the same across groups. Many authors treat calibration as a kind of
fairness constraint---in particular, to ensure that the meaning of estimated
risks do not differ across groups---and it has received considerable attention
in the fairness literature \citep[e.g.,][]{hebert2018multicalibration,
rothblum2022decision}. We note, though, that miscalibration is equivalent to
blindness in practice. In particular, when estimation error is small, risk
estimates that are allowed to depend on group membership are calibrated;
conversely, risk estimates that are blind to group membership typically are
miscalibrated---an empirical phenomenon shown and discussed in
Figure~\ref{fig:anticlassification} below. Because of this close relationship,
we do not treat calibration as a separate fairness constraint, but we do discuss
calibration and its relationship to blinding in detail in
Sections~\ref{sssec:blinding} and~\ref{sec:design}.

In contrast to blinding---in which race and other protected attributes are
barred from being an explicit input to a decision rule---the causal versions of
this idea consider both the direct and indirect effects of protected attributes
on decisions~\citep{%
  wang2019equal, kusner2017counterfactual, nabi2018fair, wu2019pc,
  mhasawade2021causal, kilbertus2017avoiding, zhang2018fairness,
  zhang2016causal%
}.
For example, even if decisions only directly depend on test scores, race may
indirectly impact decisions through its effects on educational opportunities,
which in turn influence test scores. In this vein, a decision rule is deemed
fair if, at a high level, decisions for individuals are the same in ``(a) the
actual world and (b) a counterfactual world where the individual belonged to a
different demographic group''~\citep{kusner2017counterfactual}.\footnote{%
  Conceptualizing a general causal effect of an immutable characteristic such as
  race or gender is rife with challenges, the greatest of which is expressed by
  the mantra, ``no causation without manipulation''
  \citep{holland1986statistics}. In particular, analyzing race as a causal
  treatment requires one to specify what exactly is meant by ``changing an
  individual's race" from, for example, White to Black~\citep{gaebler2020causal,
  hu2020s}. Such difficulties can sometimes be addressed by considering a change
  in the \emph{perception} of race by a decision
  maker~\citep{greiner2011causal}---for instance, by changing the name listed on
  an employment application~\citep{bertrand2004emily}, or by masking an
  individual's appearance~\citep{%
    goldin2000orchestrating, grogger2006testing, pierson2020large,
    chohlas2021blind%
  }.
}
This idea can be formalized by requiring that decisions remain the same in
expectation even if one's protected characteristics are counterfactually
altered, a condition known as counterfactual
fairness~\citep{kusner2017counterfactual}.

\begin{definition}
\label{defn:counterfactual-fairness}
  \emph{Counterfactual fairness} holds when
  \begin{equation}
  \label{eq:counterfactual_fairness}
    \EE[D(a') \mid X] = \EE[D \mid X],
  \end{equation}
  where \(D(a')\) denotes the decision
  when one's protected attributes are counterfactually altered to be any \(a'
  \in \mathcal{A}\).
\end{definition}
In our running college admissions example, this means that for each group of
observationally identical applicants (i.e., those with the same values of \(X\),
meaning identical race and test score), the proportion of students who are
actually admitted is the same as the proportion who would be admitted if their
race were counterfactually altered.

Counterfactual fairness aims to limit all direct and indirect effects of
protected traits on decisions. In a generalization of this criterion---termed
path-specific fairness~\citep{%
  chiappa2019path, nabi2018fair, zhang2016causal, wu2019pc%
}---one
allows protected traits to influence decisions along certain causal paths but
not others. For example, one may wish to allow the direct consideration of race
by an admissions committee to implement an affirmative action policy, while also
guarding against any indirect influence of race on admissions decisions that may
stem from cultural biases in standardized tests~\citep{williams1983some}.

\begin{figure}
  \begin{center}
    \begin{tikzpicture}[xscale = 3, yscale = 3, align = center]
      \node at (0,0) (race) {\(A\)\\{\scriptsize Race}};
      \node at (1,0) (educ) {\(E\)\\{\scriptsize Education}};
      \node at (2,0) (test) {\(T\)\\{\scriptsize Test Score}};
      \node at (3/2, -2/3) (medi) {\(M\)\\{\scriptsize Preparation}};
      \node at (3,0) (deci) {\(D\)\\{\scriptsize Decision}};
      \node at (4, 0) (pass) {\(Y\)\\{\scriptsize Graduation}};

      \draw[->, color = red, line width=0.25mm] (race) to (educ);
      \draw[->, bend left = 30, line width=0.25mm] (race) to (deci);
      \draw[->, line width=0.25mm] (educ) to (medi);
      \draw[->, color = red, line width=0.25mm] (educ) to (test);
      \draw[->, line width=0.25mm] (medi) to (test);
      \draw[->, bend right = 20, line width=0.25mm] (medi) to (pass);
      \draw[->, color = red, line width=0.25mm] (test) to (deci);
      \draw[->, line width=0.25mm] (deci) to (pass);

    \end{tikzpicture}
  \end{center}
  \caption{%
    A causal DAG illustrating a hypothetical process for college admissions.
    Under path-specific fairness, one may require, for example, that race does
    not affect decisions along the path highlighted in red.
  }
\label{fig:dag}
\end{figure}

The formal definition of path-specific fairness requires specifying a causal DAG
describing relationships between attributes (both observed covariates and latent
variables), decisions, and outcomes. In our running example of college
admissions, we imagine that each individual's observed covariates are the result
of the process illustrated by the causal DAG in Figure~\ref{fig:dag}. In this
graph, an applicant's race \(A\) influences the educational opportunities \(E\)
available to them prior to college; and educational opportunities in turn
influence an applicant's level of college preparation, \(M\), as well as their
score on a standardized admissions test, \(T\), such as the SAT. We assume the
admissions committee only observes an applicant's race and test score so that
\(X = (A, T)\), and makes their decision \(D\) based on these attributes.
Finally, whether or not an admitted student subsequently graduates (from any
college), \(Y\), is a function of both their preparation and whether they were
admitted.\footnote{%
  In practice, the racial composition of an admitted class may itself influence
  degree attainment, if, for example, diversity provides a net benefit to
  students~\citep{page2007making}. Here, for simplicity, we avoid consideration
  of such peer effects.
}

To formalize path-specific fairness, we start by defining, for the decision
\(D\), path-specific counterfactuals, a general concept in causal
DAGs~\citep[cf.{}][]{pearl2001direct}. Suppose \(\mathcal{G} = (\mathcal{V},
\mathcal{U}, \mathcal{F})\) is a causal model with nodes \(\mathcal{V}\),
exogenous variables \(\mathcal{U}\), and structural equations \(\mathcal{F}\)
that define the value at each node \(V_j\) as a function of its parents
\(\wp(V_j)\) and its associated exogenous variable \(U_j\). (See, for example,
\citet{pearl2009causal} for further details on causal DAGs.) Let \(V_1, \dots,
V_m\) be a topological ordering of the nodes, meaning that \(\wp(V_j) \subseteq
\{V_1, \dots, V_{j-1}\}\) (i.e., the parents of each node appear in the ordering
before the node itself). Let \(\Pi\) denote a collection of paths from node
\(A\) to \(D\). Now, for two possible values \(a\) and \(a'\) for the variable
\(A\), the path-specific counterfactuals \(D_{\Pi,a,a'}\) for the decision \(D\)
are generated by traversing the list of nodes in topological order, propagating
counterfactual values obtained by setting \(A = a'\) along paths in \(\Pi\), and
otherwise propagating values obtained by setting \(A = a\). (In Algorithm 1 in
the Appendix, we formally define path-specific counterfactuals for an arbitrary
node---or collection of nodes---in the DAG.)

To see this idea in action, we work out an illustrative example, computing
path-specific counterfactuals for the decision \(D\) along the single path \(\Pi
= \{A \rightarrow E \rightarrow T \rightarrow D\}\) linking race to the
admissions committee's decision through test score, highlighted in red in
Figure~\ref{fig:dag}. We describe the distribution of \(D_{\Pi,a,a'}\)
generatively, formally showing how to produce a draw from this distribution. To
start, we draw values \(U_E^*\), \(U_M^*\), \(U_T^*\), \(U_D^*\) of the
exogenous variables. Now, the first column in Table~\ref{tab:dag} corresponds to
draws \(V^*\) for each node \(V\) in the DAG, where we set \(A\) to \(a\), and
then propagate that value as usual. The second column corresponds to draws
\(\overline{V}^*\) of path-specific counterfactuals, where we set \(A\) to
\(a'\), and then propagate the counterfactuals only along the path \(A
\rightarrow E \rightarrow T \rightarrow D\). In particular, the value for the
test score \(\overline{T}^*\) is computed using the value of \(\overline{E}^*\)
(since the edge \(E \rightarrow T\) is on the specified path) and the value of
\(M^*\) (since the edge \(M \rightarrow T\) is not on the path). As a result of
this process, we obtain a draw \(\overline{D}^*\) from the distribution of
\(D_{\Pi,a,a'}\).

\begin{table}[t]
  \begin{align*}
    A^{*} &= a                        & \overline{A}^* &= a' \\
    E^{*} &= f_E(A^{*}, U_E^*)        & \overline{E}^* &= f_E(\overline{A}^*, U_E^*), \\
    M^{*} &= f_M(E^{*}, U_M^*)        & \overline{M}^* &= f_M(E^{*}, U_M^*), \\
    T^{*} &= f_T(E^{*}, M^{*}, U_T^*) & \overline{T}^* &= f_T(\overline{E}^*, M^{*}, U_T^*) \\
    D^{*} &= f_D(A^{*}, T^{*}, U_D^*) & \overline{D}^* &= f_D(A^{*}, \overline{T}^*, U_D^*)
  \end{align*}
    \caption{%
      Computing path-specific counterfactuals for the DAG in
      Figure~\ref{fig:dag}. The first column corresponds to draws \(V^*\) for
      each node \(V\), where we set \(A\) to \(a\), and then propagate that
      value as usual. The second column corresponds to draws \(\overline{V}^*\)
      of path-specific counterfactuals, where we set \(A\) to \(a'\), and then
      propagate the counterfactuals only along the path \(A \rightarrow E
      \rightarrow T \rightarrow D\).
  }
\label{tab:dag}
\end{table}

Path-specific fairness formalizes the intuition that the influence of a
sensitive attribute on a downstream decision may, in some circumstances, be
considered ``legitimate'' (i.e., it may be acceptable for the attribute to
affect decisions along certain paths in the DAG). For instance, an admissions
committee may believe that the effect of race \(A\) on admissions decisions
\(D\) which passes through college preparation \(M\) is legitimate, whereas the
effect of race along the path \(A \rightarrow E \rightarrow T \rightarrow D\),
which may reflect access to test prep or cultural biases of the tests, rather
than actual academic preparedness, is illegitimate. In that case, the admissions
committee may seek to ensure that the proportion of applicants they admit from a
certain race group remains unchanged if one were to counterfactually alter the
race of those individuals along the path \(\Pi = \{A \rightarrow E \rightarrow T
\rightarrow D\}\).

\begin{definition}
\label{defn:ps}
  Let \(\Pi\) be a collection of paths, and, for some function \(\omega\) on
  \(\C X\), let \(W = \omega(X)\) describe a reduced set of the covariates
  \(X\). \emph{Path-specific fairness}, also called \emph{\(\Pi\)-fairness},
  holds when, for any \(a' \in \mathcal{A}\),
  \begin{equation}
  \label{eq:path_specific_fairness}
    \EE[D_{\Pi, A, a'} \mid W] = \EE[D \mid W].
  \end{equation}
\end{definition}

In the definition above, rather than a particular counterfactual level \(a\),
the baseline level of the path-specific effect is \(A\), i.e., an individual's
actual (non-counterfactually altered) group membership (e.g., their actual
race). We have implicitly assumed that the decision variable \(D\) is a
descendant of the covariates \(X\). In particular, without loss of generality,
we assume \(D\) is defined by the structural equation \(f_D(x, u_D) =
\mathbb{1}_{u_D \leq d(x)}\), where the exogenous variable \(U_D \sim
\unif(0,1)\), so that \(\Pr(D = 1 \mid X = x) = d(x)\). If \(\Pi\) is the set of
all paths from \(A\) to \(D\), then \(D_{\Pi,A,a'} = D(a')\), in which case, for
\(W = X\), path-specific fairness is the same as counterfactual fairness.

\section{Equitable Decisions in the Absence of Externalities}
\label{sec:absence}

In many decision-making settings, the decision maker is free to make the optimal
decision for each individual, without consideration of spillover effects or
other externalities. For instance, in our diabetes screening example, one could,
in principle, screen all patients if that course of action were medically
advisable.

To investigate notions of fairness in these settings, we first introduce a
framework for utilitarian decision analysis. Specifically, we consider in this
section situations in which there is broad agreement on the utility of different
potential courses of action. (In the subsequent section, we consider cases where
stakeholders disagree on the precise form of the utility.) In this setting,
``threshold rules'' maximize utility. We then describe the statistical
phenomenon of inframarginality, a property that is endemic to fairness
definitions that seek to enforce some form of classification parity. In
particular, we discuss, both informally and mathematically, why inframarginality
almost surely---in a measure theoretic sense---renders optimal decision making
incompatible with classification parity. Finally, we discuss blinding. In
parallel to our discussion of classification parity, we see that in many
important settings, the information loss associated with, e.g., removing
protected information from a predictive model, results in less efficient
decision making without compensatory benefits. Moreover, in general, we see that
the more stringent the standard of masking---e.g., removing not only direct but
also indirect effects of protected attributes---the greater the potential harm
that results from enforcing it.

\subsection{Utility, Risk, and Threshold Rules}
\label{sec:thresholds}

A natural way to analyze a decision, such as deciding whether an
individual should be screened for diabetes, is to consider the costs and
benefits of various possible outcomes under different courses of action. For
instance, a patient screened for diabetes who does not have the disease still
has to bear the risks, discomfort, and inconvenience associated with the blood
test itself, while a patient who is \emph{not} screened but does in fact have
the disease loses out on the opportunity to start treatment.

In general, the benefit of making decision \(D = 1\) \emph{over} \(D = 0\) when
the outcome \(Y\) equals \(y\) can be represented by \(v(y)\). For instance, in
our diabetes example, \(v(1)\) represents the net benefit of screening over not
screening when the patient has diabetes; and \(-v(0)\) is the net cost of
screening when the patient does not have diabetes, including both monetary and
non-monetary costs, such as discomfort and loss of time.\footnote{%
    For ease of exposition, we assume that costs and benefits are identical
    across individuals; in reality, these could vary, e.g., depending on age.
    When utilities vary by person, the optimal decision rule is to screen only
    those with positive individual utility, in line with our subsequent
    discussion.
}
Let \(r(x) = \Pr(Y = 1
\mid X = x)\) be the \emph{risk} of \(Y\) equalling 1 when \(X = x\). Then the
expected benefit of making decision \(D = 1\) over \(D = 0\) for an individual
with covariates \(X = x\) is
\begin{align*}
  u(x)
    &=\B E[v(Y) \mid X = x]\\
    &= r(x) \cdot v(1) + [1 - r(x)] \cdot v(0).
\end{align*}
Here, for ease of interpretation, we restrict our utility to be of the form
\(u(x) = \B E[v(Y) \mid X = x]\) for some function \(v\), and we also assume
there is no budget constraint (i.e., \(b = 1\)). In Section~\ref{sec:presence},
we allow the utility \(u(x)\) to be an arbitrary function on \(\C
X\) and consider \(b < 1\), which induces the trade-offs in decisions that are
central to our later discussion.

The \emph{aggregate} expected utility of a decision policy \(d(x)\)---relative
to the baseline policy of taking action \(D = 0\) for all individuals---is then
given by \(u(d) = \B E[d(X) \cdot u(X)]\). We say a decision policy \(d^*(x)\)
is utility-maximizing if
\begin{equation*}
  u(d^*) = \max_{d} u(d).
\end{equation*}

It is better, in expectation, for an individual with covariates \(X = x\) to
take action \(D = 1\) instead of \(D = 0\) when \(u(x) > 0\); that is,
when\footnote{%
  We assume, without loss of generality, that \(v(1) > v(0)\). If \(v(1) <
  v(0)\), we can take \(Y' = 1 - Y\) as our outcome of interest; relative to
  \(Y'\), the inequality will be reversed. If \(v(1) = v(0)\), then the outcome
  is irrelevant. In this degenerate case, the higher utility decision depends on
  the sign of \(v(1)\) alone, and not the risk.
}
\begin{equation}
\label{eq:threshold}
  r(x) > \frac {v(0)} {v(0) - v(1)}.
\end{equation}
Thus, the decision with the maximum utility can be determined by comparing an
individual's risk against a particular risk threshold \(t\), defined by the
right-hand side of Eq.~\eqref{eq:threshold}. We refer to this kind of policy as
a \emph{threshold policy}. In particular, we see that a utility-maximizing
decision for each individual---i.e., \(d(x) = 1\) if \(r(x) > t\) and \(d(x) =
0\) if \(r(x) \leq t\)---is also a decision policy that maximizes aggregate
utility, so there is no conflict between doing what is best from each individual
person's perspective and what is best for the population as a whole.

While our framing in terms of expected utility is suitably general, threshold
policies can be simpler to interpret when we reparameterize in terms of more
familiar quantities. In the diabetes screening example, if the patient does not
have diabetes, the cost of action \(D = 1\) over \(D = 0\) is \(-v(0) = c_{\D
{Test}}\), i.e., the cost (monetary and non-monetary) of the test. If the
patient does have diabetes, the benefit of \(D = 1\) over \(D = 0\) is \(v(1) =
b_{\D {Treat}} - c_{\D {Test}}\), i.e., the benefit of treatment minus the cost
of the test. Rewriting Eq.~\eqref{eq:threshold} in terms of these quantities
gives
\begin{equation*}
    t = \frac {c_{\D {Test}}} {b_{\D {Treat}}}.
\end{equation*}
In particular, if the benefit of early treatment of diabetes is 50 times greater
than the cost of performing the diagnostic test, one would ideally screen
patients who have at least a 2\% chance of developing the disease.

Threshold rules are a natural approach to decision making in a variety of
settings. In our running medical example, a threshold rule corresponds to
screening patients with a sufficiently high risk of having diabetes. A threshold
rule---with the optimally chosen threshold---ensures that only the patients at
highest risk of having diabetes take the test, thereby optimally balancing the
costs and benefits of screening. Indeed, in many medical examples, from
diagnosis to treatment, there are no significant externalities. As a result,
deviating from utility-maximizing threshold policies can only force individuals
to experience greater costs---in the form of unnecessary tests or untreated
illness---in expectation, without compensatory benefits. We return to the
problem of optimal (and equitable) decision-making in the presence of
externalities in Section~\ref{sec:presence}.

\subsection{The Problem of Inframarginality}
\label{sec:inframarginality}

In the setting that we have been considering, threshold policies guarantee
optimal choices are made for each individual. However, as we now show, threshold
policies in general violate various versions of classification parity, such as
demographic parity and equalized false positive rates. This incompatibility
highlights a critical limitation of classification parity as a fairness
criterion, as enforcing the definition often requires making decisions that harm
individuals without any clear compensating benefits.

\begin{figure}[t]
  \begin{center}
    \includegraphics{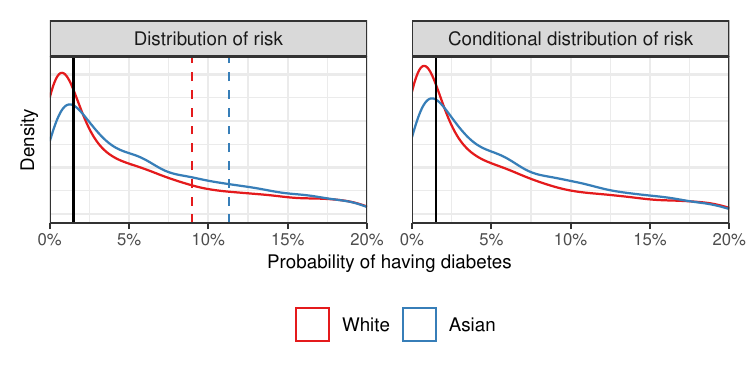}
  \end{center}
  \caption{%
    A graphical illustration of the incompatibility between threshold policies
    and classification parity, based on the National Health and Nutrition
    Survey. \emph{Left:}~The distribution of diabetes risk for White Americans
    and Asian Americans, with the dashed vertical lines corresponding to the
    overall incidence rate within each group. At a screening threshold of 1.5\%
    (indicated by the solid black line), the screening rate for Asian Americans
    is higher than for White Americans, violating demographic parity.
    \emph{Right:}~The distribution of diabetes risk among individuals who do
    \emph{not} have diabetes. Since the proportion of Asian Americans above the
    screening threshold is greater than the proportion of White Americans above
    the threshold, the false positive rate for Asian Americans is greater than
    the false positive rate for White Americans.
  }
\label{fig:risk-dist}
\end{figure}

To help build intuition for this phenomenon, we consider the empirical
distribution of diabetes risk among White and Asian patients. Following
\citet{agarwal2022diabetes}, we base our risk estimates on age, BMI, and race,
using a sample of approximately 15,000 U.S. adults aged 18--70 interviewed as
part of the National Health and Nutrition Survey \citep[NHANES;][]{nhanes}. The
resulting risk distributions are shown in the left-hand panel of
Figure~\ref{fig:risk-dist}. The dashed vertical lines show the group means, and
indicate that the incidence of diabetes is higher among Asian Americans (11\%)
than among White Americans (9\%).\footnote{%
  The precise shapes of the risk distributions depend on the set of covariates
  used to estimate outcomes, but the means of the distributions correspond to
  the overall incidence of diabetes in each group, and, in particular, are
  unaffected by the choice of covariates. It is thus necessarily the case that
  the risk distributions will differ across groups in this example, regardless
  of which covariates are used.
}
This difference in base rates is also reflected in the heavier tail of the risk
distribution among Asian individuals.

Drawing on recommendations from the United States Preventative Screening Task
Force, \cite{agarwal2022diabetes} suggest screening patients with at least a
1.5\% risk of diabetes, irrespective of race. We depict this risk threshold by
the solid black vertical line in the plot. Based on that recommendation, 81\% of
Asian Americans and 69\% of White Americans are to the right of the threshold
and should be screened---violating demographic parity. If, hypothetically, we
were to raise the screening threshold to 2.2\% for Asian Americans and lower the
threshold to 1\% for White Americans, 75\% of people in both groups would be
screened, satisfying demographic parity.\footnote{%
  \citet{corbett2017algorithmic} show that group-specific threshold policies are
  utility-maximizing under the constraint of satisfying various notions of
  classification parity, including demographic parity and equality of false
  positive rates.
}
The cost of doing so, however, would be failing to screen some Asian Americans
who have a relatively high risk of diabetes, and subjecting some relatively
low-risk White Americans to a procedure that is medically inadvisable given
their low likelihood of having diabetes. In an effort to satisfy demographic
parity, we would have harmed members from both groups.

This example illustrates a similar incompatibility between threshold policies
and equalized false positive rates. In our setting, the false positive rate for
a group is the screening rate among those in the group who do not in reality
have diabetes. To visualize the race-specific false positive rates, the
right-hand panel of Figure~\ref{fig:risk-dist} shows the distribution of
diabetes risk among those individuals who do not have diabetes. (Because the
overall prevalence of diabetes is low, the conditional distribution displayed in
the right-hand panel is nearly identical to the unconditional distribution
displayed in the left-hand panel.) The false positive rate for each group is the
proportion of people in the group falling to the right of the 1.5\% screening
threshold. In this case, the false positive rate is 79\% for Asian Americans and
67\% for White Americans---violating equalized false positive rates. As before,
we could alter the screening guidelines to equalize false positive rates, but
doing so requires deviating from our threshold policy, in which case we would
end up screening some individuals who are relatively low-risk and not screening
others who are relatively high-risk.

In this example, the incompatibility between threshold policies and
classification parity stems from the fact that the risk distributions differ
across groups. This general phenomenon is known as the problem of
\emph{inframarginality} in the economics and statistics literature, and has
long been known to plague tests of discrimination in human
decisions~\citep{%
  simoiu2017problem, ayres2002outcome, galster1993, carr1993, knowles2001,
  engel2008, anwar2006, pierson2018fast%
}.
Common legal and economic understandings of fairness are concerned with what
happens at the \emph{margin} (e.g., whether the same standard is applied to all
individuals)---a point we return to in Section~\ref{sec:forward}. What happens
at the margin also determines whether decisions maximize social welfare, with
the optimal threshold set at the point where the marginal benefits equal
marginal costs. However, popular error metrics assess behavior away from the
margin, hence they are called \emph{infra}-marginal statistics. As a result,
when risk distributions differ, standard error metrics are often poor proxies
for individual equity or social well-being.

In general, we expect any two non-random subgroups of a population to differ on
a variety of social and economic dimensions, which in turn is likely to yield
risk distributions that differ across groups. As a result, as our running
diabetes example shows, the optimal decision policy---which maximizes each
patient's own well-being---will likely violate various measures of
classification parity. Thus, to the extent that formal measures of fairness are
violated, that tells us more about the shapes of the risk distributions than
about the quality of decisions or the utility delivered to members of any group.
This intuition can be made precise, in the sense that for \emph{almost every}
risk distribution, the optimal decision policy violates the various notions of
classification parity considered here.

The notion of \emph{almost every} distribution that we use here was formalized
by \citet{christensen1972sets}, \citet{hunt1992prevalence},
\citet{anderson2001genericity}, and others \citep[cf.{}][for a
review]{ott2005prevalence}. Suppose, for a moment, that combinations of
covariates and outcomes take values in a finite set of size \(m\). Then the
space of joint distributions on covariates and outcomes can be represented by
the unit \((m-1)\)-simplex: \(\Delta^{m-1} = \{p \in \mathbb{R}^{m} \mid p_i
\geq 0 \ \text{and} \ \sum_{i=1}^m p_i = 1\}\). Since \(\Delta^{m-1}\) is a
subset of an \((m-1)\)-dimensional hyperplane in \(\mathbb{R}^m\), it inherits
the usual Lebesgue measure on \(\mathbb{R}^{m-1}\). In this finite-dimensional
setting, \emph{almost every} distribution means a subset of distributions that
has full Lebesgue measure on the simplex. Given a property that holds for almost
every distribution in this sense, that property holds almost surely under any
probability distribution on the space of distributions that is described by a
density on the simplex. We use a generalization of this basic idea that extends
to infinite-dimensional spaces, allowing us to consider distributions with
arbitrary support. (See the Appendix for further details.)

\begin{theorem}
\label{thm:fpr}
  Let \(t\) be the optimal decision threshold, as in Eq.~\eqref{eq:threshold}.
  If \(0 < t < 1\), then for almost every collection of group-specific risk
  distributions which have densities on \([0,1]\), no utility-maximizing
  decision policy satisfies demographic parity or equalized false positive
  rates.
\end{theorem}

The proof of Theorem~\ref{thm:fpr}, which formalizes the informal discussion
above, is given in Appendix~\ref{app:fpr}. At a high level, the constraints of
classification parity are sensitive to even small perturbations in the
underlying risk distributions. As a result, any particular collection of risk
distributions is unlikely to satisfy the constraints. For simplicity, we have
been considering settings in which the decision \(D\) does not impact the
outcome \(Y\). However, this basic style of argument extends to causal settings,
showing that threshold policies are almost surely, in the measure theoretic
sense, incompatible with counterfactual predictive parity, counterfactual
equalized odds, and conditional principal fairness---definitions of fairness
that we consider in depth in Section~\ref{sec:presence}, in the more complex
setting of having a budget \(b < 1\).

\subsection{The Problem with Fairness through Unawareness}
\label{sec:anti-classification}

We now consider notions of fairness, both causal and non-causal, that aim to
limit the effects of attributes on decisions. As above, we show the inherent
incompatibility of these definitions with optimal decision making. We note,
though, that while blinding can lead to suboptimal decisions---and, in some
cases, harm marginalized groups---the legal, political, and social benefits of,
for example, race-blind and gender-blind algorithms may outweigh their costs in
certain instances~\citep{cerdena2020race, coots2023reevaluating}.

\subsubsection{Blinding}
\label{sssec:blinding}

A common starting point for designing an ostensibly fair algorithm is to exclude
protected characteristics from the statistical model. This strategy ensures that
decisions have no explicit dependence on group membership. For instance, in the
case of estimating diabetes risk, one could use only BMI and age---rather than
including race, as we did above. However, excluding race from models of diabetes
risk can ultimately harm both White and Asian patients.

\begin{figure}[t]
  \begin{center}
    \begin{subfigure}{2.5in}
      \includegraphics[width=2.5in]{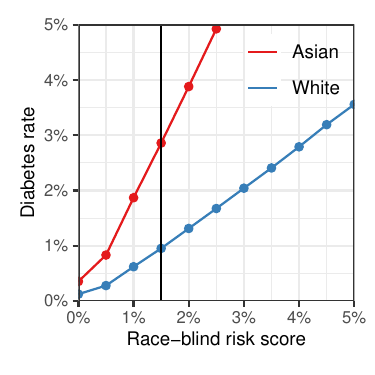}
      \caption{Diabetes}
    \label{fig:anti-classification-diabetes}
    \end{subfigure}
    \begin{subfigure}{2.5in}
      \includegraphics[width=2.5in]{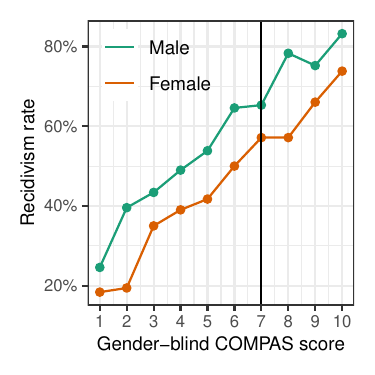}
      \caption{Recidivism}
    \label{fig:anti-classification-compas}
    \end{subfigure}
  \end{center}
  \caption{%
    Calibration plots showing the effect of removing protected attributes from
    risk models when estimating the risk of diabetes (left) and recidivism
    (right). Because Asian patients with the same BMI and age have a higher rate
    of diabetes, the race-blind model underestimates their risk of having
    diabetes. Similarly, because women reoffend at lower rates than men with
    similar criminal histories, the gender-blind COMPAS score overstates the
    recidivism risk for women.
  }
\label{fig:anticlassification}
\end{figure}

In Figure~\ref{fig:anti-classification-diabetes}, we compare the actual diabetes
rate to estimated diabetes risk resulting from the race-blind risk model.
\citet{agarwal2022diabetes} showed that Asian patients have higher incidence of
diabetes than White patients with comparable age and BMI. As a result, the
race-blind model systematically underestimates risk for Asian patients and
systematically overestimates risk for White individuals. In particular, applying
a nominal 1.5\% screening threshold under the race-blind model amounts to
effectively applying a 1\% screening threshold to White patients and a 3\%
screening threshold to Asian patients. Thus, by using race-blind risk scores, we
subject relatively low-risk White patients to screening, and fail to screen
Asian patients who have a relatively high risk for having diabetes. A race-aware
model would ensure that nominal risk thresholds correspond to observed incidence
rates across race groups.

This phenomenon---which we call miscalibration across subgroups---is not unique
to diabetes screening. Consider, for instance, the case of pretrial recidivism
predictions. Shortly after an individual is arrested in the United States, a
judge must often determine conditions of release pending future court
proceedings. In many jurisdictions across the country, these pretrial decisions
are informed by statistical risk estimates of the likelihood the individual
would be arrested or convicted of a future crime if released. After adjusting
for factors such as criminal history, age, and substance use, women have been
found to reoffend less often than men in many
jurisdictions~\citep{skeem2016gender,demichele2018public}. Consequently,
gender-blind risk assessments are miscalibrated, meaning that they tend to
overstate the recidivism risk of women and understate the recidivism risk of
men.

Figure~\ref{fig:anti-classification-compas} illustrates this point, plotting the
observed recidivism rate for men and women in Broward County, Florida as a
function of their gender-blind COMPAS risk scores---a commonly used risk
assessment tool \citep{bao2021s}. In particular, women with a COMPAS score of
seven recidivate about 55\% of the time, whereas men with the same score
recidivate about 65\% of the time. Said differently, women with a score of
seven recidivate approximately as often as men with a score of five, and this
two-point differential persists across the range of scores. By acknowledging the
predictive value of gender in this setting, one could create a decision rule
that detains fewer people (particularly women) while achieving the same public
safety benefits. Conversely, by ignoring this information and basing decisions
solely on the gender-blind risk assessments, one would effectively be subjecting
women to a more stringent risk standard---and potentially harsher
penalties---than men.

As in the case of classification parity, one cannot typically remove protected
attributes from the risk predictions without decreasing utility
\citep[cf.{}][]{manski2022using}; however, the reduction in utility is not
always as large as one might expect \citep{coots2023reevaluating}. In concrete
terms, in our running diabetes example, basing decisions on race-blind risk
estimates necessarily means screening some patients who would have preferred not
to be screened had they been given race-aware risk estimates, and, conversely,
not screening some patients who would have preferred to be screened had they
been given the more complete estimates. We state this result formally below.

\begin{theorem}
\label{thm:blinding}
    Suppose \(0 < t < 1\), where \(t\) is the optimal decision threshold on the
    risk scale, as in Eq.~\eqref{eq:threshold}. Let \(\pi : \C X_u \times \C A
    \to \C X_u\) denote restriction to the unprotected covariates. Let \(\rho(x)
    = \Pr(Y = 1 \mid \pi(X) = \pi(x))\) denote the risk estimated using the
    blinded covariates. Suppose that \(r(x)\) and \(\rho(x)\) have densities on
    \([0,1]\) that are positive in a neighborhood of \(t\). Further suppose that
    there exists \(\epsilon > 0\) such that the conditional variance \(\D
    {Var}(r(X) \mid \rho(X)) > \epsilon\) a.s., where \(r(x)\) is the risk
    estimated from the full set of covariates. Then no blind policy is
    utility-maximizing.
\end{theorem}

The proof of Theorem~\ref{thm:blinding} is given in Appendix~\ref{app:blinding}.
In short, when race, gender, or other protected traits add predictive value---a
condition codified in our assumption that the conditional variance be greater
than \(\epsilon\)---excluding these attributes will in general decrease utility,
both for individuals and in the aggregate.

Basing decisions on blinded risk scores can harm individuals and communities,
for example by failing to flag relatively high-risk Asian patients for diabetes
screening. But it is also important to consider potential harms stemming from
the use of race- and gender-specific risk tools. In medicine, for instance, one
might worry that race-specific risk assessments could encourage doctors and the
public-at-large to advance spurious and pernicious arguments about inherent
differences between race groups. In reality, the differences in diabetes risk we
see are likely due to a complex mix of factors, both environmental and genetic,
and should not be misinterpreted as indicating any causal effects of race.
Indeed, even ``race'' itself is a thorny, socially influenced, concept, that
elides easy definition. Similarly, the use of gender-specific recidivism
estimates could reduce trust in the criminal justice system, giving the
impression that individuals are held to different standards based on their
gender. (Though, as we have seen above, \emph{blinded} risk assessments can
likewise---and perhaps more persuasively---be said to subject individuals to
different standards based on their race and gender.) In some circumstances,
race- and gender-specific risk estimates are even prohibited by law---a topic
we return to in Section~\ref{sec:law}. For these reasons, risk assessments in
medicine, criminal justice, and beyond have generally avoided using race,
gender, and other sensitive demographic attributes. Ultimately, when
constructing risk assessment tools, it is important to acknowledge and carefully
balance both the costs and benefits of blinding in any given circumstance.

\subsubsection{Counterfactual and Path-Specific Fairness}

As discussed in Section~\ref{sec:defn}, counterfactual and path-specific
fairness are generalizations of simple blinding that attempt to account for both
the direct and indirect effects of protected attributes on decisions. Because
the constraints are more stringent, the resulting decrease in utility is
proportionally greater. In particular, in some common settings, path-specific
fairness with \(W = X\) constrains decisions so severely that the only allowable
policies are constant (i.e., \(d(x_1) = d(x_2)\) for all \(x_1, x_2 \in \C X\)).
For instance, in our running admissions example, path-specific fairness requires
admitting all applicants with the same probability, irrespective of academic
preparation or group membership.

To build intuition for this result, we sketch the argument for a finite
covariate space \(\C X\). Given a policy \(d\) that satisfies path-specific
fairness, select \(x^* \in \arg \max_{x \in \C X} d(x)\). By the definition of
path-specific fairness, for any \(a \in \C A\),
\begin{equation}
  \begin{aligned}
  \label{eq:ps-sketch}
    d(x^*) & = \EE[D_{\Pi, A, a} \mid X = x^*] \\
      & = \sum_{x \in \alpha^{-1}(a)} \ d(x) \cdot \Pr(X_{\Pi, A,
      a} = x \mid X = x^*).
  \end{aligned}
\end{equation}
That is, the probability of an individual with covariates \(x^*\) receiving a
positive decision must be the average probability of the individuals with
covariates \(x\) in group \(a\) receiving a positive decision, weighted by the
probability that an individual with covariates \(x^*\) in the real world
\emph{would} have covariates \(x\) counterfactually.

Next, we suppose that there exists an \(a' \in \C A\) such that \(\Pr(X_{\Pi,
A, a'} = x \mid X = x^*) > 0\) for all \(x \in \alpha^{-1}(a')\). In this case,
because \(d(x) \leq d(x^*)\) for all \(x \in \C X\), Eq.~\eqref{eq:ps-sketch}
shows that in fact \(d(x) = d(x^*)\) for all \(x \in \alpha^{-1}(a')\).

Now, let \(x'\) be arbitrary. Again, by the definition of path-specific
fairness, we have that
\begin{align*}
  d(x') & = \EE[D_{\Pi, A, a'} \mid X = x'] \\
    & = \ \sum_{\mathclap{x \in \alpha^{-1}(a')}} \ d(x) \cdot \Pr(X_{\Pi, A,
    a'} = x \mid X = x') \\
    &=  \ \sum_{\mathclap{x \in \alpha^{-1}(a')}} \ d(x^*) \cdot \Pr(X_{\Pi, A,
    a'} = x \mid X = x^*), \\
    &= d(x^*),
\end{align*}
where we use in the third equality the fact \(d(x) = d(x^*)\) for all \(x \in
\alpha^{-1}(a')\), and in the final equality the fact that \(X_{\Pi, A, a'}\) is
supported on \(\alpha^{-1}(a')\).

Theorem~\ref{thm:path_specific} formalizes and extends this argument to more
general settings, where \(\Pr(X_{\Pi, A, a'} = x \mid X = x^*)\) is not
necessarily positive for all \(x \in \alpha^{-1}(a')\). The proof of
Theorem~\ref{thm:path_specific} is in the Appendix, along with extensions to
continuous covariate spaces and a more complete characterization of \(\Pi\)-fair
policies for finite \(\C X\).

\begin{theorem}
\label{thm:path_specific}
  Suppose \(\C X\) is finite and \(\Pr(X = x) > 0\) for all \(x \in \C X\).
  Suppose \(Z = \zeta(X)\) is a random variable such that:
  \begin{enumerate}
    \item \(Z = Z_{\Pi, A, a'}\) for all \(a' \in \C A\),
    \item \(\Pr(X_{\Pi, A, a'} = x' \mid X = x) > 0\) for all \(a' \in \C A\)
      such that \(\alpha(x') = a'\), \(\alpha(x) \neq a'\), and \(x, x' \in \C
      X\) such that \(\zeta(x) = \zeta(x')\).
  \end{enumerate}
  Then, for any \(\Pi\)-fair policy \(d\), with \(W = X\), there exists a
  function \(f\) such that \(d(X) = f(Z)\), i.e., \(d\) is constant across
  individuals having the same value of \(Z\).
\end{theorem}

The first condition of Theorem~\ref{thm:path_specific} holds for any reduced set
of covariates \(Z\) that is not causally affected by changes in \(A\) (e.g.,
\(Z\) is not a descendant of \(A\)). The second condition requires that among
individuals with covariates \(x\), a positive fraction have covariates \(x'\) in
a counterfactual world in which they belonged to another group \(a'\). Because
\(\zeta(x)\) is the same in the real and counterfactual worlds---since \(Z\) is
unaffected by \(A\), by the first condition---we only consider \(x'\) such that
\(\zeta(x') = \zeta(x)\) in the second condition.

In our admissions example, this result shows that, under mild conditions,
causally-fair policies require admitting all applicants with equal probability.
In particular, suppose that among students with a given test score, a positive
fraction achieve any other test score in the counterfactual world in which their
race is altered---as, for instance, we might expect if the individual-level
causal effects are drawn from an (appropriately discretized) normal
distribution. In this case, the empty set of reduced covariates---formally
encoded by setting \(\zeta\) to a constant function---satisfies the conditions
of Theorem~\ref{thm:path_specific}. The theorem then implies that under any
\(\Pi\)-fair policy, every applicant is admitted with equal probability. (We
motivated our admissions example by assuming that only a fraction \(b < 1\) of
applicants could be admitted; however, Theorem~\ref{thm:path_specific} holds
irrespective of the budget, and, in particular, when \(b = 1\), and so we
discuss this result together with our others on unconstrained decision making as
a natural extension of blinding.)

Even when decisions are not perfectly uniform lotteries,
Theorem~\ref{thm:path_specific} suggests that enforcing \(\Pi\)-fairness can
lead to unexpected outcomes. For instance, suppose we modify our admissions
example to additionally include age as a covariate that is causally unconnected
to race---as some past work has done. In that case, \(\Pi\)-fair policies would
admit students based on their age alone, irrespective of test score or race.
Although in some cases such restrictive policies might be desirable, this strong
structural constraint implied by \(\Pi\)-fairness appears to be a largely
unintended consequence of the mathematical formalism.

The conditions of Theorem~\ref{thm:path_specific} are relatively mild, but do
not hold in every setting. Suppose that in our admissions example it were the
case that \(T_{\Pi, A, a_0} = T_{\Pi, A, a_1} + c\) for some constant
\(c\)---that is, suppose the effect of intervening on race is a constant change
to an applicant's test score. Then the second condition of
Theorem~\ref{thm:path_specific} would no longer hold for a constant \(\zeta\).
Indeed, any multiple-threshold policy in which \(t_{a_0} = t_{a_1} + c\) would
be \(\Pi\)-fair. In practice, though, such deterministic counterfactuals would
seem to be the exception rather than the rule. For example, it seems reasonable
to expect that test scores would depend on race in complex ways that induce
considerable heterogeneity. Lastly, we note that \(W \neq X\) in some variants
of path-specific fairness \citep[e.g.,][]{zhang2018fairness, nabi2018fair}, in
which case Theorem~\ref{thm:path_specific} does not apply. Although, in that
case, path-specific fairness is still typically incompatible with optimal
decision-making, as shown in Theorem~\ref{thm:dist}.

\section{Equitable Decisions in the Presence of Externalities}
\label{sec:presence}

We have thus far considered cases where there is largely agreement on the
utility of different decision policies. In that setting, we showed that
maximizing utility is at odds with various mathematical formalizations of
fairness. We further argued that these results illustrate weaknesses in the
formalizations themselves, since deviating from utility-maximizing polices in
that setting can harm both individuals and groups---as seen in our diabetes
screening example.

Agreement on the utility, however, is perhaps the exception rather than the
rule. One could indeed argue that the value of mathematical formalizations of
fairness is their ability to arbitrate between competing definitions of utility.
Here we critically examine that perspective. We show, in analog to our previous
results, that even when it is unclear how to balance competing priorities,
enforcing existing fairness constraints typically leads to worse outcomes on
each dimension. For instance, in our running college admissions example,
policies constrained to satisfy various fairness constraints will typically
require admitting a student body that is both less academically prepared and
less diverse, relative to alternative policies that violate these mathematical
fairness definitions.

We start, in Section~\ref{sec:geometry}, by examining our college admissions
example in detail, illustrating in geometric terms how existing fairness
definitions can lead to problematic admissions policies. Then, in
Section~\ref{sec:theory}, we develop our formal theory of equitable decision
making in the presence of externalities. The mathematics necessary to establish
our key results are significantly deeper than what we have needed thus far, but
our high-level message is the same: enforcing several formal notions of fairness
leads to policies that can paradoxically harm the very groups that they were
designed to protect.

\subsection{The Geometry of Fair Decision Making}
\label{sec:geometry}

To build intuition about the limitations of popular definitions of fairness, we
return to our running example on college admissions. In that setting, we imagine
an admissions committee debating the merits of different admissions policies. In
particular, we imagine disagreement within the committee over how best to
balance two competing objectives: academic preparation (operationalized, e.g.,
in terms of the high school grades and standardized test scores of admitted
students) and class diversity (e.g., the number of admitted applicants from
marginalized groups).

We assume that our hypothetical committee members all agree that more (total)
academic preparedness and more class diversity are better. Thus, in the absence
of any resource constraints (with \(b = 1\), as is approximated in some online
courses), the university could admit all applicants, maximizing both the number
of admitted students from marginalized groups and also the total academic
preparedness of the admitted class. But given limits on the number of students
who can be admitted (i.e., \(b < 1\)), one must make difficult choices on whom
to admit, with reasonable and expected disagreement on how much to trade one
dimension for another. The trade-offs in decision making are most acute when the
budget \(b < 1\), and for this reason we focus here on that case.

In light of these trade-offs, one might turn to the myriad formal fairness
criteria we have discussed to ensure admissions decisions are equitable. Many of
the fairness definitions we consider make reference to a distinguished outcome
\(Y\). In our example, we can imagine this outcome corresponds to college degree
attainment, an \emph{ex post} measure of academic preparedness. In the case of
causal fairness definitions, we could take \(Y(1)\) to mean degree attainment if
the student were admitted, and \(Y(0)\) to be degree attainment if the student
were not admitted, with the understanding that a student who is not admitted
could potentially attend and graduate from another university. For example,
satisfying counterfactual predictive parity requires that among rejected
applicants, the proportion who would have attained a college degree, had they
been accepted, is equal across race groups. In these cases, we imagine academic
preparedness is some student-level measure that connects observables
\(X\)---upon which the committee must make their admissions decisions---to
(potential) outcomes \(Y\). For example, the ``academic index'' \(m(x)\) might
be a prediction of \(Y(1)\) given \(X\) based on historical data, or, more
generally, could encode committee preferences for both academic preparation and
participation in extracurricular activities, among other factors.

The key point of our informal discussion thus far is that we assume committee
members would like to enact an admissions policy \(d\) that balances two
competing objectives. First, they would like a policy that leads to large
\(m(x)\), i.e., they would like \(\B E[m(X) \cdot d(X)]\) to be big, where
\(m(x)\) is some quantity that may, for example, encode academic preparedness
and other preferences. Second, the committee would like large diversity, i.e.,
they would like \(\B E[\B 1_{\alpha(X) = a_1} \cdot d(X)]\) to be big, where
\(a_1\) corresponds to some target group of interest. All committee members
would like more of each dimension, but, given the budget constraint, it is in
general impossible to maximize both dimensions simultaneously, leading to the
inherent trade-offs we consider in this section.

We now explore the consequences of imposing additional fairness constraints on
our college admissions example, as given by the causal DAG in
Figure~\ref{fig:dag}, via a simulation study of one million hypothetical
applicants, for one quarter of whom (\(b = \tfrac 1 4\)) seats are allocated. In
particular, in the hypothetical pool of applicants we consider, applicants in
the target race group \(a_1\) have, on average, fewer educational opportunities
than those applicants in group \(a_0\), which leads to lower average academic
preparedness, as well as lower average test scores. We define the ``academic
index'' \(m(x)\) of applicants to be the estimated probability that an applicant
will graduate if admitted, based on their observed test score and race. See
Appendix~\ref{appendix:example} for additional details, including the specific
structural equations we use in the simulation.

\begin{figure}[t]
  \begin{center}
    \includegraphics[width=0.9\linewidth,clip]{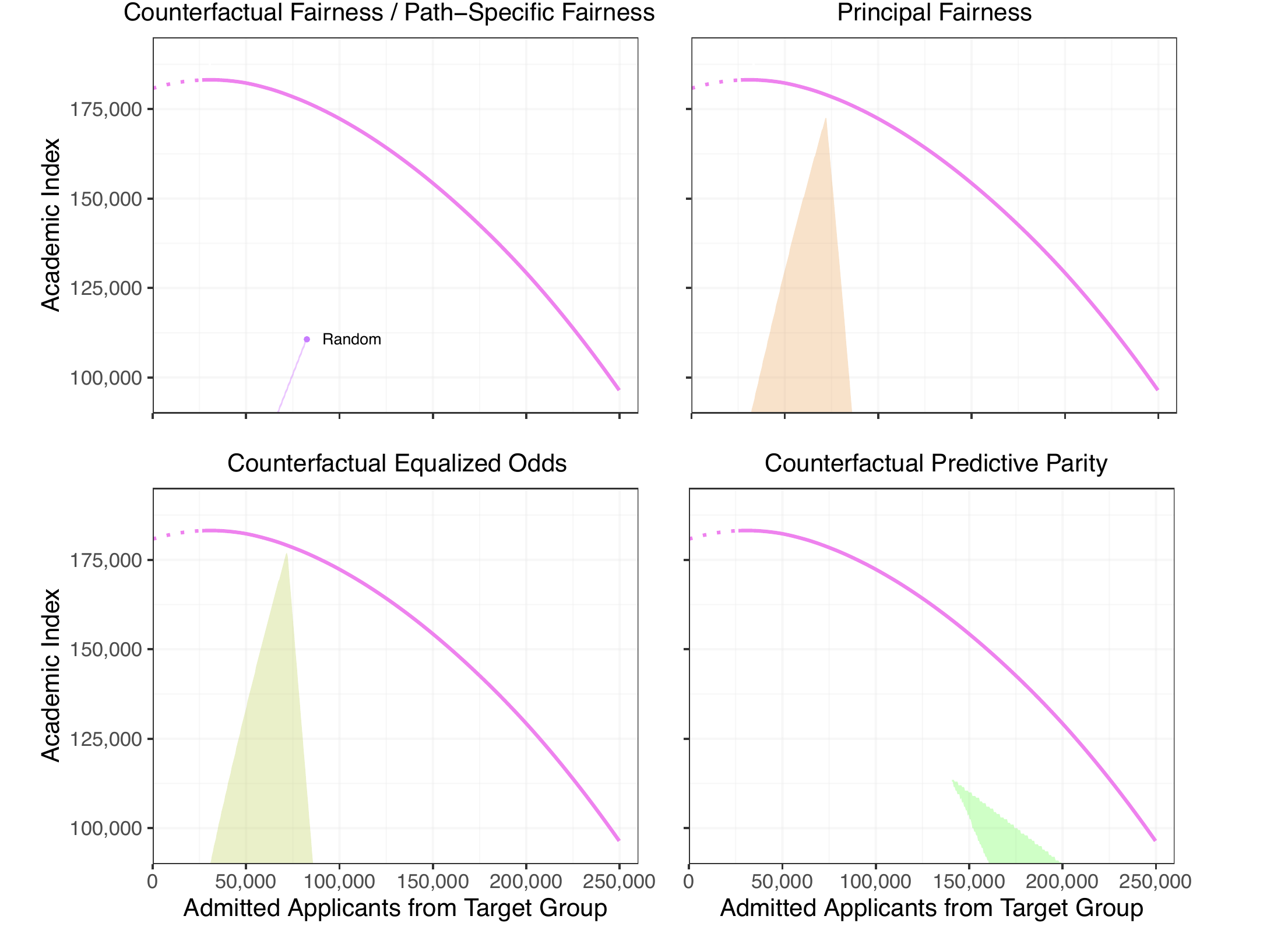}
  \end{center}
  \caption{%
    The geometry of fairness constraints, in an illustrative example of college
    admissions. Points under the purple curve correspond to all feasible
    policies---given the budget constraint---whereas the shaded regions
    correspond to feasible policies that satisfy various formal definitions of
    fairness. (For path-specific fairness, we set \(\Pi\) equal to the single
    path \(A \rightarrow E \rightarrow T \rightarrow D\) highlighted in
    Figure~\ref{fig:dag}, and set \(W = X\).) For each definition, the
    constrained policies lie strictly under the purple curve, meaning there are
    alternative, unconstrained, feasible policies that simultaneously achieve
    greater student-body diversity (indicated by the \(x\)-axis) and greater
    academic preparedness (indicated by the \(y\)-axis). The solid segments of
    the purple lines correspond to policies on the Pareto frontier---for which
    one cannot simultaneously increase both diversity and academic
    preparedness. The points labeled ``random'' in the upper left-hand corner
    correspond to policies generated by random lotteries in which each
    individual is admitted with equal probability, in accordance with
    Theorem~\ref{thm:path_specific}.
  }
\label{fig:frontier}
\end{figure}

Each of the panels in Figure~\ref{fig:frontier} illustrates the geometry of
fairness constraints for five different formal notions of fairness described in
Section~\ref{sec:defn}: counterfactual fairness, path-specific fairness,
principal fairness, counterfactual equalized odds, and counterfactual predictive
parity. The vertical axes of each panel correspond to aggregate academic index
and the horizontal axes to the number of admitted applicants from the target
group. The purple lines trace out the boundary of the set of feasible policies,
with points on or below the curves achievable by policies that adhere to the
budget constraint. Policies lying strictly below the purple curves (or,
similarly, on the dashed segments of the purple curves) are ``Pareto
dominated,'' meaning that one can find feasible alternatives that are larger on
both of the depicted axes (i.e., academic index and diversity). Since we have
assumed committee members prefer higher values on each dimension, their
effective choice set consists of those policies on the solid purple
segments---the ``Pareto frontier.'' Committee members may still disagree over
which policy on the frontier to adopt. But for any policy not on the frontier,
there is a feasible policy above and to the right of it, which is thus preferred
by every member of the committee.

Finally, the shaded regions indicate the set of feasible policies constrained to
satisfy each of the fairness definitions. (In Appendix~\ref{sec:construction},
we show that these feasibility regions can be computed by solving a series of
linear programs.) In each case, the constrained regions do not intersect the
Pareto frontier, and so there is an alternative, unconstrained feasible policy
that simultaneously achieves more student-body diversity and an overall higher
academic index. For example, in the case of policies satisfying counterfactual
or path-specific fairness, shown in the upper left panel, the set of feasible
policies lie on a single line segment. That structure follows from
Theorem~\ref{thm:path_specific}, since the only policies satisfying either of
these notions of fairness in our setting are ones that admit all students with a
constant probability, irrespective of their covariates. While not as extreme,
the other fairness definitions similarly restrict the space of feasible policies
in severe ways, as shown in the remaining panels. These results illustrate that
constraining decision-making algorithms to satisfy popular definitions of
fairness can have unintended consequences, and may even harm the very groups
they were ostensibly designed to help.

Our discussion in this section aimed to highlight the geometry of ``fair''
decision policies and their consequences in the context of a simple motivating
example. We next show that these qualitative findings are guaranteed to hold
much more generally.

\subsection{A Formal Theory of Fairness in the Presence of Externalities}
\label{sec:theory}

Our simulation above showed that policies satisfying one of the mentioned
fairness definitions are suboptimal, in the sense that they constrain one to a
portion of the feasible region in which policies could be improved along both
dimensions of interest. As was the case in the absence of trade-offs in
Section~\ref{sec:inframarginality}, the phenomenon occurring in our simulation
is true much more generally. To understand why, we begin by isolating and
formalizing the relevant mathematical properties of our example. To generalize
our setting in Section~\ref{sec:absence}, we consider arbitrary utility
functions of the form \(u : \C X \to \B R\). As before, for a function \(u\) and
decision policy \(d\), we write \(u(d) = \EE[d(X) \cdot u(X)]\) to denote the
expected utility of decision policy \(d(x)\) under the utility \(u\). An
important constraint on the admissions committee was the fact that their
admissions decisions could not, in expectation, exceed the budget.

\begin{definition}
  For a budget \(b\), we say a decision policy \(d(x)\) is \emph{feasible} if
  \(\EE[d(X)] \leq b\).
\end{definition}

A key feature of the college admissions example is that despite some level of
uncertainty regarding the ``true'' utility---i.e., exactly how to trade off
between its objectives---the committee knows what its objectives are: to
increase the academic index and diversity of the incoming class. One way to
encode this kind of uncertainty is to consider a set \(\C U\) consisting of all
``reasonable'' ways of trading off between the objectives. While the utilities
need not be the same, they should be consistent, in the sense that conditional
on an applicant's group membership, all of the utilities should ``agree'' that a
higher academic index is better.

\begin{definition}
\label{defn:eight}
  We say that a set of utilities \(\mathcal{U}\) is \emph{consistent modulo
  \(\alpha\)} if, for any \(u, u' \in \mathcal{U}\):
  \begin{enumerate}
    \item For any \(x\), \(\sgn(u(x)) = \sgn(u'(x))\);
    \item For any \(x_1\) and \(x_2\) such that \(\alpha(x_1) = \alpha(x_2)\),
      \(u(x_1) > u(x_2)\) if and only if \(u'(x_1) > u'(x_2)\).
  \end{enumerate}
\end{definition}

A second relevant feature of the admissions problem is that certain policies
were strictly better from the admissions committee's perspective, despite their
uncertainty about the exact form of their utility. The notion that one policy is
better than another \emph{regardless} of the exact form of the utility is
formalized by Pareto dominance.

\begin{definition}
\label{defn:pareto}
  Suppose \(\mathcal{U}\) is a collection of utility functions. A decision
  policy \(d\) is \emph{Pareto dominated} if there exists a feasible alternative
  \(d'\) such that \(u(d') \geq u(d)\) for all \(u \in \mathcal{U}\), and there
  exists \(u' \in \mathcal{U}\) such that \(u'(d') > u'(d)\). A policy \(d\) is
  \emph{strongly Pareto dominated} if there exists a feasible alternative \(d'\)
  such that \(u(d') > u(d)\) for all \(u \in \mathcal{U}\). A policy \(d\) is
  \emph{Pareto efficient} if it is feasible and not Pareto dominated, and the
  \emph{Pareto frontier} is the set of Pareto efficient policies.
\end{definition}

As discussed above and in Section~\ref{sec:thresholds}, in the absence of
trade-offs, optimal decision policies take the simple form of threshold
policies. The existence of trade-offs broadens the range of forms a Pareto
efficient policy can take. Even so, for consistent collections of utilities, the
Pareto efficient policies take a closely related form.

\begin{proposition}
\label{prop:threshold}
  Suppose \(\mathcal{U}\) is a set of utilities that is consistent modulo
  \(\alpha\). Then any Pareto efficient decision policy \(d\) is a
  \emph{multiple-threshold policy}. That is, for any \(u \in \mathcal{U}\),
  there exist group-specific constants \(t_{a} \geq 0\) such that, a.s.:
  \begin{equation}
    d(x)=
    \begin{cases}
      1 & u(x) > t_{\alpha(x)}, \\
      0 & u(x) < t_{\alpha(x)}. \\
    \end{cases}
  \end{equation}
\end{proposition}
The proof of Proposition~\ref{prop:threshold} is in the Appendix.\footnote{%
  \label{fn:thresholds}%
  In the statement of the proposition, we do not specify
  what happens at the thresholds \(u(x) = t_{\alpha(x)}\) themselves, as one can
  typically ignore the exact manner in which decisions are made at the
  threshold. Specifically, given a multiple-threshold policy \(d\), we can
  construct a standardized multiple-threshold policy \(d'\) that is constant
  within group at the threshold (i.e., \(d'(x) = c_{\alpha(x)}\) when \(u(x) =
  t_{\alpha(x)}\)), and for which: (1) \(\EE[d'(X)|A] = \EE[d(X)|A]\); and (2)
  \(u(d') = u(d)\). In our running example, this means we can standardize
  multiple-threshold policies so that applicants at the threshold are admitted
  with the same group-specific probability.
}

\subsubsection{Fairness definitions with many constraints}
\label{sec:many_constraints}

All of the definitions we study in this section prominently feature causal
quantities, but the important quality driving our analysis in this section is
that each definition imposes many constraints. For instance, counterfactual
equalized odds requires that
\begin{equation*}
  \Pr(D = 1 \mid A = a, Y(1) = y) = \Pr(D = 1 \mid Y(1) = y)
\end{equation*}
for every outcome \(y\).

Theorem~\ref{thm:dist} shows that for \emph{almost every} joint
distribution of \(X\), \(Y(0)\), and \(Y(1)\) such that \(u(X)\) has a density,
any feasible decision policy satisfying counterfactual equalized odds or
conditional principal fairness is Pareto dominated. Similarly, for almost every
joint distribution of \(X\) and \(X_{\Pi, A, a}\), we show that feasible
policies satisfying path-specific fairness---including counterfactual
fairness---are Pareto dominated. (The analogous statements for counterfactual
predictive parity, equalized false positive rates, and demographic parity are
not true; we return to this point in Section~\ref{sec:few_constraints}.) That
is, we show that, for a typical joint distribution, any feasible policy
satisfying the fairness definitions enumerated above cannot have the form of a
multiple-threshold policy. To prove this result, we make relatively mild
restrictions on the set of distributions and utilities we consider to exclude
degenerate cases, as formalized by Definition~\ref{defn:fine}.

\begin{definition}
\label{defn:fine}
  Let \(\C G\) be a collection of functions from \(\C Z\) to \(\B R^d\) for some
  set \(\C Z\). We say that a distribution of \(Z\) on \(\C Z\) is \emph{\(\C
  G\)-fine} if \(g(Z)\) has a density for all \(g \in \C G\).
\end{definition}

In particular, \(\C U\)-fineness ensures that the distribution of \(u(X)\) has a
density. In the absence of \(\C U\)-fineness, corner cases can arise in which an
especially large number of policies may be Pareto efficient, in particular when
\(u(X)\) has large atoms and \(X\) can be used to predict the potential outcomes
\(Y(0)\) and \(Y(1)\) even after conditioning on \(u(X)\). See
Proposition~\ref{prop:counterexample} in the Appendix for details.

\begin{theorem}
\label{thm:dist}
  Suppose \(\C U\) is a set of utilities consistent modulo \(\alpha\). Further
  suppose that for all \(a \in \C A\) there exist a \(\C U\)-fine distribution
  of \(X\) and a utility \(u \in \C U\) such that \(\Pr(u(X) > 0, A = a) > 0\),
  where \(A = \alpha(X)\). Then,
  \begin{itemize}
    \item For almost every \(\C U\)-fine distribution of \(X\) and \(Y(1)\), any
      feasible decision policy satisfying counterfactual equalized odds is
      strongly Pareto dominated.
    \item If \(|\img(\omega)| < \infty\) and there exists a \(\C U\)-fine
      distribution of \(X\) such that \(\Pr(A = a, W = w) > 0\) for all \(a \in
      \C A\) and \(w \in \img(\omega)\), where \(W = \omega(X)\), then, for
      almost every \(\C U\)-fine joint distribution of \(X\), \(Y(0)\), and
      \(Y(1)\), any feasible decision policy satisfying conditional principal
      fairness is strongly Pareto dominated.
    \item If \(|\img(\omega)| < \infty\) and there exists a \(\C U\)-fine
      distribution of \(X\) such that \(\Pr(A = a, W = w_i) > 0\) for all \(a
      \in \C A\) and some distinct \(w_0, w_1 \in \img(\omega)\), then, for
      almost every \(\C U^{\C A}\)-fine joint distributions of \(A\) and the
      counterfactuals \(X_{\Pi, A, a'}\), any feasible decision policy
      satisfying path-specific fairness is strongly Pareto dominated.\footnote{%
        Here, \(u^\C A : (x_a)_{a \in \C A} \mapsto (u(x_a))_{a \in \C A}\) and
        \(\C U^{\C A}\) is the set of \(u^{\C A}\) for \(u \in \C U\), i.e.,
        component-wise application of \(u\) to elements of \(\mathcal
        X^{\mathcal A}\). In other words, the requirement is that the joint
        distribution of the \(u(X_{\Pi, A, a})\) has a density.
      }
  \end{itemize}
\end{theorem}

The proof of Theorem~\ref{thm:dist} is given in the Appendix. At a high level,
the proof proceeds in three steps, which we outline below using the example of
counterfactual equalized odds. First, we show that for almost every fixed \(\C
U\)-fine joint distribution \(\mu\) of \(X\) and \(Y(1)\) there is at most one
policy \(d^*(x)\) satisfying counterfactual equalized odds that is not strongly
Pareto dominated. To see why, note that for any specific \(y_0\), since
counterfactual equalized odds requires that \(D \indep A \mid Y(1) = y_0\),
setting the threshold for one group determines the thresholds for all the
others; the budget constraint then can be used to fix the threshold for the
original group. Second, we construct a ``slice'' around \(\mu\) such that for
any distribution \(\nu\) in the slice, \(d^*(x)\) is still the only policy that
can potentially lie on the Pareto frontier while satisfying counterfactual
equalized odds. We create the slice by strategically perturbing \(\mu\) only
where \(Y(1) = y_1\), for some \(y_1 \neq y_0\). This perturbation moves mass
from one side of the thresholds of \(d^*(x)\) to the other. Due to
inframarginality, this perturbation typically breaks the balance requirement \(D
\indep A \mid Y(1) = y_1\) for almost every \(\nu\) in the slice. Finally, we
appeal to the notion of prevalence to stitch the slices together, showing that
for almost every distribution, any policy satisfying counterfactual equalized
odds is strongly Pareto dominated. Analogous versions of this general argument
apply to the cases of conditional principal fairness and path-specific
fairness.\footnote{%
  This argument does not depend in an essential way on the definitions being
  causal. In Corollary~\ref{cor:eo} in the Appendix, we show an analogous result
  for the non-counterfactual version of equalized odds.
}
We note that the conditions of Theorem~\ref{thm:dist} are sufficient, rather
than necessary, meaning that the conclusion of the theorem may---and, indeed, we
expect will---hold even in some cases where the conditions are not satisfied. In
particular, we note that this proof technique prevents the conditions of
Theorem~\ref{thm:dist} from holding when \(A\) factors through \(W\) and, in
particular, when \(W = X\). Although, when \(X = W\),
Theorem~\ref{thm:path_specific} shows that under slightly different conditions,
a much stronger result holds.

To bring our discussion full circle, we now map Theorem~\ref{thm:dist} onto the
motivation offered in Section~\ref{sec:geometry}. Recall that the admissions
committee knew that given the opportunity, it preferred policies that increased
both the overall academic index of its admitted class, and policies that
resulted in more students being admitted from the target group. In other words,
we imagine that members of the admissions committee have utilities \(u^*\) of
the form\footnote{%
  Strictly speaking, we are saying that members of the admissions committee,
  rather than having an aggregate utility---which, as we have considered so far,
  has the form \(\B E[u(X) \cdot d(X)]\)---has a utility on aggregate
  \emph{outcomes}.
}
\begin{equation}
\label{eq:agg_util}
  u^*(d) = v \left(\B E[m(X) \cdot d(X)], \B E[\B 1_{\alpha(X) = a_1} \cdot
  d(X)] \right),
\end{equation}
where, as above, \(m(x)\) denotes the academic index of an applicant with
covariates \(X = x\), and \(v\) increases in both coordinates.
Corollary~\ref{cor:committee} establishes the inherent incompatibility of such
preferences with the formal fairness criteria we have been considering.

\begin{corollary}
\label{cor:committee}
    Consider a utility of the form given in Eq.~\eqref{eq:agg_util},
    where \(v\) is monotonically increasing in both coordinates and \(m(x) \geq
    0\). Then, under the same hypotheses as in
    Theorem~\ref{thm:dist},\footnote{%
        The full statement is given in Appendix~\ref{app:committee}.
    }
    for almost every joint distribution, no utility-maximizing decision-policy
    satisfies counterfactual equalized odds, conditional principal fairness, or
    path-specific fairness.
\end{corollary}

Lastly, while, in general, one's decision policy can depend only on the
covariates known at the time of the decision, in some cases, the restriction
that \(u(x)\) be a function of \(x \in \C X\) alone may be too restrictive; the
connection between an individual having covariates \(X = x\) and our utility may
depend also on the relationship between \(X\) and \(Y\). For instance, in the
admissions example, the admissions committee may value high test scores and
extracurriculars not, e.g., as \emph{per se} measures of academic merit, but
rather instrumentally insofar as they are connected to whether an applicant will
eventually graduate. However, allowing \(u\) to depend on both \(x\) and \(y\)
greatly complicates the underlying geometry of the problem. Proving
Theorem~\ref{thm:dist} in this more general setting remains an open problem.
However, intuition from finite-dimensions---where more powerful
measure-theoretic tools are available---suggests that the result remains true in
the more general setting. For example, Proposition~\ref{prop:main_cee} presents
a version of this result over a natural, finite-dimensional family of
distributions.

\begin{proposition}
\label{prop:main_cee}
  Suppose \(\C A = \{a_0, a_1\}\), and consider the family \(\C U\) of utility
  functions of the form
  \begin{equation*}
    u(x) = r(x) + \lambda \cdot \B 1_{\alpha(x) = a_1},
  \end{equation*}
  indexed by \(\lambda \geq 0\), where \(r(x) = \EE[Y(1) \mid X = x]\). For
  almost every \((\alpha_0, \beta_0, \alpha_1, \beta_1) \in \B R^4_{>0}\), if
  the conditional distributions of \(r(X)\) given \(A\) are beta distributed
  with
  \begin{equation*}
    r(X) \mid A = a_i \sim \bbeta(\alpha_i, \beta_i),
  \end{equation*}
  then any policy satisfying counterfactual equalized odds is strongly
  Pareto dominated.
\end{proposition}

\subsubsection{Fairness definitions with few constraints}
\label{sec:few_constraints}

We conclude this analysis by considering equalized false positive rates,
demographic parity, and counterfactual predictive parity. These fairness notions
are less demanding than the notions considered above, in that they introduce
only ``one'' additional constraint, e.g., that \(Y(1) \indep A \mid D = 0\), in
the case of counterfactual predictive parity. Since the budget introduces a
second constraint, and the form of a multiple-threshold policy allows for a
degree of freedom in each group, the number of constraints and the number of
degrees of freedom are equal---as opposed to the causal fairness definitions
covered by Theorem~\ref{thm:dist}, in which the constraints outnumber the
degrees of freedom. As such, it is possible in some instances to have a policy
on the Pareto frontier that satisfies these conditions; though see
Section~\ref{sec:case-study} for discussion about why such policies are still
often at odds with broader goals.

However, it is not \emph{always} possible to find a point on the Pareto frontier
satisfying these definitions. In Proposition~\ref{prop:pred_parity}, we show
that counterfactual predictive parity cannot lie on the Pareto frontier in some
common cases, including our example of college admissions. In that setting, when
the target group has lower average graduation rates---a pattern that often
motivates efforts to actively increase diversity---decision policies constrained
to satisfy counterfactual predictive parity are Pareto dominated. The proof of
the proposition is in Appendix~\ref{app:pred_parity}.

\begin{proposition}
\label{prop:pred_parity}
  Suppose \(\C A = \{a_0, a_1\}\), and consider the family \(\C U\) of utility
  functions of the form
  \begin{equation*}
    u(x) = r(x) + \lambda \cdot \B 1_{\alpha(x) = a_1},
  \end{equation*}
  indexed by \(\lambda \geq 0\), where \(r(x) = \EE[Y(1) \mid X = x]\). Suppose
  the conditional distributions of \(r(X)\) given \(A\) are beta distributed,
  i.e.,
  \begin{equation*}
    r(X) \mid A = a \sim \bbeta(\mu_a, v),
  \end{equation*}
  with \(\mu_{a_0} > \mu_{a_1}\) and \(v > 0\).\footnote{%
    Here we parameterize the beta distribution in terms of its mean \(\mu\) and
    sample size \(v\). In terms of the common, alternative \(\alpha\)-\(\beta\)
    parameterization, \(\mu = \alpha / (\alpha + \beta)\) and \(v = \alpha +
    \beta\).
  }
  Then any policy satisfying counterfactual predictive parity is strongly Pareto
  dominated.
\end{proposition}

\section{A Path Forward}
\label{sec:forward}

We have thus far worked to clarify some of the statistical limitations of
existing mathematical definitions of fairness. We have argued that in many cases
of interest, these definitions can ultimately do more harm than good, hurting
even those individuals that these notions of fairness were ostensibly designed
to help.

We end on a more optimistic note, charting out a potential path toward designing
more equitable algorithms. To do so, we start, in Section~\ref{sec:law} by
reviewing conceptions of discrimination in law and economics, and, in
particular, we contrast process-oriented and outcome-oriented notions of
fairness. Whereas the computer science literature is dominated by
process-oriented, \emph{deontological} definitions of fairness, we see more
promise in adopting an outcome-oriented, \emph{consequentialist} approach
represented by the utilitarian analysis we have described above. In
Section~\ref{sec:design}, we enumerate and discuss four issues that we feel are
critical in developing equitable algorithms: (1) balancing inherent trade-offs
in decision problems; (2) assessing calibration; (3) selecting the inputs and
targets of prediction; and (4) designing data collection strategies. Finally, in
Section~\ref{sec:case-study}, we illustrate how to grapple with these
considerations in a case study of complex medical care, motivated by work
from~\citet{obermeyer2019dissecting}.

\subsection{Competing Notions of Ethical Decision Making: Process vs. Outcomes}
\label{sec:law}

There are many distinct but related understandings of ethical decision making in
law, economics, philosophy, and beyond. One key dimension on which we organize
these notions is the extent to which they consider the \emph{process} through
which decisions are made versus the \emph{outcomes} that those decisions render.

The dominant legal doctrine of discrimination in the United States treats
explicit race- and gender-based decisions with heightened scrutiny. The Equal
Protection Clause of the U.S. Constitution's Fourteenth Amendment restricts
government agencies from adopting policies that explicitly reference legally
protected categories, and myriad federal and state \emph{disparate treatment}
statutes similarly constrain a variety of private actors. Conversely, policies
that do not explicitly consider legally protected traits---or obvious
proxies---are generally deemed not to violate disparate treatment principles.
Formally, it is lawful to use legally protected attributes in a limited way to
further a compelling government interest, but, in practice, such exceptions are
few and far between. Until recently, the prime example of a race-conscious
policy passing legal muster was affirmative action in college
admissions~\citep{fisher2016}. However, in 2023, the U.S. Supreme Court barred
the explicit consideration of race in admissions decisions \citep{2023students}.

Disparate treatment doctrine has evolved over time, and reflects ongoing debates
about the role of \emph{classification} (use of protected traits, a
process-oriented, deontological notion) versus \emph{subordination} (subjugation
of disadvantaged groups, an outcome-oriented notion) in discrimination cases
\citep{fiss1976groups}. Some legal scholars have argued that courts, even when
formally applying anti-classification criteria, are often sympathetic to the
potential effects of judgments on social stratification, indicating tacit
concern for anti-subordination~\citep{%
  colker1986anti, balkin2003american, siegel2003equality%
}.
Others, though, have noted that such judicial support for anti-subordination
appears to be waning~\citep{nurse2014anti}. At a high level, we thus view modern
disparate treatment law as primarily interested in process over outcomes, though
these debates illustrate that the two concepts cannot be perfectly separated.

In contrast to process-oriented disparate treatment principles, the economics
literature distinguishes between two outcome-focused, consequentialist
rationales for explicitly considering race, gender, and other protected traits:
taste-based and statistical. With \emph{taste-based
discrimination}~\citep{becker1957}, decision makers act as if they have a
preference or ``taste'' for bias, sacrificing profit to avoid certain
transactions. This includes, for example, an employer who forfeits financial
gain by failing to hire exceptionally qualified minority applicants. But, in
contrast to legal reasoning, the economic argument against taste-based
discrimination is not that decisions are based on race \emph{per se}, but rather
because consideration of race leads to worse outcomes: a loss of profit. With
\emph{statistical discrimination}~\citep{arrow1973,phelps1972}, decision makers
explicitly consider protected attributes in order to optimally achieve some
non-prejudicial goal. For example, profit-maximizing auto insurers may charge a
premium to male drivers to account for gender differences in accident rates.
Despite their differing outcome-based justifications, both taste-based and
statistical discrimination are often considered legally problematic as they
explicitly consider race, in violation of disparate treatment laws.\footnote{%
  In the case of auto insurance specifically, some states (including California,
  Hawaii, Massachusetts, Maine, Michigan, North Carolina, and
  Pennsylvania)---though not all---have barred the use of gender in pricing
  policies.
}

As the above insurance example and our running diabetes example illustrate, one
might consider it acceptable to base decisions in part on legally protected
traits when doing so leads to good outcomes. Conversely, whereas
process-oriented disparate treatment principles generally deem race-blind
policies acceptable, one might declare such blind policies problematic if they
lead to bad outcomes. Indeed, under the statutory \emph{disparate impact}
standard, a practice may be deemed discriminatory if it has an unjustified
adverse effect on legally protected groups, even in the absence of explicit
categorization~\citep{barocas2016big}.\footnote{%
  The legal doctrine of disparate impact stems largely from federal statutes,
  not constitutional law, and applies only in certain contexts, such as
  employment (via Title VII of the 1964 Civil Rights Act) and housing (via the
  Fair Housing Act of 1968). Apart from federal statutes, some states have
  passed more expansive disparate impact laws, including Illinois and
  California. The distinction between statutory and constitutional rules is
  particularly relevant here, as there is debate among scholars over whether
  disparate impact laws violate the Equal Protection Clause and are thus
  unconstitutional~\citep{primus2003equal}. There is also debate over whether
  disparate impact law is motivated primarily by an interest in banning bad
  \emph{outcomes}, or seeks to provide an alternative pathway for ferreting out
  bad \emph{intent}, when actors may mask animus with race-neutral
  policies~\citep{watson}. But, regardless of its underlying justification,
  disparate impact law is formally focused on outcomes, not intent or
  classification, and so we view it as an outcomes-focused principle.
}
The disparate impact doctrine was formalized in \emph{Griggs v. Duke Power Co.},
a 1971 U.S. Supreme Court case. In 1955, the Duke Power Company mandated that
employees have a high school diploma to be considered for promotion, which, in
practice, severely limited the eligibility of Black employees. The Court found
that this facially race-neutral requirement had little relation to job
performance, and accordingly deemed it to have an unjustified---and
illegal---disparate impact. The Court noted that the employer's motivation for
instituting the policy was irrelevant to its decision; even if enacted without
discriminatory purpose, the policy was deemed discriminatory in its effects and
hence illegal. However, disparate impact law does not prohibit \emph{all} group
differences produced by a policy---the law only prohibits \emph{unjustified}
disparities. For example, if, hypothetically, the high-school diploma
requirement in \emph{Griggs} were shown to be necessary for job success, the
resulting disparities would be legal.

On the spectrum from process- to outcome-based understandings of discrimination,
we view the formal, axiomatic fairness definitions described in
Section~\ref{sec:defn} as reflecting a largely process-based orientation.
Blinding and its more stringent causal variants---counterfactual fairness and
path-specific fairness---can be viewed as descendants of disparate treatment
considerations, as they seek to remove the effects of race and other protected
attributes on decisions. The remaining definitions---for example, those that aim
to equalize error rates across groups---do explicitly reference an outcome
\(Y\), but they do so in a way that seems largely disconnected from the
consequences one might naturally consider. As we have argued, whether error
rates are equal across groups has more to do with the structure of
group-specific risk distributions than with whether decisions lead to good or
bad outcomes for group members. For example, in our college admissions example,
enforcing various formal notions of fairness would, in theory, typically lead to
student bodies that are both less diverse and less academically prepared than
those resulting from feasible alternatives not constrained to satisfy these
notions.

One might, on principle, favor certain process-based understandings of
discrimination over outcome-based notions. One might even adopt a
meta-consequentialist position, and argue that procedural considerations (e.g.,
ensuring medical decisions are blind to race) engender trust and in turn bring
about better downstream outcomes. In many cases, though, the ethical
underpinnings of popular mathematical definitions of fairness have not been
clearly articulated. Absent such justification, we advocate for an approach that
more directly engages with the real-world costs and benefits of different
decision policies, a perspective that we outline in more detail in the remaining
sections.

\subsection{Designing Equitable Algorithms}
\label{sec:design}

A key advantage of the dominant axiomatic approach to algorithmic fairness is
that it can be readily applied across contexts, with little domain-specific
knowledge. One can build automated tests to check whether any predictive
algorithm satisfies various formal fairness desiderata, and even automatically
modify algorithms to ensure that they do satisfy a specific fairness
criterion~\citep[e.g.,][]{cotter2019two, weerts2023fairlearn}. But, as we have
argued, this approach is often at odds with improving well-being, including for
disadvantaged groups. A particularly pernicious risk of automated, axiomatic
approaches is that they can make invisible the cost to well-being: automatically
constraining algorithms to be ``fair'' can lead one to overlook unconstrained
alternatives that are clearly preferable. We have instead called for a more
careful analysis of the consequences, good and bad, of different decision
policies, selecting the appropriate course of action based on the specific
context. This is admittedly hard to do---and does not easily scale---but there
are general principles that we believe are helpful to keep in mind when
navigating this terrain. Below we enumerate and discuss four of them.

\subsubsection{Contending with inherent trade-offs}

There are inherent trade-offs in many important decision problems. For instance,
in our college admissions example, one must balance academic preparedness with
student-body diversity. Although one cannot generally circumvent these
trade-offs, we believe it is useful to explicitly enumerate the primary
dimensions of interest and to acknowledge the trade-offs between them. In some
cases, like our stylized admissions example, one might be able to explicitly
calculate the Pareto frontier shown in Figure~\ref{fig:frontier}, in which case
it often makes sense to focus on those policies lying on the frontier. In many
cases, it won't be possible to compute the frontier. Still, by listing and
discussing trade-offs, even informally, one can reduce the risk of adopting
clearly problematic policies, like those that typically result from uncritically
constraining decisions to satisfy formal fairness criteria.

In this sense, designing equitable algorithms is akin to designing equitable
policy writ large. One might accordingly adapt democratic mechanisms used to
draft and enact legislation to algorithm design. For example, adopting such a
policy-oriented perspective, \citet{L2BF} and \citet{koenecke2023} surveyed a
diverse sample of Americans to elicit preferences on how best to balance
competing objectives in programs that algorithmically allocate government
benefits.

\subsubsection{Assessing Calibration}

When designing or auditing a risk assessment algorithm, it is important to check
whether predictions are \emph{calibrated}, meaning that risk scores correspond
to the same observed level of risk across groups. In general, the relationship
between predictors and outcomes may plausibly differ across groups, leading to
miscalibrated risk estimates---what \citet{ayres2002outcome} calls the problem
of \emph{subgroup validity}. Figure~\ref{fig:anticlassification} shows instances
of risk scores for diabetes and recidivism that are miscalibrated across race
and gender, respectively. For example, a nominal 1.5\% diabetes risk---based on
age and BMI---corresponds to an actual, observed diabetes rate of approximately
1\% among White patients and 3\% among Asian patients. Similarly, among
individuals receiving a COMPAS risk score of 7---based on criminal history and
related factors---about 55\% of women recidivate, compared to 65\% of men. These
miscalibrated risk scores can result in inequitable decisions. For instance, a
policy to screen patients with a nominal diabetes risk of 1.5\% or above---in
line with existing medical recommendations~\citep{agarwal2022diabetes}---would
overscreen White patients and underscreen Asian patients, harming individuals in
both groups.

Calibration can often be visually assessed by plotting predicted risk against
average outcomes, as in
Figure~\ref{fig:calibration}~\citep[cf.{}][]{arrieta2022metrics}. For a simple,
more quantitative, measure, we recommend regressing observed outcomes against
risk estimates and group membership. A coefficient of approximately zero on
group membership suggests risk estimates correspond to similar average outcomes
across groups, with deviations from zero indicating the degree of
miscalibration.

In practice, miscalibration can often be rectified by training group-specific
risk models, or, roughly equivalently, including group membership in a single
risk model fit across groups. For example, diabetes risk models that include
race, and recidivism risk models that include gender, are approximately
calibrated. The relatively new literature on multicalibration has introduced new
computational techniques to ensure predictions are simultaneously calibrated
across many different subgroups~\citep{hebert2018multicalibration}. Of course,
including protected traits in risk models raises additional legal and ethical
challenges. In some cases, it may be possible to reduce or eliminate
miscalibration by incorporating additional, non-protected covariates.
Regardless, we believe it is important to check the calibration of risk scores
to make informed decisions about if and how to address any observed disparities.

Calibration is an important necessary condition to ensure risk estimates
correspond to actually observable levels of risk across groups. But it is not
sufficient. Indeed, even calibrated risk scores can encode and reinforce deeply
discriminatory policies. To see this, imagine a bank that wants to discriminate
against Black applicants. Further suppose that: (1) within ZIP code, White and
Black applicants have similar default rates; and (2) Black applicants live in
ZIP codes with relatively high default rates. Then the bank can surreptitiously
discriminate against Black borrowers by basing estimates of default risk only on
an applicant's ZIP code, ignoring all other relevant information. Such scores
would be calibrated (White and Black applicants with the same score would
default equally often), and the bank could use these scores to justify denying
loans to nearly all Black applicants. The bank, however, would be sacrificing
profit by refusing loans to creditworthy Black applicants,\footnote{%
  These applicants are creditworthy in the sense that they would have been
  issued a loan had the bank used all the information it had available to
  determine their risk.
}
and thus engaging in taste-based discrimination. This discriminatory lending
strategy is indeed closely related to the historical (and illegal) practice of
redlining, and illustrates the limitations of calibration as a measure of
equity.

 \begin{figure}[t]
  \begin{center}
    \begin{subfigure}{2.5in}
      \includegraphics{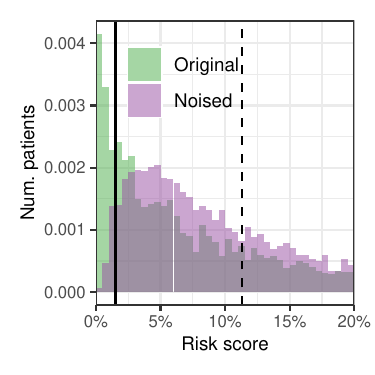}
    \end{subfigure}
    \begin{subfigure}{2.5in}
      \includegraphics{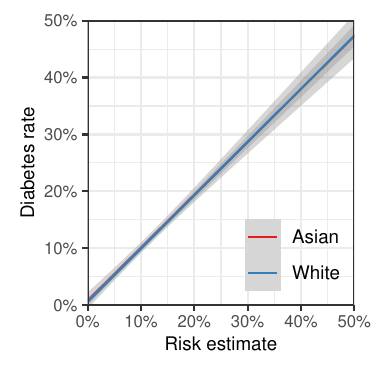}
    \end{subfigure}
  \end{center}
  \caption{%
    Calibration is insufficient to prevent discrimination. \emph{Left:}~The
    distribution in green shows diabetes risk for Asian patients based on
    accurately collected age and BMI, and the distribution in purple shows
    estimates when the risk model is trained on noisy inputs. Estimates under
    the noisy model concentrate around the mean (dashed vertical line), pushing
    more Asian patients above the screening threshold (solid vertical line).
    \emph{Right:}~A calibration plot comparing noisy risk estimates for Asian
    patients and accurate risk estimates for White patients. The calibrated risk
    scores can mask both intentional discrimination and inadvertent errors.
  }
\label{fig:redlining}
\end{figure}

Figure~\ref{fig:redlining} shows another example of calibrated scores masking
disparities. In the left-hand panel, we plot in green the distribution of
diabetes risk for Asian patients, as estimated from age, BMI, and race. In
purple, we plot the distribution of estimated risk when age and BMI are
imperfectly measured for Asian patients in the training data. Training the risk
model on noisy features pushes risk estimates toward the mean. As a result,
based on the noisy risk model, 97\% of Asian patients are above the 1.5\%
screening threshold, compared to 81\% of Asian patients under the more
accurately estimated model---leading to more medically unnecessary screening
under the noisy model. Importantly, however, the noisy model is still
calibrated, as shown in the right-hand panel of Figure~\ref{fig:redlining},
where we compare risk scores for Asian patients estimated from the noisy
predictors and risk scores for White patients estimated from the accurately
estimated information. In theory, a malicious algorithm designer could generate
such calibrated but inaccurate scores to intentionally harm Asian patients. In
practice, this pattern could equally arise from negligence rather than malice.
These examples illustrate the importance of considering all available data when
constructing statistical risk estimates; assessments that either intentionally
or inadvertently ignore predictive information may facilitate discriminatory
decisions while satisfying calibration---though, as we discuss below, even this
intuitive heuristic of ``use all the data'' has its limitations.

\subsubsection{Selecting the target of prediction}

In constructing algorithmic risk scores, a key ingredient is the target of
prediction. In practice, though, there is often a mismatch between our true
outcome of interest and the available data---an occurrence we call \emph{label
bias}~\citep{label-bias}. As with the other issues we discuss, there is
typically no perfect solution to this problem, but there are ways to mitigate
it.

For example, in pretrial risk assessment, we would often like to estimate the
likelihood a defendant would commit a crime if released. But there are two key
difficulties with this goal. First, though we might want to measure crime
conducted by defendants awaiting trial, we typically only observe crime that
results in a conviction or an arrest. These observable outcomes, however, are
imperfect proxies for the underlying criminal act. Further, heavier policing in
communities of color might lead to Black and Hispanic defendants being arrested,
and later convicted, more often than White defendants who commit the same
offense~\citep{lum2016predict}. Poor outcome data might thus cause one to
systematically underestimate the risk posed by White defendants. The second,
related, issue is that our target of interest is a \emph{counterfactual}
outcome; it corresponds to what would have happened had a defendant been
released. In reality, we only observe what actually happened conditional on the
judge's actual detention decision.

One way to reduce label bias in this case is to adjust the target of interest.
For example, criminologists have found that arrests for violent crime---as
opposed to drug crime---may suffer from less racial bias.\footnote{%
  \citet{dalessio2003race} find evidence that White offenders are even somewhat
  more likely than Black offenders to be arrested for certain categories of
  crime, including robbery, simple assault, and aggravated assault. Measurements
  of minor criminal activity, like drug offenses, are more problematic. For
  example, there is evidence that drug arrests in the United States are biased
  against Black and Hispanic individuals, with racial minorities who commit drug
  crimes substantially more likely to be arrested than White individuals who
  commit the same offenses~\citep{ramchand2006racial}. Although this pattern is
  well known, many existing risk assessment tools still consider arrests or
  convictions for \emph{any} new criminal activity---including drug
  crimes---which may lead to biased estimates. As another example of label bias,
  auto insurance rates are determined in part by a driver's record of receiving
  speeding tickets, but disparities in police enforcement mean that tickets are
  biased proxies of dangerous driving behavior~\citep{cai2022measuring}.
}
In particular, \citet{skeem2015risk} note that the racial distribution of
individuals arrested for violent offenses is in line with the racial
distribution of offenders inferred from victim reports, and is also in line with
self-reported offending data. In other cases, like lending, where one may seek
to estimate default rates, the measured outcome (e.g., failure to pay)
corresponds more closely to the event of interest. The problem of estimating
counterfactuals can likewise be partially addressed in some applications. In the
pretrial setting, \citet{angwin2016} measure recidivism rates in the first
two-year period during which a defendant is not incarcerated; this is not
identical to the desired counterfactual outcome---since the initial detention
may be criminogenic, for example---but it seems like a reasonable estimation
strategy. Further, unaided human decisions often exhibit considerable
randomness, a fact that can be exploited to facilitate statistical estimation of
counterfactual outcomes~\citep{simplerules,kleinberg2017human}. More generally,
a spate of recent work at the intersection of machine learning and causal
inference~\citep{mullainathan2017machine,hill2011bayesian,jung2018algorithmic}
offers hope for more gains in counterfactual estimation.

\subsubsection{Collecting training data}

A final issue we discuss is collecting suitable training data for risk
assessment algorithms to mitigate the effects of sample bias. Ideally, one would
train algorithms on data sets that are broadly representative of the populations
on which they are ultimately applied---though there are subtleties to this
heuristic that we describe below. While often challenging in practice, failure
to train on representative data can lead to unintended, and potentially
discriminatory, consequences. For example, \citet{buolamwini2018gender} found
that commercial facial analysis tools struggle to correctly classify the gender
of dark-skinned individuals---and of dark-skinned women in particular---a
disparity likely attributable to the relative dearth of dark-skinned faces in
facial analysis data sets. Similarly, \citet{koenecke2020racial} found that
several popular automated speech recognition systems were significantly worse at
transcribing Black speakers than White speakers, likely due to insufficient data
from speakers of African American Vernacular English (AAVE), a variety of
English spoken by many Black Americans.

The problems of non-representative data can be even more acute in the case of
risk assessment algorithms, especially when the target of interest is a causal
quantity. For instance, in our running college admissions example, we seek to
estimate how a student would (counterfactually) perform if admitted. Historical
data are typically the result of past, potentially biased, decisions, and so may
not fully generalize. Imagine, for example, that predictions of college
performance are informed by where an applicant went to high school, and that,
historically, only applicants from certain high schools were accepted---and we
consequently only see outcomes for students from those high schools. Then we
would expect less accurate predictions for students from those absent high
schools. In general, regression to the mean could attenuate estimates for high
achieving students who differ from those previously accepted, potentially
reinforcing existing admissions practices.

As a general heuristic, we believe it is advisable to train models on
representative data, but, as \citet{cai2022adaptive} note, the optimal sampling
strategy depends on the statistical structure of the problem and the
group-specific costs of collecting data. Interestingly, the value of
representative data collection strategies depends in part on the degree to which
race, gender, and other protected attributes are predictive. In theory, if
protected attributes are not predictive, one could build an accurate risk model
using only examples from one particular group (e.g., White men). Given enough
examples of White men, the model would learn the relationship between features
and risk, which by our assumption would generalize to the entire population.
This phenomenon highlights a tension in informal discussions of fairness,
with some advocating both for representative training data and for the
exclusion of protected attributes. However, representative data are often most
important precisely when protected attributes add information, in which case
their use is arguably more justified. Even if protected attributes are not
predictive, representative data can still help in two additional ways. First, a
representative sample ensures that the full support of features is present at
training time, as it is possible that the distribution of features varies across
groups, even if the connection between features and outcomes does not. We note,
though, that one might have adequate support even without a representative
sample in many real-world settings, particularly when models are trained on
large data sets and the feature space is relatively low dimensional. Second, a
representative sample can help with model validation, allowing one to assess the
potential effects of group imbalance on model fit. In particular, without a
representative sample, it can be difficult to determine whether a model trained
on a single group generalizes to the entire population.

In many settings, one may be able to gather better training data with greater
investment of time and money. For example, in our diabetes example one could aim
to collect more complete medical records, a process that may be both costly and
logistically difficult. In theory, this additional information may lead to
welfare gains, and policymakers must accordingly evaluate the relative costs and
benefits to all groups of exerting this extra effort when designing algorithms.
Fortunately, in practice, there are often diminishing returns to information,
with a relatively short list of key features providing most of the predictive
power~\citep{simplerules}, at least partially mitigating this concern.

As with the other issues we have discussed, there is no universal solution to
data collection. It might, for example, simply be prohibitive in the short run
to train models on the data set one would ideally like to use. Nevertheless, as
in all situations, one must carefully weigh the potential costs and benefits of
adopting a necessarily imperfect risk assessment algorithm relative to the other
possible options. In particular, even an imperfect algorithm may in some
circumstances be better than leaving decisions to similarly imperfect humans who
have their own biases.

\subsection{Case Study: An Algorithm to Allocate Limited Medical Resources}
\label{sec:case-study}

We conclude by illustrating the principles discussed above with a real-world
example: an algorithm for referring patients into a ``high-risk care
management'' program, previously considered by \cite{obermeyer2019dissecting}.
The care management program more effectively aids patients with complex medical
needs, in principle both improving outcomes for patients and reducing costs to
the medical system for patients who are enrolled. But the program has limited
capacity, which we formalize by assuming that only 2\% of patients can be
enrolled (i.e., we set \(b = \tfrac 1 {50}\)). For our analysis, we use the
data released by \citeauthor{obermeyer2019dissecting},\footnote{%
    \citeauthor{obermeyer2019dissecting}\ released a synthetic data set closely
    mirroring the real data set, available at:
    \url{https://gitlab.com/labsysmed/dissecting-bias}. In contrast to \(b =
    2\%\), which we adopt to better illustrate some of the statistical phenomena
    we discuss in this section, \citeauthor{obermeyer2019dissecting}\ use a
    budget of \(b = 3\%\).
}
which contain demographic variables, cost information, comorbidities, biomarker
and medication details, and health outcomes for a population of approximately
43,000 White and 5,600 Black primary care patients at an academic hospital from
2013--2015.

As a first step toward identifying patients to enroll in the program, one could
train a model predicting the healthcare resources a patient is likely to require
over the next year. Sicker patients require more care and, consequently, incur
greater healthcare costs. Thus, our initial approach is to predict how likely a
patient is to be ``high cost''---which we operationalize as being in the top
decile of healthcare expenditures in a given year---based on the available
information. (One of the main contributions of \cite{obermeyer2019dissecting} is
highlighting that healthcare costs is a problematic outcome due to label bias, a
point we return to shortly.)

\begin{figure}
  \begin{center}
    \includegraphics{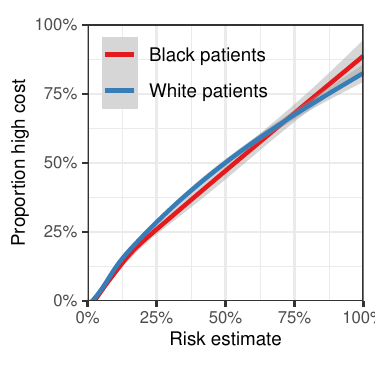}
  \end{center}
  \caption{
    Calibration of a race-blind model predicting whether a patient is
    ``high-cost''. The lack of a gap between the proportion of patients who are
    actually high-cost and the proportion predicted to be high-cost for both
    groups indicates that the model is well calibrated across race groups.
  }
\label{fig:calibration}
\end{figure}

As discussed above, it is useful to assess the calibration of risk assessment
algorithms across groups. In particular, while calibration across race is
largely guaranteed if race is included as a predictor in the statistical model,
race-blind models are often preferred, particularly in healthcare, in part to
avoid perceptions of bias. As a result, we assess the calibration of a
race-blind model trained on all available information except for race, shown in
Figure~\ref{fig:calibration}. Unlike in the diabetes screening example
considered in Section~\ref{sec:inframarginality}, the race-blind healthcare cost
predictions are calibrated across race groups, meaning that risk estimates
largely match observed costs across groups. It accordingly appears that race
provides little marginal predictive power in this example, assuaging potential
concerns with its omission.

We turn next to assessing the effects of applying formal fairness criteria to
our enrollment decisions. Often, in healthcare contexts, a false
negative---e.g., failing to screen for a disease when it is present---is more
consequential than a false positive---e.g., screening for a disease when it is
not present. For this reason, one might seek to ensure enrollment decisions are
fair by mandating false negative rates be equal across race groups
\citep[e.g.,][]{seyyed2021underdiagnosis},\footnote{%
    The authors describe their metric of concern as the ``false positive rate,''
    where the positive prediction is understood as a ``no finding'' label, e.g.,
    not having a disease. In our example, we follow the more common convention
    that a ``positive'' classification is assigned to the rare event (i.e.,
    being high cost), and so we call this metric the ``false negative rate.''
}
i.e., requiring that \(A \indep D \mid Y = 1\). In our example, equalizing false
negative rates means that among patients who ultimately incur high medical costs
(\(Y = 1\)), the same proportion (\(\B E[D]\)) are referred into the program
across race groups (\(A\)).

\begin{figure}[t]
  \begin{center}
    \begin{subfigure}{2.5in}
       \includegraphics{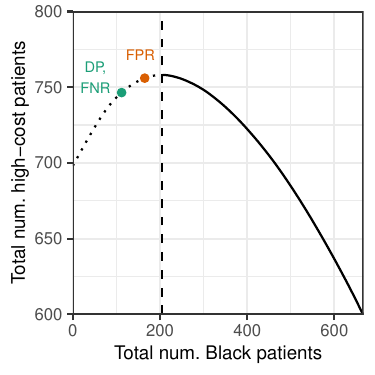}
       \caption{Full population}
    \label{fig:pareto_full}
    \end{subfigure}
    \begin{subfigure}{2.5in}
      \includegraphics{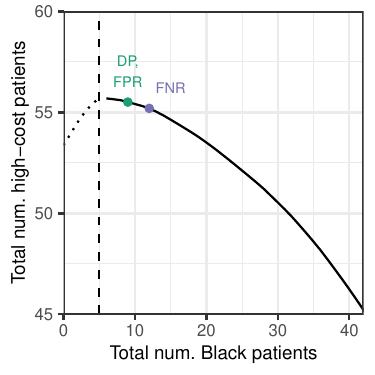}
      \caption{Women, aged 25--34}
    \label{fig:pareto_sub}
    \end{subfigure}
  \end{center}
  \caption{%
    Enforcing formal fairness criteria can harm marginalized groups. Feasible
    regions for admissions policies to a high-risk care management program,
    where the dashed line indicates the number of Black patients admitted by
    the policy admitting the maximal number of high-cost patients.
    \emph{Left:}~The Pareto frontier for all patients in the population. Because
    more Black patients incur high medical costs in the population as a whole,
    equalizing false negative rates (FNR)---as well as false positive rates
    (FPR), or enforcing demographic parity (DP)---results in fewer Black
    patients being admitted than under the policy that maximizes the total
    number of high-cost patients admitted. (Equalized false negative rates and
    demographic parity are achieved at the same point.) \emph{Right:}~The Pareto
    frontier for the subpopulation of women between the ages of 25 and 34.
    Within this subpopulation, Black patients incur lower medical costs, and so
    equalizing false positive rates, false negative rates, or achieving
    demographic parity all result in more Black patients being admitted to the
    high-risk management program than the policy that admits the maximum number
    of high-cost patients.
  }
\label{fig:pareto}
\end{figure}

In our setting, approximately 1,000 patients can be referred into the care
management program---2\% of the roughly 50,000 patients in our data set. When we
equalize false-negative rates, 747 of the enrolled patients ultimately incur
high costs, and 113 enrolled patients are Black.\footnote{%
  Following \citet{corbett2017algorithmic}, we equalize false-negative rates in
  a manner that maximizes the number of enrolled patients who ultimately incur
  high costs.
}
However, an unconstrained decision rule (i.e., one that enrolls the patients
most likely to incur high costs) enrolls both more high-cost patients (758) and
more Black patients (205). In this example, we end up providing worse care to
Black patients when we constrain our algorithm to satisfy the formal,
mathematical fairness criterion.

Instead of applying such mathematical fairness criteria, we advocate for
directly weighing the costs and benefits of different decision policies. In
Figure~\ref{fig:pareto_full}, we show the Pareto frontier for our example,
tracing out policies that optimally trade off the demographic composition of the
enrolled population with the number of enrolled patients who in reality incur
high costs. The green point corresponds to equalizing false negative rates, and
is to the left of the dashed vertical line that corresponds to the unconstrained
decision rule---visually illustrating how constraining our algorithm leads to
fewer resources for Black patients. Also shown on the plot are points
corresponding to demographic parity and equal false positive rates, both of
which likewise lead to fewer resources for Black patients.\footnote{%
  As discussed in Section~\ref{sec:presence}, with the exception of
  counterfactual predictive parity, all of the remaining fairness definitions
  given in Section~\ref{sec:defn} are known \emph{a priori} to restrict one to
  enrollment policies that will lower \emph{both} the number of Black patients
  enrolled as well as the number of truly high-cost patients enrolled.
}

The result in Figure~\ref{fig:pareto_full} stems from the false negative rate
for Black patients being lower than the false negative rate for White patients
in the unconstrained algorithm---a pattern we expect since Black patients are
more likely than White patients to incur high medical costs in our data.
Equalizing false negative rates thus means raising the enrollment bar for Black
patients and lowering the bar for White patients. In light of this example, one
might argue for applying formal fairness criteria only when error rates for
racial minorities are higher than for White individuals.
Figure~\ref{fig:pareto_sub} repeats our analysis above for the subset of women
between the ages of 25 and 34, a subpopulation in which Black patients have
higher error rates than White patients. In this case, equalizing false positive
rates or false negative rates, or enforcing demographic parity all result in
more Black patients being admitted into the program. It is, however, unclear why
one should adopt those particular error-rate equalizing policies over any other.
The policies on the Pareto frontier (i.e., on the curve to the right of the
dashed line) are all arguably reasonable to consider. It is admittedly difficult
to determine which of these policies to adopt, but we believe it is best to
confront this challenge head on, recognizing the hard trade-offs inherent to the
problem.

\begin{figure}[t]
  \begin{center}
    \begin{subfigure}{2.5in}
      \includegraphics{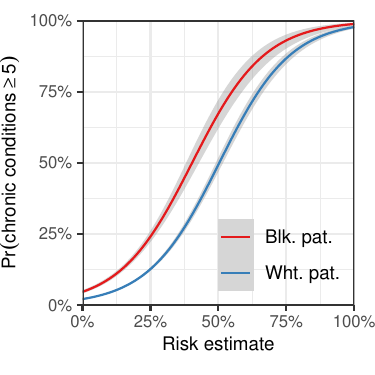}
      \caption{Label bias}
    \label{fig:label_bias}
    \end{subfigure}
    \begin{subfigure}{2.5in}
      \includegraphics{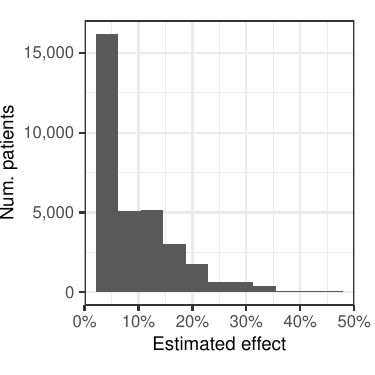}
      \caption{Heterogeneous treatment effects}
    \label{fig:treatment_effects}
    \end{subfigure}
  \end{center}
  \caption{%
    The target of prediction impacts equity. \emph{Left:}~The probability of
    having complex medical needs (i.e., at least five active chronic conditions)
    for Black and White patients as a function of their estimated likelihood of
    incurring high medical costs, reproducing an analysis by
    \citet{obermeyer2019dissecting}. The large gap across groups indicates that
    Black patients have greater medical need than White patients with similar
    anticipated healthcare costs. \emph{Right:}~Distribution of the estimated
    change in the probability that an individual will have complex medical needs
    after enrolling in the care management program, showing that the extent to
    which enrollment reduces complex medical needs varies considerably across
    individuals.
  }
\end{figure}

We conclude our case study by considering label bias, the primary concern
identified by~\citet{obermeyer2019dissecting} in this context. As those authors
noted, medical cost is a poor proxy for medical need, and so allocating
healthcare resources to minimize anticipated costs can lead to severe
disparities. Replicating an analysis by \citeauthor{obermeyer2019dissecting},
Figure~\ref{fig:label_bias} shows that among patients with similar likelihood of
incurring high-cost medical care, Black patients are considerably more likely
than White patients to have complex medical needs, operationalized as having
five or more active chronic conditions. This gap is likely a consequence of
worse access to healthcare among Black patients, due to a mix of socioeconomic
factors and discrimination. To the extent that care management programs aim to
aid the sickest patients---as opposed to simply reducing costs---targeting
resources based on anticipated costs can lead to inefficient and inequitable
outcomes. \citeauthor{obermeyer2019dissecting}\ accordingly suggest switching
the target of prediction from medical costs to health status.

It is possible to achieve further gains by recognizing resource allocation as an
inherently causal problem. In particular, one may seek to enroll patients in the
program in a manner that maximizes \(\B E[Y(D)]\), where \(Y(D)\) is the
potential health outcome under the enrollment decision. To do so, we could
prioritize patients by their estimated treatment effect \(\hat{Y}_i(1) -
\hat{Y}_i(0)\), rather than an estimate \(\hat{Y}_i\) of their future health
status that ignores the causal effect of the program.\footnote{%
  In many analyses that do not explicitly grapple with the causal effects of
  interventions, the estimand \(Y\) is further corrupted by the fact that some
  patients are expected to be enrolled in the program. As a result, some of the
  sickest patients may not be prioritized for care, since their expected outcome
  already incorporates the fact that they would have been enrolled, leading to a
  prediction paradox. To avoid this situation, one could explicitly estimate and
  prioritize patients by \(Y_i(0)\), the potential outcome in the absence of
  care. In this case, however, the allocation decisions do not necessarily lead
  to the largest health gains, as the patients likely to be sickest in the
  absence of care are not typically the same as those likely to benefit the most
  from the program.
}
A proper causal analysis is, in general, a complex topic requiring careful
treatment beyond the scope of this article. Nonetheless,
Figure~\ref{fig:treatment_effects} shows the distribution of \(\hat{Y}_i(1) -
\hat{Y}_i(0)\), as estimated with a simple regression model. The plot suggests
that there is considerable predictable heterogeneity in the extent to which
enrollment in the care management program causally improves health. In
particular, we find that the estimated treatment effect is only weakly
correlated with the number of chronic conditions a patient currently exhibits
(\(r = 0.05\)). Consequently, directly targeting resources to those most likely
to benefit could yield large health improvements. Once the types of label bias
we have discussed above have been identified, it may be possible to re-train
predictive models to better align decision-making algorithms with policy goals.

\section{Conclusion}

From medicine to criminal justice, practitioners are increasingly turning to
statistical risk assessments to help guide and improve human decisions.
Algorithms can avoid many of the implicit and explicit biases of human decisions
makers, but they can also exacerbate historical inequities if not developed with
care. Policymakers, in response, have rightly demanded that these high-stakes
decision systems be designed and audited to ensure outcomes are equitable. The
research community has responded to the challenge, coalescing around several
formal mathematical definitions of fairness. However, as we have aimed to
articulate, these popular measures of fairness suffer from significant
statistical limitations. Indeed, adopting these measures as algorithmic design
principles can often harm the groups that these measures were designed to
protect.

In contrast to the dominant axiomatic approach to algorithmic fairness, we
advocate for a more consequentialist orientation~\citep{%
  nyarko2021breaking, L2BF, cai2022adaptive, liang2021algorithmic%
}.
Most importantly, we stress the importance of grounding technical and policy
discussions of fairness in terms of real-world quantities. For example, in the
pretrial domain, one might consider a risk assessment's short and long-term
impacts on public safety and the size of the incarcerated population, as well as
a tool's alignment with principles of due process. In lending, one could
similarly consider a risk assessment's immediate and equilibrium effects on
community development and the sustainability of a loan program. Formal
mathematical measures of fairness only indirectly address such issues, and can
inadvertently lead discussions astray. Of course, it is not always clear how
best to quantify or to balance the relevant costs and benefits of proposed
algorithmic interventions. In some cases, it may be possible to conduct
randomized controlled trials; in other cases, the best one can do is hypothesize
about an algorithm's potential effects. Regardless, we believe a more explicit
focus on consequences is necessary to make progress.

We further recommend decoupling the statistical problem of risk assessment from
the policy problem of designing interventions. At their best, predictive
algorithms estimate the likelihood of events under different scenarios; they
cannot dictate policy. An algorithm might (correctly) infer that a defendant has
a 20\% chance of committing a violent crime if released, but that fact does not,
in and of itself, determine a course of action. For example, detention is not
the only alternative to release, as one could take any number of rehabilitative
interventions \citep{barabas2018interventions}. Even if detention is deemed an
appropriate intervention, one must still determine what threshold would
appropriately balance public safety with the social and financial costs of
detention. One might even decide that society's goals are best achieved by
setting different thresholds for different groups. For example, a policymaker
might reason that, all else being equal, the social costs of detaining a single
parent are higher than the social costs of detaining an individual without
children, and thus decide to apply different thresholds to the two groups. When
policymakers consider these options and others, we believe the primary role of a
risk assessment tool is, as its name suggests, to estimate risk. This view,
however, is at odds with requiring that algorithms satisfy popular fairness
criteria. Such constrained algorithms typically do not reflect the best
available estimates of risk, and thus implicitly conflate the statistical and
policy problems.

Fair machine learning still has much left to accomplish and there are several
important avenues of research that could benefit from new statistical and
computational insights. From mitigating measurement error and sample bias, to
understanding externalities and equilibrium effects, to eliciting and
aggregating preferences to arbitrate between competing algorithms, there is much
work to be done. But the benefits are equally large. When carefully designed and
evaluated, statistical algorithms have the potential to dramatically improve
both the efficacy and equity of consequential decisions. As these algorithms are
increasingly deployed in all walks of life, it will become ever more important
to ensure they are fair.

\acks{%
  We thank Guillaume Basse, Sander Beckers, Hana Chockler, Alex Chohlas-Wood,
  Madison Coots, Avi Feller, Josh Grossman, Joe Halpern, Jennifer Hill, Aziz
  Huq, David Kent, Keren Ladin, Julian Nyarko, Emma Pierson, Ravi Sojitra, and
  Michael Zanger-Tishler for helpful conversations. This paper is based on work
  by \citet{corbett2018measure} and \citet{nilforoshan2022causal}. H.N was
  supported by a Stanford Knight-Hennessy Scholarship and an NSF Graduate
  Research Fellowship under Grant No. DGE-1656518. J.G was supported by a
  Stanford Knight-Hennessy Scholarship. R.S. was supported by the NSF Program on
  Fairness in AI in Collaboration with Amazon under the award ``FAI: End-to-End
  Fairness for Algorithm-in-the-Loop Decision Making in the Public Sector,'' no.
  IIS-2040898. S.G. was supported by a grant from the Harvard Data Science
  Initiative. Any opinions, findings, conclusions, or recommendations expressed
  in this material are those of the authors and do not necessarily reflect the
  views of the NSF or Amazon. Reproduction materials are available at
  \url{https://github.com/jgaeb/measure-mismeasure}.
}

\appendix
\renewcommand{\thesection}{\Alph{section}}
\renewcommand{\thesubsection}{\Alph{section}.\arabic{subsection}}
\renewcommand{\thesubsubsection}
  {\Alph{section}.\arabic{subsection}.\arabic{subsubsection}}

\addappheadtotoc

\section{Path-specific Counterfactuals}
\label{sec:path-specific-counterfactuals}

Constructing policies which satisfy path-specific fairness requires computing
path-specific counterfactual values of features. In
Algorithm~\ref{alg:counterfactuals}, we describe the formal construction of
path-specific counterfactuals \(Z_{\Pi,a,a'}\), for an arbitrary variable \(Z\)
(or collection of variables) in the DAG. To generate a sample \(Z_{\Pi,a,a'}^*\)
from the distribution of \(Z_{\Pi,a,a'}\), we first sample values \(U^*_j\) for
the exogenous variables. Then, in the first loop, we traverse the DAG in
topological order, setting \(A\) to \(a\) and iteratively computing values
\(V_j^*\) of the other nodes based on the structural equations in the usual
fashion. In the second loop, we set \(A\) to \(a'\), and then iteratively
compute values \(\overline{V_j}^*\) for each node. \(\overline{V_j}^*\) is
computed using the structural equation at that node, with value
\(\overline{V_{\ell}}^*\) for each of its parents that are connected to it along
a path in \(\Pi\), and the value \(V^*_{\ell}\) for all its other parents.
Finally, we set \(Z_{\Pi,a,a'}^*\) to \(\overline{Z}^*\).

\begin{algorithm2e}[ht]
\KwData{\(\mathcal{G}\) (topologically ordered), \(\Pi\), \(a\), and \(a'\)}
\KwResult{A sample \(Z_{\Pi, a, a'}^*\) from \(Z_{\Pi, a, a'}\)}
  \vspace{2mm}
  Sample values \(\{U_{j}^{*}\}\) for the exogenous variables
  \vspace{2mm}
  \tcc{Compute counterfactuals by setting \(A\) to \(a\)}
    \For{\(j = 1, \ldots, m\)}{
      \eIf{\(V_j = A\)}{
        \(V_j^* \gets a\)
      }{
        \(\wp(V_{j})^{*} \gets \{V_{\ell}^* \mid V_{\ell} \in \wp(V_j)\}\) \\
        \(V_{j}^{*} \gets f_{V_{j}}(\wp(V_{j})^{*}, U_{j}^{*})\)\\
      }
    }
  \vspace{2mm}
  \tcc{%
    Compute counterfactuals by setting \(A\) to \(a'\) and propagating values
    along paths in \(\Pi\)
  }
  \For{\(j = 1, \ldots, m\)}{
    \eIf{\(V_j = A\)}{
      \(\overline{V}_j^* \gets a'\)
    }{
      \For{\(V_k \in \wp(V_j)\)}{
        \eIf{edge \((V_k, V_j)\) lies on a path in \(\Pi\)}{
          \(V_{k}^{\dagger} \gets \overline{V}_{k}^*\)
        }{
          \(V_{k}^{\dagger} \gets V_{k}^*\)
        }
      }
    \(\wp(V_{j})^{\dagger} \gets \{V_{\ell}^{\dagger} \mid V_{\ell} \in
    \wp(V_j)\}\) \\ \(\overline{V}_j^* \gets f_{V_j}(\wp(V_j)^{\dagger},
    U_j^*)\)
    }
  }
  \vspace{2mm}
  \(Z_{\Pi, a, a'}^* \gets \overline{Z}^*\)
  \caption{Path-specific counterfactuals}
\label{alg:counterfactuals}
\end{algorithm2e}

\section{Constructing Causally Fair Policies}
\label{sec:construction}

Our aim is to identify the feasible region of expected outcomes attainable via
policies which are constructed to satisfy various causal fairness constraints.

First, consider the problem of finding decision policies that maximize expected
utility, subject to satisfying a given definition of causal fairness, as well as
the outcome and budget constraints. Specifically, letting \(\mathcal{C}\) denote
the family of all decision policies that satisfy one of the causal fairness
definitions listed above, a utility-maximizing policy \(d^*\) is given by
\begin{align}
  \label{eq:opt}
  \begin{split}
    d^* \in \arg\max_{d \in \mathcal{C}} & \quad \EE[d(X) \cdot u(X)] \\
    \text{s.t.} &\quad o_1 - \epsilon \leq \EE[d(X) \cdot
                \mathbb{1}_{\alpha(X) = a_1} ] \leq o_1 + \epsilon \\
    \text{s.t.} &\quad o_2 - \epsilon \leq \EE[d(X) \cdot \EE[Y(1) \mid X]]
                \leq o_2 + \epsilon \\
    \text{s.t.} &\quad \EE[d(X)] \leq b. \\
  \end{split}
\end{align}

We prove that this optimization problem can be efficiently solved as a single
linear program---in the case of counterfactual equalized odds, conditional
principal fairness, counterfactual fairness, and path-specific fairness---or as
a series of linear programs in the case of counterfactual predictive parity.

\begin{theorem}
\label{thm:lp}
  Consider the optimization problem given in Eq.~\eqref{eq:opt}.
  \begin{enumerate}
    \item If \(\mathcal{C}\) is the class of policies that satisfies
      counterfactual equalized odds or conditional principal fairness, and the
      distribution of \((X, Y(0), Y(1))\) is known and supported on a finite set
      of size \(n\), then a utility-maximizing policy constrained to lie in
      \(\mathcal{C}\) can be constructed via a linear program with \(O(n)\)
      variables and constraints.
    \item If \(\mathcal{C}\) is the class of policies that satisfies
      path-specific fairness (including counterfactual fairness), and the
      distribution of \((X, D_{\Pi, A, a})\) is known and supported on a finite
      set of size \(n\), then a utility-maximizing policy constrained to lie in
      \(\mathcal{C}\) can be constructed via a linear program with \(O(n)\)
      variables and constraints.
    \item Suppose \(\mathcal{C}\) is the class of policies that satisfies
      counterfactual predictive parity, that the distribution of \((X, Y(1))\)
      is known and supported on a finite set of size \(n\), and that the
      optimization problem in Eq.~\eqref{eq:opt} has a feasible solution.
      Further suppose \(Y(1)\) is supported on \(k\) points, and let
      \(\Delta^{k-1} = \{p \in \mathbb{R}^{k} \mid p_i \geq 0 \ \text{and} \
      \sum_{i=1}^k p_i = 1\}\) be the unit \((k-1)\)-simplex. Then one can
      construct a set of linear programs \(\mathcal{L} = \{L(v)\}_{v \in
      \Delta^k}\), with each having \(O(n)\) variables and constraints, such
      that the solution to one of the LPs in \(\mathcal{L}\) is a
      utility-maximizing policy constrained to lie in \(\mathcal{C}\).
  \end{enumerate}
\end{theorem}

Before moving on to the proof of Theorem~\ref{thm:lp}, we note that since the
constraints of the linear programs are convex, the feasible regions in
Figure~\ref{fig:frontier} can be determined by solving the convex feasibility
problem where we impose the additional convex constraint that the expected
outcomes---in our admissions example, the aggregate academic index and number of
admitted applicants from the target group---lie within some distance
\(\epsilon\) of a given point. Performing a grid search over all points then
determines the feasible regions.

\begin{proof}
  Let \(\C X = \{x_1, \ldots, x_m\}\); then, we seek decision variables \(d_i\),
  \(i = 1, \ldots, m\), corresponding to the probability of making a positive
  decision for individuals with covariate value \(x_i\). Therefore, we require
  that \(0 \leq d_i \leq 1\).

  Letting \(p_i = \Pr(X = x_i)\) denote the mass of \(X\) at \(x_i\), note that
  the objective function, as well as the outcome and budget constraints are all
  linear in the decision variables.

\begin{enumerate}
  \item The objective function \(\EE[d(X) \cdot u(X)]\) equals \(\sum_{i=1}^m
    d_i \cdot u(x_i) \cdot p_i\)
  \item The budget constraint \(\EE[d(X)] \leq b.\) constraint equals \(\sum_{i
    = 1}^m d_i \cdot p_i \leq b\)
  \item The first outcome constraint \(o_1 - \epsilon \leq \EE[d(X) \cdot
    \mathbb{1}_{\alpha(X) = a_1} ] \leq o_1 + \epsilon\) equals \(o_1 - \epsilon
    \leq \sum_{i = 1}^m \mathbb{1}_{\alpha(x_i) = a_1} \cdot d_i \cdot p_i ]
    \leq o_1 + \epsilon\)
  \item The second outcome constraint \( o_2 - \epsilon \leq \EE[d(X) \cdot
    \EE[Y(1) \mid X]] \leq o_2 + \epsilon\) equals \(o_2 - \epsilon \leq \sum_{i
    = 1}^m \EE[Y(1) \mid X = x_i] \cdot d_i \cdot p_i \leq o_2 + \epsilon\)
\end{enumerate}
  We now show that each of the causal fairness definitions can be enforced via
  linear constraints. We do so in three parts as listed in theorem.

  \medskip\noindent\emph{Theorem~\ref{thm:lp} Part 1.}\qquad
  First, we consider counterfactual equalized odds. A decision policy satisfies
  counterfactual equalized odds when \(D \indep A \mid Y(1).\) Since \(D\) is
  binary, this condition is equivalent to the expression \(\EE[d(X) \mid A = a,
  Y(1) = y] = \EE[d(X) \mid Y(1) = y]\) for all \(a \in \C A\) and \(y \in \C
  Y\) such that \(\Pr(Y(1) = y) > 0\). Expanding this expression and replacing
  \(d(x_j)\) by the corresponding decision variable \(d_j\), we obtain that
  \begin{equation*}
    \sum_{i = 1}^m d_i \cdot \Pr(X = x_i \mid A = a, Y(1) = y) = \sum_{i = 1}^m
    d_i \cdot \Pr(X = x_i \mid Y(1) = y)
  \end{equation*}
  for each \(a \in \C A\) and each of the finitely many values \(y \in \C Y\)
  such that \(\Pr(Y(1) = y) > 0\). These constraints are linear in the \(d_i\)
  by inspection.

  Next, we consider conditional principal fairness. A decision policy satisfies
  conditional principal fairness when \(D \indep A \mid Y(0), Y(1), W\), where
  \(W = \omega(X)\) denotes a reduced set of the covariates \(X\). Again, since
  \(D\) is binary, this condition is equivalent to the expression \( \EE[d(X)
  \mid A = a, Y(0) = y_0, Y(1) = y_1, W = w] = \EE[d(X) \mid Y(0) = y_0, Y(1) =
  y_1, W = w]\) for all \(y_0\), \(y_1\), and \(w\) satisfying \(\Pr(Y(0) = y_0,
  Y(1) = y_1, W = w) > 0\). As above, expanding this expression and replacing
  \(d(x_j)\) by the corresponding decision variable \(d_j\) yields linear
  constraints of the form
  \begin{equation*}
    \sum_{i = 1}^m d_i \cdot \Pr(X = x_i \mid A = a, S = s) = \sum_{j = 1}^m d_i
    \cdot \Pr(X = x_i \mid S = s)
  \end{equation*}
  for each \(a \in \C A\) and each of the finitely many values of \(S = (Y(0),
  Y(1), W)\) such that \(s = (y_0, y_1, w) \in \C Y \times \C Y \times \C W\)
  satisfies \(\Pr(Y(0) = y_0, Y(1) = y_1, W = w) > 0\). Again, these constraints
  are linear by inspection.

  \medskip\noindent\emph{Theorem~\ref{thm:lp} Part 2.}\qquad
  Suppose a decision policy satisfies path-specific fairness for a given
  collection of paths \(\Pi\) and a (possibly) reduced set of covariates \(W =
  \omega(X),\) meaning that for every \(a' \in \C A\), \(\EE[D_{\Pi, A, a'}
  \mid W ] = \EE[D \mid W]\).

  Recall from the definition of path-specific counterfactuals that
  \[
    D_{\Pi, A, a'} = f_{D}(X_{\Pi, A, a'}, U_{D}) = \mathbb{1}_{U_{D} \leq
    d(X_{\Pi, A, a'})},
  \]
  where \(U_D \indep \{X_{\Pi, A, a}, X\}\). Since \(W = \omega(X)\),
  \(U_D \indep \{X_{\Pi, A, a}, W\}\), it follows that
  \begin{align*}
    \EE[D_{\Pi, A, a'} &\mid W = w] \\
      &= \sum_{i = 1}^m \EE[D_{\Pi, A, a'} \mid X_{\Pi, A, a} = x_i, W = w]
        \cdot \Pr(X_{\Pi, A, a} = x_i \mid W = w) \\
      &= \sum_{i = 1}^m \EE[\B 1_{U_D \leq d(X_{\Pi, A, a'})}
        \mid X_{\Pi, A, a} = x_i, W = w] \cdot \Pr(X_{\Pi, A, a'} = x_i \mid W =
        w) \\
      &= \sum_{i = 1}^m d(X_{\Pi, A, a'}) \cdot \Pr(X_{\Pi, A, a'} = x_i \mid W
        = w) \\
      &= \sum_{i = 1}^m d_i \cdot \Pr(X_{\Pi, A, a'} = x_i \mid W = w).
  \end{align*}
  An analogous calculation yields that \(\EE[D \mid W = w] = \sum_{i = 1}^m d_i
  \cdot \Pr(X = x_i \mid W = w)\). Equating these expressions gives
  \begin{equation*}
    \sum_{i = 1}^m d_i \cdot \Pr(X = x_i \mid W = w) = \sum_{i = 1}^m d_i \cdot
    \Pr(X_{\Pi, A, a'} = x_i \mid W = w)
  \end{equation*}
  for each \(a' \in \C A\) and each of the finitely many \(w \in \C W\) such
  that \(\Pr(W = w) > 0\). Again, each of these constraints is linear by
  inspection.

  \medskip\noindent\emph{Theorem~\ref{thm:lp} Part 3.}\qquad
  A decision policy satisfies counterfactual predictive parity if \(Y(1) \indep
  A \mid D = 0,\) or equivalently, \(\Pr(Y(1) = y \mid A = a, D = 0) = \Pr(Y(1)
  \mid D = 0)\) for all \(a \in \C A\). We may rewrite this expression to
  obtain:
  \begin{align*}
    \dfrac{\Pr(Y(1) = y, A = a, D = 0)}{\Pr(A = a, D = 0 )} = C_{y},
  \end{align*}
  where \(C_{y} = \Pr(Y(1) = y \mid D = 0)\).

  Expanding the numerator on the left-hand side of the above equation yields
  \begin{equation*}
    \Pr(Y(1) = y, A = a, D = 0) = \sum_{i=1}^m [1 - d_i] \cdot \Pr(Y(1) = y, A =
    a, X = x_i)
  \end{equation*}

  Similarly, expanding the denominator yields
  \begin{equation*}
    \Pr(Y(1) = y, D = 0) = \sum_{i=1}^m [1 - d_i] \cdot \Pr(Y(1) = y, X = x_i).
  \end{equation*}
  for each of the finitely many \(y \in \C Y\). Therefore, counterfactual
  predictive parity corresponds to
  \begin{equation}
  \label{eq:cpp_constraint}
    \sum_{i=1}^m [1 - d_i] \cdot \Pr(Y(1) = y, A = a, X = x_i) = C_y \cdot
    \sum_{i=1}^m [1 - d_i] \cdot \Pr(Y(1) = y, X = x_i),
  \end{equation}
  for each \(a \in \C A\) and \(y \in \C Y\). Again, these constraints are
  linear in the \(d_i\) by inspection.

  Consider the family of linear programs \(\C L = \{L(v)\}_{v \in \Delta^k}\)
  where the linear program \(L(v)\) has the same objective function, outcome
  constraint, and budget constraint as before, together with additional
  constraints for each \(a \in \C A\) as in Eq.~\eqref{eq:cpp_constraint}, where
  \(C_{y_i} = v_i\) for \(i = 1, \ldots, k\).

  By assumption, there exists a feasible solution to the optimization problem in
  Eq.~\eqref{eq:opt}, so the solution to at least one program in \(\C L\) is a
  utility-maximizing policy that satisfies counterfactual predictive parity.
\end{proof}

\section{A Stylized Example of College Admissions}
\label{appendix:example}

In the example that we consider in Section~\ref{sec:geometry}, the exogenous
variables in the DAG, \( \C U = \{u_A, u_D, u_E, u_M, u_T, u_Y\}\), are
independently distributed as follows:
\begin{align*}
  U_A, U_D, U_Y & \sim \unif(0, 1),\\
  U_E, U_M, U_T & \sim \C N(0, 1).
\end{align*}
For fixed constants \(\mu_A\), \(\beta_{E,0}\), \(\beta_{E,A}\),
\(\beta_{M,0}\), \(\beta_{M,E}\), \(\beta_{T,0}\), \(\beta_{T,E}\),
\(\beta_{T,M}\), \(\beta_{T,B}\), \(\beta_{T,u}\), \(\beta_{Y,0}\),
\(\beta_{Y,D}\), we define the endogenous variables \( \C V = \{A, E, M,
T, D, Y\}\) in the DAG by the following structural equations:
\begin{align*}
  f_A(u_A)
    &=
    \begin{cases}
      a_1 & \text{if} \ u_A \leq \mu_A \\
      a_0 & \text{otherwise}
    \end{cases}, \\
  f_E(a, u_E)
    &= \beta_{E,0} + \beta_{E,A} \cdot \B 1(a = a_1) + u_E, \\
  f_M(e, u_M)
    &= \beta_{M,0} + \beta_{M,E} \cdot e + u_M, \\
  f_T(e, m, u_T)
    &= \beta_{T,0} + \beta_{T,E} \cdot e \\
    & \hspace{5mm} + \beta_{T,M} \cdot m + \beta_{T,B} \cdot e \cdot m +
    \beta_{T,u} \cdot u_T, \\
  f_D(x, u_D)
    &= \B 1(u_D \leq d(x)), \\
  f_Y(m, u_Y, \delta)
    &= \B 1(u_Y \leq \logit^{-1}(\beta_{Y,0} + m + \beta_{Y, D} \cdot \delta)),
\end{align*}
where \(\logit^{-1}(x) = (1 + \exp(-x))^{-1}\) and \(d(x)\) is the decision
policy. In our example, we use constants \(\mu_A = \tfrac 1 3\), \(\beta_{E,0} =
1\), \(\beta_{E,A} = -1\), \(\beta_{M,0} = 0\), \(\beta_{M,E} = 1\),
\(\beta_{T,0} = 50\), \(\beta_{T,E} = 4\), \(\beta_{T,M} = 4\), \(\beta_{T,u} =
7\), \(\beta_{T,B} = 1\), \(\beta_{Y,0} = - \tfrac 1 2\), \(\beta_{Y,D}= \tfrac
1 2\). We also assume a budget \(b=\frac{1}{4}\).

\section{Proof of Theorem~\ref{thm:blinding}}
\label{app:blinding}

We begin with the following simple lemma.

\begin{lemma}
\label{lem:rev-cheby}
    Suppose \(\C F\) is a sub--\(\sigma\)-algebra of measurable sets, and
    suppose \(X\) is a non-negative bounded random variable with \(X \leq b\)
    a.s. Then
    \[
        \frac {\D {Var}(X \mid \C F)} b \leq \B E[X \mid \C F].
    \]
\end{lemma}

\begin{proof}
    Note that since \(0 \leq X \leq b\) a.s.,
    \[
        \B E[X^2 \mid \C F] \leq \B E[X \mid \C F] \cdot b.
    \]
    By Jensen's inequality, \(\D {Var}(X \mid \C F) \leq \B E[X^2 \mid \C F]\),
    and so, it follows that, a.s.,
    \[
        \frac {\D {Var}(X \mid \C F)} b \leq \B E[X \mid \C F],
    \]
    as desired.
\end{proof}

Ignoring the conditioning, Lemma~\ref{lem:rev-cheby} can be interpreted as
saying that the minimum of a bounded random variable cannot be too close to the
mean. This fact enables us to prove Theorem~\ref{thm:blinding}.

\begin{proof}[Proof of Theorem~\ref{thm:blinding}]
    We wish to show that no event \(E\) of the form \(\{r(X) \geq t\} \subseteq
    E \subseteq \{r(X) > t\}\) is \(\pi\)-measurable. To that end, it suffices
    to show that \(\B E[E \mid \pi(X)] \notin \{0,1\}\) with positive
    probability.

    Since \(\Pr(r(X) = t) = 0\), \(\B 1_{r(X) \geq t} = \B 1_{r(X) > t}\) a.s.,
    and so it suffices to show that \(\B E[\B 1_{r(X) > t} \mid \pi(X)] \notin
    \{0,1\}\) with positive probability.

    Consider the set of covariates \(x = (x_u, a)\) such that \(\rho(x)\) lies
    in the interval \((t, t + \epsilon)\), where, without loss of generality, we
    assume \(t + \epsilon < 1\). Since \(\B E[r(X) \mid X_u = x_u] = \rho(x) >
    t\), it follows immediately that for these \(x\), \(\Pr(r(X) > t \mid X_u) >
    0\). Let \(I(x)\) denote the essential infimum of the conditional
    distribution of \(r(X) \mid \rho(X) = \rho(x)\).

    Then, we can apply Lemma~\ref{lem:rev-cheby} to \(r(X) - I(X)\), using the
    fact that \(0 \leq r(X) - I(X) \leq 1\) a.s., to obtain that \(\rho(X) -
    I(X) > \epsilon\) a.s. It follows that the event
    \begin{equation}
    \label{eq:cntrdct}
        \{\rho(X) \in (t, t + \epsilon), \Pr(r(X) < t \mid \rho(X)) > 0\}
    \end{equation}
    has positive probability. We can conclude from this that the related event
    \[
        \{\rho(X) \in (t, t + \epsilon), \Pr(r(X) < t \mid \pi(X)) > 0\}
    \]
    cannot have probability zero, since, by the tower law, we would consequently
    have that, a.s.,
    \[
        0 = \B 1_{\rho(X) \in (t, t + \epsilon)} \cdot \Pr(r(X) < t \mid
        \pi(X)),
    \]
    and so, taking conditional expectations with respect to \(\rho(X)\), a.s.,
    \[
        0 = \B 1_{\rho(X) \in (t, t + \epsilon)} \cdot \Pr(r(X) < t \mid
        \rho(X)),
    \]
    which contradicts Eq.~\eqref{eq:cntrdct}.

    Therefore, for \(x\) such that \(\rho(x) \in (t, t + \epsilon)\), \(\Pr(r(X)
    < t \mid X_u = x_u) > 0\) and \(\Pr(r(X) > t \mid X_u = x_u) > 0\). Since
    \(\Pr(\rho(X) \in (t, t + \epsilon)) > 0\), it follows that \(\B E[\B
    1_{r(X) > t} \mid \pi(X)] \notin \{0,1\}\) with positive probability, as
    desired.
\end{proof}

\section{Proof of Proposition~\ref{prop:threshold}}

We begin by more formally defining (multiple) threshold policies. We assume,
without loss of generality, that \(\Pr(A = a) > 0\) for all \(a \in \C A\)
throughout.

\begin{definition}
  Let \(u(x)\) be a utility function. We say that a policy \(d(x)\) is a
  \emph{threshold policy} with respect to \(u\) if there exists some \(t\) such
  that
  \begin{equation*}
    d(x) =
      \begin{cases}
        1 & u(x) > t, \\
        0 & u(x) < t,
      \end{cases}
  \end{equation*}
  and \(d(x) \in [0, 1]\) is arbitrary if \(u(x) = t\). We say that \(d(x)\) is
  a \emph{multiple threshold policy} with respect to \(u\) if there exist
  group-specific constants \(t_a\) for \(a \in \C A\) such that
  \begin{equation*}
    d(x) =
      \begin{cases}
        1 & u(x) > t_{\alpha(x)}, \\
        0 & u(x) < t_{\alpha(x)},
      \end{cases}
  \end{equation*}
  and \(d(x) \in [0, 1]\) is arbitrary if \(u(x) = t_{\alpha(x)}\).
\end{definition}

\begin{remark}
\label{rmk:unique_threshold}
  In general, it is possible for different thresholds to produce threshold
  policies that are almost surely equal. For instance, if \(u(X) \sim
  \bern(\tfrac 1 2)\), then the policies \(\B 1_{u(X) > p}\) are almost surely
  equal for all \(p \in [0, 1)\). Nevertheless, we speak in general of
  \emph{the} threshold associated with the threshold policy \(d(X)\) unless
  there is ambiguity.
\end{remark}

We first observe that if \(\C U\) is consistent modulo \(\alpha\), then whether
a decision policy \(d(x)\) is a multiple threshold policy does not depend on our
choice of \(u \in \C U\).

\begin{lemma}
\label{lem:canonical}
  Let \(\C U\) be a collection of utilities consistent modulo \(\alpha\), and
  suppose \(d : \C X \to [0,1]\) is a decision rule. If \(d(x)\) is a multiple
  threshold rule with respect to a utility \(u^* \in \C U\), then \(d(x)\) is a
  multiple threshold rule with respect to every \(u \in \C U\). In particular,
  if \(d(x)\) can be represented by non-negative thresholds over \(u^*\), it can
  be represented by non-negative thresholds over any \(u \in \C U\).
\end{lemma}

\begin{proof}
  Suppose \(d(x)\) is represented by thresholds \(\{t_a^*\}_{a \in \C A}\) with
  respect to \(u^*\). We construct the thresholds \(\{t_a\}_{a \in \C A}\)
  explicitly.

  Fix \(a \in \C A\) and suppose that there exists \(x^* \in \alpha^{-1}(a)\)
  such that \(u^*(x^*) = t_a^*\). Then set \(t_a = u(x^*)\). Now, if \(u(x) >
  t_a = u(x^*)\) then, by consistency modulo \(\alpha\), \(u^*(x) > u^*(x^*) =
  t_a^*\). Similarly if \(u(x) < t_a\) then \(u^*(x) < t_a^*\). We also note
  that by consistency modulo \(\alpha\), \(\sgn(t_a) = \sgn(u(x^*)) =
  \sgn(u^*(x^*)) = \sgn(t_a^*)\).

  If there is no \(x^* \in \alpha^{-1}(a)\) such that \(u^*(x^*) = t_a^*\), then
  let
  \begin{equation*}
    t_a = \inf_{x \in S_a} u(x)
  \end{equation*}
  where \(S_a = \{x \in \alpha^{-1}(a) \mid u^*(x) > t_a^* \}\). Note that since
  \(\sgn(u(x)) = \sgn(u^*(x))\) for all \(x\) by consistency modulo \(\alpha\),
  if \(t_a^* \geq 0\), it follows that \(t_a \geq 0\) as well.

  We need to show in this case also that if \(u(x) > t_a\) then \(u^*(x) >
  t_a^*\), and if \(u(x) < t_a\) then \(u^*(x) < t_a^*\). To do so, let \(x \in
  \alpha^{-1}(a)\) be arbitrary, and suppose \(u(x) > t_a\). Then, by
  definition, there exists \(x' \in \alpha^{-1}(a)\) such that \(u(x) > u(x') >
  t_a\) and \(u^*(x') > t_a^*\), whence \(u^*(x) > u^*(x') > t_a^*\) by
  consistency modulo \(\alpha\). On the other hand, if \(u(x) < t_a\), it
  follows by the definition of \(t_a\) that \(u^*(x) \leq t_a^*\); since
  \(u^*(x) \neq t_a^*\) by hypothesis, it follows that \(u^*(x) < t_a^*\).

  Therefore, it follows in both cases that for
  \(x \in \alpha^{-1}(a)\), if \(u(x) > t_a\) then \(u^*(x) > t_a^*\), and if
  \(u(x) < t_a\) then \(u^*(x) < t_a^*\). Therefore
  \begin{equation*}
    d(x)=
    \begin{cases}
      1 & \text{if} \ u(x) > t_{\alpha(x)}, \\
      0 & \text{if} \ u(x) < t_{\alpha(x)}, \\
    \end{cases}
  \end{equation*}
  i.e., \(d(x)\) is a multiple threshold policy with respect to \(u\). Moreover,
  as noted above, if \(t_a^* \geq 0\) for all \(a \in \C A\), then \(t_a \geq
  0\) for all \(a \in \C A\).
\end{proof}

We now prove the following strengthening of Prop.~\ref{prop:threshold}.

\begin{lemma}
\label{lem:threshold}
  Let \(\C U\) be a collection of utilities consistent modulo \(\alpha\). Let
  \(d(x)\) be a feasible decision policy that is not a.s.\ a multiple threshold
  policy with non-negative thresholds with respect to \(\C U\), then \(d(x)\) is
  strongly Pareto dominated.
\end{lemma}

\begin{proof}
  We prove the claim in two parts. First, we show that any policy that is not a
  multiple threshold policy is strongly Pareto dominated. Then, we show that any
  multiple threshold policy that cannot be represented with non-negative
  thresholds is strongly Pareto dominated.

  If \(d(x)\) is not a multiple threshold policy, then there exists a \(u \in \C
  U\) and \(a^* \in \C A\) such that \(d(x)\) is not a threshold policy when
  restricted to \(\alpha^{-1}(a^*)\) with respect to \(u\).

  We will construct an alternative policy \(d'(x)\) that attains strictly
  greater utility on \(\alpha^{-1}(a^*)\) and is identical elsewhere. Thus,
  without loss of generality, we assume there is a single group, i.e.,
  \(\alpha(x) = a^*\). The proof proceeds heuristically by moving some of the
  mass below a threshold to above a threshold to create a feasible policy with
  improved utility.

  Let \(b = \EE[d(X)]\). Define
  \begin{align*}
    m^{\low}(t)
      &= \EE[d(X) \cdot \B 1_{u(X) < t}], \\
    m^{\up}(t)
      &= \EE[(1 - d(X)) \cdot \B 1_{u(X) > t}].
  \end{align*}
  We show that there exists \(t^*\) such that \(m^{\up}(t^*) > 0\) and
  \(m^{\low}(t^*) > 0\). For, if not, consider
  \begin{equation*}
    \tilde t = \inf \{t \in \B R : m^{\up}(t) = 0\}.
  \end{equation*}
  Note that \(d(X) \cdot \B 1_{u(X) > \tilde t} = \B 1_{u(X) > \tilde t}\) a.s.
  If \(\tilde{t} = -\infty\), then by definition \(d(X) = 1\) a.s., which is a
  threshold policy, violating our assumption on \(d(X)\). If \(\tilde t >
  -\infty\), then for any \(t' < \tilde t\), we have, by definition that
  \(m^{\up}(t') > 0\), and so by hypothesis \(m^{\low}(t') = 0\). Therefore
  \(d(X) \cdot \B 1_{u(X) < \tilde t} = 0\) a.s., and so, again, \(d(X)\) is a
  threshold policy, contrary to hypothesis.

  Now, with \(t^*\) as above, for notational simplicity, let \(m^{\up} =
  m^{\up}(t^*)\) and \(m^{\low} = m^{\low}(t^*)\) and consider the alternative
  policy
  \begin{equation*}
    d'(x) =
      \begin{cases}
        (1 - m^{\up}) \cdot d(x) & u(x) < t^*, \\
        d(x) & u(x) = t^*, \\
        1 - (1 - m^{\low}) \cdot (1 - d(x))
        & u(x) > t^*.
      \end{cases}
  \end{equation*}
  Then it follows by construction that
  \begin{align*}
     \EE[d'(X)]
      &= (1 - m^{\up}) \cdot m^{\low} + \EE[d(X) \cdot \B 1_{u(X) = t^*}] +
      \Pr(u(X) > t^*) - (1 - m^{\low}) \cdot m^{\up} \\
      &= m^{\low} + \EE[d(X) \cdot \B 1_{u(X) = t^*}] + \Pr(u(X) > t^*) -
      m^{\up} \\
      &= \EE[d(X) \cdot \B 1_{u(X) < t^*}] + \EE[d(X) \cdot \B 1_{u(X) = t^*}] +
      \EE[\B 1_{u(X) > t^*}] - \EE[(1 - d(X)) \cdot \B 1_{u(X) >
      t^*}] \\
      &= \EE[d(X)] \\
      &= b,
  \end{align*}
  so \(d'(x)\) is feasible. However,
  \begin{equation*}
    d'(x) - d(x)
      = m^{\low} \cdot (1 - d(x)) \cdot \B 1_{u(x) > t^*} - m^{\up} \cdot d(x)
        \cdot \B 1_{u(x) < t^*},
  \end{equation*}
  and so
  \begin{align*}
    \EE[(d'(X) - d(X)) \cdot u(X)]
      &= m^{\low} \cdot \EE[(1 - d(X)) \cdot \B 1_{u(X) > t^*} \cdot u(X)] \\
      &\hspace{2cm} - m^{\up} \cdot \EE[d(X) \cdot \B 1_{u(X) < t^*} \cdot u(X)]
      \\
      &> m^{\low} \cdot t^* \cdot \EE[(1 - d(X)) \cdot \B 1_{u(X) > t^*}] -
      m^{\up} \cdot t^* \cdot \EE[d(X) \cdot \B 1_{u(X) < t^*}]
      \\
      &= t^* \cdot m^{\low} \cdot m^{\up} - t^* \cdot m^{\up} \cdot m^{\low} \\
      &= 0.
  \end{align*}
  Therefore
  \begin{equation*}
    \EE[d(X) \cdot u(X)] < \EE[d'(X) \cdot u(X)].
  \end{equation*}

  It remains to show that \(u'(d') > u'(d)\) for arbitrary \(u' \in \C U\). Let
  \begin{equation*}
    t' = \inf \{u'(x) : d'(x) > d(x)\}.
  \end{equation*}
  Note that by construction for any \(x, x' \in \C X\), if \(d'(x) > d(x)\) and
  \(d'(x') < d(x')\), then \(u(x) > t^* > u(x')\). It follows by consistency
  modulo \(\alpha\) that \(u'(x) \geq t' \geq u'(x')\), and, moreover, that at
  least one of the inequalities is strict. Without loss of generality, assume
  \(u'(x) > t' \geq u'(x')\). Then, we have that \(u(x) > t^*\) if and only if
  \(u'(x) > t'\). Therefore, it follows that
  \begin{equation*}
    \EE[(d'(X) - d(X)) \cdot \B 1_{u'(X) > t'}] = m^{\up} > 0.
  \end{equation*}
  Since \(\EE[d'(X) - d(X)] = 0\), we see that
  \begin{align*}
    \EE[(d'(X) - d(X)) \cdot u'(X)]
      &= \EE[(d'(X) - d(X)) \cdot \B 1_{u'(X) > t'} \cdot u'(X)] \\
      &\hspace{0.75cm} +\EE[(d'(X) - d(X)) \cdot \B 1_{u'(X) \leq t'} \cdot
      u'(X)] \\
      &> t' \cdot \EE[(d'(X) - d(X)) \cdot \B 1_{u'(X) > t'}] \\
      &\hspace{0.75cm}+ t' \cdot \EE[(d'(X) - d(X)) \cdot \B 1_{u'(X) \leq t'}]
      \\
      &= t' \cdot \EE[d'(X) - d(X)] \\
      &= 0,
  \end{align*}
  where in the inequality we have used the fact that if \(d'(x) > d(x)\),
  \(u'(x) > t' \), and if \(d'(x) < d(x)\), \(u'(x) \leq t'\). Therefore
  \begin{equation*}
    \EE[d(X) \cdot u'(X)] < \EE[d'(X) \cdot u'(X)],
  \end{equation*}
  i.e., \(d'(x)\) strongly Pareto dominates \(d(x)\).

  Now, we prove the second claim, namely, that a multiple threshold policy
  \(\tau(x)\) that cannot be represented with non-negative thresholds is
  strongly Pareto dominated. For, if \(\tau(x)\) is such a policy, then, by
  Lemma~\ref{lem:canonical}, for any \(u \in \C U\), \(\EE[\tau(X) \cdot \B
  1_{u(X) < 0}] > 0\). It follows immediately that \(\tau'(x) = \tau(x) \cdot \B
  1_{u(x) > 0}\) satisfies \(u(\tau') > u(\tau)\). By consistency modulo
  \(\alpha\), the definition of \(\tau'(x)\) does not depend on our choice of
  \(u\), and so \(u(\tau') > u(\tau)\) for every \(u \in \C U\), i.e.,
  \(\tau'(x)\) strongly Pareto dominates \(\tau(x)\).
\end{proof}

The following results, which draw on Lemma~\ref{lem:threshold}, are useful in
the proof of Theorem~\ref{thm:dist}.

\begin{definition}
  We say that a decision policy \(d(x)\) is \emph{budget-exhausting} if
  \begin{equation*}
    \begin{split}
      \min(b, \Pr(u(X) > 0)) & \leq \EE[d(X)]\\
      &\leq \min(b, \Pr(u(X) \geq 0)).
    \end{split}
  \end{equation*}
\end{definition}

\begin{remark}
  We note that if \(\C U\) is consistent modulo \(\alpha\), then whether or not
  a decision policy \(d(x)\) is budget-exhausting does not depend on the choice
  of \(u \in \C U\). Further, if \(\Pr(u(X) = 0) = 0\)---e.g., if the
  distribution of \(X\) is \(\C U\)-fine---then the decision policy is
  budget-exhausting if and only if \(\EE[d(X)] = \min(b, \Pr(u(X) > 0))\).
\end{remark}

\begin{corollary}
\label{cor:exhaust}
  Let \(\C U\) be a collection of utilities consistent modulo \(\alpha\). If
  \(\tau(x)\) is a feasible policy that is not a budget-exhausting multiple
  threshold policy with non-negative thresholds, then \(\tau(x)\) is strongly
  Pareto dominated.
\end{corollary}

\begin{proof}
  Suppose \(\tau(x)\) is not strongly Pareto dominated. By
  Lemma~\ref{lem:threshold}, it is a multiple threshold policy with non-negative
  thresholds.

  Now, suppose toward a contradiction that \(\tau(x)\) is not budget-exhausting.
  Then, either \(\EE[\tau(X)] > \min(b, \Pr(u(X) \geq 0))\) or \(\EE[\tau(X)] <
  \min(b, \Pr(u(X) > 0))\).

  In the first case, since \(\tau(x)\) is feasible, it follows that
  \(\EE[\tau(X)] > \Pr(u(X) \geq 0)\). It follows that \(\tau(x) \cdot \B
  1_{u(x) < 0}\) is not almost surely zero. Therefore
  \begin{equation*}
    \EE[\tau(X)] < \EE[\tau(X) \cdot \B 1_{u(X) > 0}],
  \end{equation*}
  and, by consistency modulo \(\alpha\), this holds for any \(u \in \C U\).
  Therefore \(\tau(x)\) is strongly Pareto dominated, contrary to hypothesis.

  In the second case, consider
  \begin{equation*}
    d(x) = \theta \cdot \B 1_{u(x) > 0} + (1 - \theta) \cdot \tau(x).
  \end{equation*}
  Since \(\EE[\tau(X)] < \min(b, \Pr(u(X) > 0))\) and
  \begin{equation*}
    \EE[d(X)] = \theta \cdot \Pr(u(X) > 0) + (1 - \theta) \cdot \EE[\tau(X)],
  \end{equation*}
  there exists some \(\theta > 0\) such that \(d(x)\) is feasible.

  For that \(\theta\), a similar calculation shows immediately that \(u(d) >
  u(\tau)\), and, by consistency modulo \(\alpha\), \(u'(d) > u'(\tau)\) for all
  \(u' \in \C U\). Therefore, again, \(d(x)\) strongly Pareto dominates
  \(\tau(x)\), contrary to hypothesis.
\end{proof}

\begin{lemma}
\label{lem:quantile}
  Given a utility \(u\), there exists a mapping \(T\) from \([0, 1]^{\C A}\) to
  \([-\infty, \infty]^{\C A}\) taking sets of quantiles \(\{q_a\}_{a \in \C A}\)
  to thresholds \(\{t_a\}_{a \in \C A}\) such that:
  \begin{enumerate}
    \item \(T\) is monotonically non-increasing in each coordinate;
    \item For each set of quantiles, there is a multiple threshold policy \(\tau
      : \C X \to [0, 1]\) with thresholds \(T(\{q_a\})\) with respect to \(u\)
      such that \(\EE[\tau(X) \mid A = a] = q_a\).
  \end{enumerate}
\end{lemma}

\begin{proof}
  Simply choose
  \begin{equation}
  \label{eq:monotonic}
    t_a = \inf \{s \in \B R : \Pr(u(X) > s) < q_a \}.
  \end{equation}
  Then define
  \begin{equation*}
    p_a =
      \begin{cases}
        \frac {q_a - \Pr(u(X) > t_a \mid A = a)} {\Pr(u(X) = t_a \mid A = a)}
          & \hspace{-4.13467pt} \Pr(u(X) = t_a, A = a) > 0 \\
        0
          & \hspace{-4.13467pt} \Pr(u(X) = t_a, A = a) = 0.
      \end{cases}
  \end{equation*}
  Note that \(\Pr(u(X) \geq t_a \mid A = a) \geq q_a\), since, by definition,
  \(\Pr(u(X) > t_a - \epsilon \mid A = a) \geq q_a\) for all \(\epsilon > 0\).
  Therefore,
  \begin{equation*}
    \Pr(u(X) > t_a \mid A = a) + \Pr(u(X) = t_a \mid A = a) \geq q_a,
  \end{equation*}
  and so \(p_a \leq 1\). Further, since \(\Pr(u(X) > t_a \mid A = a) \leq q_a\),
  we have that \(p_a \geq 0\).

  Finally, let
  \begin{equation*}
    d(x) =
      \begin{cases}
        1 & u(x) > t_{\alpha(x)}, \\
        p_a & u(x) = t_{\alpha(x)}, \\
        0 & u(x) < t_{\alpha(x)},
      \end{cases}
  \end{equation*}
  and it follows immediately that \(\EE[d(X) \mid A = a] = q_a\). That \(t_a\)
  is a monotonically non-increasing function of \(q_a\) follows immediately from
  Eq.~\eqref{eq:monotonic}.
\end{proof}

We can further refine Cor.~\ref{cor:exhaust} and Lemma~\ref{lem:quantile} as
follows:

\begin{lemma}
\label{lem:maximize}
  Let \(u\) be a utility. Then a feasible policy is utility maximizing if and
  only if it is a budget-exhausting threshold policy. Moreover, there exists at
  least one utility maximizing policy.
\end{lemma}

\begin{proof}
  Let \(\bar \alpha\) be a constant map, i.e., \(\bar \alpha : \C X \to \bar {\C
  A}\), where \(|\bar {\C A}| = 1\). Then \(\C U = \{u\}\) is consistent modulo
  \(\bar \alpha\), and so by Cor.~\ref{cor:exhaust}, any Pareto efficient policy
  is a budget exhausting multiple threshold policy relative to \(\C U\). Since
  \(\C U\) contains a single element, a policy is Pareto efficient if and only
  if it is utility maximizing. Since \(\bar \alpha\) is constant, a policy is a
  multiple threshold policy relative to \(\bar \alpha\) if and only if it is a
  threshold policy. Therefore, a policy is utility maximizing if and only if it
  is a budget exhausting threshold policy. By Lemma~\ref{lem:quantile}, such a
  policy exists, and so the maximum is attained.
\end{proof}

\section{Prevalence and the Proof of Theorem~\ref{thm:dist}}
\label{app:dist}

The notion of a probabilistically ``small'' set---such as the event in which an
idealized dart hits the exact center of a target---is, in finite-dimensional
real vector spaces, typically encoded by the idea of a Lebesgue null set.

Here we prove that the set of distributions such that there exists a policy
satisfying either counterfactual equalized odds, conditional principal fairness,
or counterfactual fairness that is not strongly Pareto dominated is ``small'' in
an analogous sense. The proof turns on the following intuition. Each of the
fairness definitions imposes a number of constraints. By
Lemma~\ref{lem:threshold}, any policy that is not strongly Pareto dominated is a
multiple threshold policy. By adjusting the group-specific thresholds of such a
policy, one can potentially satisfy one constraint per group. If there are more
constraints than groups, then one has no additional degrees of freedom that can
be used to ensure that the remaining constraints are satisfied. If, by chance,
those constraints \emph{are} satisfied with the same threshold policy, they are
not satisfied robustly---even a minor distribution shift, such as increasing the
amount of mass above the threshold by any amount on the relevant subpopulation,
will break them. Therefore, over a ``typical'' distribution, at most \(|\C A|\)
of the constraints can simultaneously be satisfied by a Pareto efficient policy,
meaning that typically no Pareto efficient policy fully satisfies all of the
conditions of the fairness definitions.

Formalizing this intuition, however, requires considerable care. In
Section~\ref{sec:shyness}, we give a brief introduction to a popular
generalization of null sets to infinite-dimensional vector spaces, drawing
heavily on a review article by \citet{ott2005prevalence}. In
Section~\ref{sec:roadmap} we provide a road map of the proof itself. In
Section~\ref{sec:shyness-prelims}, we establish the main hypotheses necessary to
apply the notion of prevalence to a convex set---in our case, the set of \(\C
U\)-fine distributions. In Section~\ref{sec:shyness-prelims2}, we establish a
number of technical lemmata used in the proof of Theorem~\ref{thm:dist}, and
provide a proof of the theorem itself in Section~\ref{sec:shyness-proof}. In
Section~\ref{sec:counterexample}, we show why the hypothesis of \(\C
U\)-fineness is important and how conspiracies between atoms in the distribution
of \(u(X)\) can lead to ``robust'' counterexamples.

\subsection{Shyness and Prevalence}
\label{sec:shyness}

Lebesgue measure \(\lambda_n\) on \(\B R^n\) has a number of desirable
properties:
  \begin{itemize}
    \item \textbf{Local finiteness:} For any point \(v \in \B R^n\), there
      exists an open set \(U\) containing \(x\) such that \(\lambda_n[U] <
      \infty\);
    \item \textbf{Strict positivity:} For any open set \(U\), if \(\lambda_n[U]
      = 0\), then \(U = \emptyset\);
    \item \textbf{Translation invariance:} For any \(v \in \B R^n\) and
      measurable set \(E\), \(\lambda_n[E + v] = \lambda_n[E]\).
  \end{itemize}
No measure on an infinite-dimensional, separable Banach space, such as \(L^1(\B
R)\), can satisfy these three properties \cite{ott2005prevalence}. However,
while there is no generalization of Lebesgue measure to infinite dimensions,
there is a generalization of Lebesgue null sets---called \emph{shy} sets---to
the infinite-dimensional context that preserves many of their desirable
properties.

\begin{definition}[\citet{hunt1992prevalence}]
  Let \(V\) be a completely metrizable topological vector space. We say that a
  Borel set \(E \subseteq V\) is \emph{shy} if there exists a Borel measure
  \(\mu\) on \(V\) such that:
  \begin{enumerate}
    \item There exists compact \(C \subseteq V\) such that \(0 < \mu[C] <
      \infty\),
    \item For all \(v \in V\), \(\mu[E + v] = 0\).
  \end{enumerate}
  An arbitrary set \(F \subseteq V\) is shy if there exists a shy Borel set \(E
  \subseteq V\) containing \(F\).

  We say that a set is \emph{prevalent} if its complement is shy.
\end{definition}

Prevalence generalizes the concept of Lebesgue ``full measure'' or ``co-null''
sets (i.e., sets whose complements have null Lebesgue measure) in the following
sense:

\begin{proposition}[\citet{hunt1992prevalence}]
\label{prop:shy_axioms}
  Let \(V\) be a completely metrizable topological vector space. Then:
  \begin{itemize}
    \item Any prevalent set is dense in \(V\);
    \item If \(G \subseteq L\) and \(G\) is prevalent, then \(L\) is prevalent;
    \item A countable intersection of prevalent sets is prevalent;
    \item Every translate of a prevalent set is prevalent;
    \item If \(V = \B R^n\), then \(G \subseteq \B R^n\) is prevalent if and
      only if \(\lambda_n[\B R^n \setminus G] = 0\).
  \end{itemize}
\end{proposition}

As is conventional for sets of full measure in finite-dimensional spaces, if
some property holds for every \(v \in E\), where \(E\) is prevalent, then we say
that the property holds for \emph{almost every \(v \in V\)} or that it holds
\emph{generically in \(V\)}.

Prevalence can also be generalized from vector spaces to convex subsets of
vector spaces, although additional care must be taken to ensure that a relative
version of Prop.~\ref{prop:shy_axioms} holds.

\begin{definition}[\citet{anderson2001genericity}]
\label{defn:shy_rel}
  Let \(V\) be a topological vector space and let \(C \subseteq V\) be a convex
  subset completely metrizable in the subspace topology induced by \(V\). We say
  that a universally measurable set \(E \subseteq C\) is \emph{shy in \(C\) at
  \(c \in C\)} if for each \(1 \geq \delta > 0\), and each neighborhood \(U\) of
  \(0\) in \(V\), there is a regular Borel measure \(\mu\) with compact support
  such that
  \begin{equation*}
    \supp(\mu) \subseteq \left(\delta(C - c) + c \right) \cap (U + c),
  \end{equation*}
  and \(\mu[E + v] = 0\) for every \(v \in V\).

  We say that \(E\) is \emph{shy in \(C\)} or \emph{shy relative to \(C\)} if
  \(E\) is shy in \(C\) at \(c\) for every \(c \in C\). An arbitrary set \(F
  \subseteq V\) is shy in \(C\) if there exists a universally measurable shy set
  \(E \subseteq C\) containing \(F\).

  A set \(G\) is \emph{prevalent} in \(C\) if \(C \setminus G\) is shy in \(C\).
\end{definition}

\begin{proposition}[\citet{anderson2001genericity}]
  If \(E\) is shy at some point \(c \in C\), then \(E\) is shy at every point in
  \(C\) and hence is shy in \(C\).
\end{proposition}

Sets that are shy in \(C\) enjoy similar properties to sets that are shy in
\(V\).

\begin{proposition}[\citet{anderson2001genericity}]
\label{prop:shy_axioms_rel}
  Let \(V\) be a topological vector space and let \(C \subseteq V\) be a convex
  subset completely metrizable in the subspace topology induced by \(V\). Then:
  \begin{itemize}
    \item Any prevalent set in \(C\) is dense in \(C\);
    \item If \(G \subseteq L\) and \(G\) is prevalent in \(C\), then \(L\) is
      prevalent in \(C\);
    \item A countable intersection of sets prevalent in \(C\) is prevalent in
      \(C\)
    \item If \(G\) is prevalent in \(C\) then \(G + v\) is prevalent in \(C +
      v\) for all \(v \in V\).
    \item If \(V = \B R^n\) and \(C \subseteq V\) is a convex subset with
      non-empty interior, then \(G \subseteq C\) is prevalent in \(C\) if and
      only if \(\lambda_n[C \setminus G] = 0\).
  \end{itemize}
\end{proposition}

Sets that are shy in \(C\) can often be identified by inspecting their
intersections with a finite-dimensional subspace \(W\) of \(V\), a strategy we
use to prove Theorem~\ref{thm:dist}.

\begin{definition}[\citet{anderson2001genericity}]
  A universally measurable subset \(E\) of a convex and completely metrizable
  set \(C\) is said to be \emph{\(k\)-shy in \(C\)} if there exists a
  \(k\)-dimensional subspace \(W \subseteq V\) such that
  \begin{enumerate}
    \item A translate of the set \(C\) has positive Lebesgue measure in \(W\),
      i.e., \(\lambda_W[C + v_0] > 0\) for some \(v_0 \in V\);
    \item Every translate of the set \(E\) is a Lebesgue null set in \(W\),
      i.e., \(\lambda_W[E + v] = 0\) for all \(v \in V\).
  \end{enumerate}
  Here \(\lambda_W\) denotes \(k\)-dimensional Lebesgue measure supported on
  \(W\).\footnote{%
    Note that Lebesgue measure on \(W\) is only defined up to a choice of basis;
    however, since \(\lambda[T(A)] = |\det(T)| \cdot \lambda[A]\) for any linear
    automorphism \(T\) and Lebesgue measure \(\lambda\), whether a set has null
    measure does not depend on the choice of basis.
  }
  We refer to such a \(W\) as a \emph{\(k\)-dimensional probe} witnessing the
  \(k\)-shyness of \(E\), and to an element \(w \in W\) as a
  \emph{perturbation}.
\end{definition}

The following intuition motivates the use of probes to detect shy sets. By
analogy with Fubini's theorem, one can imagine trying to determine whether a
subset of a finite-dimensional vector space is large or small by looking at its
cross sections parallel to some subspace \(W \subseteq V\). If a set \(E
\subseteq V\) is small in each cross section---i.e., if \(\lambda_W[E + v] = 0\)
for all \(v \in V\)---then \(E\) itself is small in \(V\), i.e., \(E\) has
\(\lambda_V\)-measure zero.

\begin{proposition}[\citet{anderson2001genericity}]
\label{prop:k_shy}
  Every \(k\)-shy set in \(C\) is shy in \(C\).
\end{proposition}

\subsection{Outline}
\label{sec:roadmap}

To aid the reader in following the application of the theory in
Section~\ref{sec:shyness} to the proof of Theorem~\ref{thm:dist}, we provide the
following outline of the argument.

In \textbf{Section~\ref{sec:shyness-prelims}} we establish the context to which
we apply the notion of relative shyness. In particular, we introduce the vector
space \(\B K\) consisting of the \emph{totally bounded Borel measures} on the
state space \(\C K\)---where \(\C K\) is \(\C X \times \C Y\), \(\C X \times \C
Y \times \C Y\), or \(\C A \times \C X^{\C A}\), depending on which notion of
fairness is under consideration. We further isolate the subspace \(\bb K
\subseteq \B K\) of \(\C U\)-fine totally bounded Borel measures. Within this
space, we are interested in the convex set \(\bb Q \subseteq \bb K\), the set of
\emph{\(\C U\)-fine joint probability distributions} of, respectively, \(X\) and
\(Y(1)\); \(X\), \(Y(0)\), \(Y(1)\); or \(A\) and the \(X_{\Pi, A, a}\). Within
\(\bb Q\), we identify \(\bb E \subseteq \bb Q\), the set of \(\C U\)-fine
distributions on \(\C K\) \emph{over which there exists a policy satisfying the
relevant fairness definition that is not strongly Pareto dominated}. The claim
of Theorem~\ref{thm:dist} is that \(\bb E\) is shy relative to \(\bb Q\).

To ensure that relative shyness generalizes Lebesgue null measure in the
expected way---i.e., that Prop.~\ref{prop:shy_axioms_rel}
holds---Definition~\ref{defn:shy_rel} has three technical requirements: (1) that
the ambient vector space \(V\) be a topological vector space; (2) that the
convex set \(C\) be completely metrizable; and (3) that the shy set \(E\) be
universally measurable. In \textbf{Lemma~\ref{lem:banach}}, we observe that \(\B
K\) is a complete topological vector space under the total variation norm, and
so is a Banach space. We extend this in \textbf{Cor.~\ref{cor:banach}}, showing
that \(\bb K\) is also a Banach space. We use this fact in
\textbf{Lemma~\ref{lem:convex}} to show that \(\bb Q\) is a completely
metrizable subset of \(\bb K\), as well as convex. Lastly, in
\textbf{Lemma~\ref{lem:e_closed}}, we show that the set \(\bb E\) is closed, and
therefore universally measurable.

In \textbf{Section~\ref{sec:shyness-prelims2}}, we develop the machinery needed
to construct a probe \(\bb W\) for the proof of Theorem~\ref{thm:dist} and prove
several lemmata simplifying the eventual proof of the theorem. To build the
probe, it is necessary to construct measures \(\mu_{\max,a}\) with maximal
support on the utility scale. This ensures that if any two threshold policies
produce different decisions on \emph{any} \(\mu \in \bb K\), they will produce
different decisions on typical perturbations. The construction of the
\(\mu_{\max,a}\), is carried out in \textbf{Lemma~\ref{lem:mtu}} and
\textbf{Cor.~\ref{cor:maximal}}. Next, we introduce the basic style of argument
used to show that a subset of \(\bb Q\) is shy in \textbf{Lemma~\ref{lem:probe}}
and \textbf{Lemma~\ref{lem:condition}}, in particular, by showing that the set
of \(\mu \in \bb Q\) that give positive probability to an event \(E\) is either
prevalent or empty. We use then use a technical lemma,
\textbf{Lemma~\ref{lem:uncountable_sum}}, to show, in effect, that a generic
element of \(\bb Q\) has support on the utility scale wherever a given fixed
distribution \(\mu \in \bb Q\) does. \textbf{In Defn.~\ref{defn:overlap}}, we
introduce the concept of overlapping and splitting utilities, and show in
\textbf{Lemma~\ref{lem:overlap}} that this property is generic in \(\bb Q\)
unless there exists a \(\omega\)-stratum that contains no positive-utility
observables \(x\). Lastly, in \textbf{Lemma~\ref{lem:simple}}, we provide a mild
simplification of the characterization of finitely shy sets that makes the proof
of Theorem~\ref{thm:dist} more straightforward.

Finally, in \textbf{Section~\ref{sec:shyness-proof}}, we give the proof of
Theorem~\ref{thm:dist}. We divide the proof into three parts. In the first part,
we restrict our attention to the case of counterfactual equalized odds, and show
in detail how to combine the lemmata of the previous section to construct the
(at most) \(2 \cdot |\C A|\)-dimensional probe \(\bb W\). In the second part we
consider two distinct cases. The argument in both cases is conceptually
parallel. First, we argue that the balance conditions of counterfactual
equalized odds encoded by Eq.~\eqref{eq:counterfactual_equalized_odds} must be
broken by a typical perturbation in \(\bb W\). In particular, we argue that for
a given base distribution \(\mu\), there can be at most one budget-exhausting
multiple threshold policy that can---although need not necessarily---satisfy
counterfactual equalized odds. We show that the form of this policy cannot be
altered by an appropriate perturbation in \(\bb W\), but that the conditional
probability of a positive decision will, in general, be altered in such a way
that Eq.~\eqref{eq:counterfactual_equalized_odds} can only hold for a
\(\lambda_{\bb W}\)-null set of perturbations. In the final section, we lay out
modifications that can be made to the proof given for counterfactual equalized
odds in the first two parts that adapt the argument to the cases of conditional
principal fairness and path-specific fairness. In particular, we show how to
construct the probe \(\bb W\) in such a way that the additional conditioning on
the reduced covariates \(W = \omega(X)\) in
Eqs.~\eqref{eq:conditional_principal_fairness}~and~\eqref{eq:path_specific_fairness}
does not affect the argument.

\subsection{Convexity, Complete Metrizability, and Universal Measurability}
\label{sec:shyness-prelims}

In this section, we establish the background requirements of
Prop.~\ref{prop:k_shy} for the setting of Theorem~\ref{thm:dist}. In particular,
we exhibit the \(\C U\)-fine distributions as a convex subset of a topological
vector space, the set of totally bounded \(\C U\)-fine Borel measures. We show
that the \(\C U\)-fine probability distributions form a completely metrizable
subset in the topology it inherits from the space of totally bounded measures.
Lastly, we show that the set of regular distributions under which there exists a
Pareto efficient policy satisfying one of the three fairness criteria is closed,
and therefore universally measurable.

\subsubsection{Background and notation}

We begin by establishing some notational conventions. We let \(\C K\) denote the
underlying state space over which the distributions in Theorem~\ref{thm:dist}
range. Specifically, \(\C K = \C X \times \C Y\) in the case of counterfactual
equalized odds; \(\C K = \C X \times \C Y \times \C Y\) in the case of
conditional principal fairness; and \(\C K = \C A \times \C X^{\C A}\) in the
case of path-specific fairness. We note that since \(\C X \subseteq \B R^k\) for
some \(k\) and \(Y \subseteq \B R\), \(\C K\) may equivalently be considered a
subset of \(\B R^n\) for some \(n \in \B N\), with the subspace topology (and
Borel sets) inherited from \(\B R^n\).\footnote{%
  In the case of path-specific fairness, we can equivalently think of \(\C A\)
  as a set of integers indexing the groups.
}

We recall the definition of totally bounded measures.

\begin{definition}
  Let \(\C M\) be a \(\sigma\)-algebra on \(V\), and let \(\mu\) be a countably
  additive \((V, \C M)\)-measure. Then, we define
  \begin{equation}
    |\mu|[E] = \sup \sum_{i = 1}^\infty |\mu[E_i]|
  \end{equation}
  where the supremum is taken over all countable partitions \(\{E_i\}_{i \in \B
  N}\), i.e., collections such that \(\bigcup_{i=1}^\infty E_i = E\) and \(E_i
  \cap E_j = \emptyset\) for \(j \neq i\). We call \(|\mu|\) the \emph{total
  variation of \(\mu\)}, and the \emph{total variation norm of \(\mu\)} is
  \(|\mu|[V]\).

  We say that \(\mu\) is \emph{totally bounded} if its total variation norm is
  finite, i.e., \(|\mu|[V] < \infty\).
\end{definition}

\begin{lemma}
\label{lem:tot_var}
  If \(\mu\) is totally bounded, then \(|\mu|\) is a finite positive measure on
  \((V, \C M)\), and \(|\mu[E]| \leq |\mu|[E]\) for all \(E \in \C M\).
\end{lemma}

See Theorem~6.2 in \citet{rudin1987real} for proof.

We let \(\B K\) denote the set of totally bounded Borel measures on \(\C K\). We
note that, in the case of path specific fairness, which involves the joint
distributions of counterfactuals, \(X\) is not defined directly. Rather, the
joint distribution of the counterfactuals \(X_{\Pi, A, a'}\) and \(A\) defines
the distribution of \(X\) through consistency, i.e., what would have happened to
someone if their group membership were changed to \(a' \in \C A\) is what
actually happens to them if their group membership \emph{is} \(a'\). More
formally, \(\Pr(X \in E \mid A = a') = \Pr(X_{\Pi, A, a'} \in E \mid A = a')\)
for all Borel sets \(E \subseteq \C X\). (See \S~3.6.3 in
\citet{pearl2009causality}.)

For any \(\mu \in \B K\), we adopt the following notational conventions. If we
say that a property holds \(\mu\)-a.s., then the subset of \(\C K\) on which the
property fails has \(|\mu|\)-measure zero. If \(E \subseteq \C K\) is a
measurable set, then we denote by \(\mu \rest_E\) the restriction of \(\mu\) to
\(E\), i.e., the measure defined by the mapping \(E' \mapsto \mu[E \cap E']\).
We let \(\EE_\mu[f] = \int_{\C K} f \, \dx \mu\), and for measurable sets \(E\),
\(\Pr_\mu(E) = \mu[E]\).\footnote{%
  To state and prove our results in a notationally uniform way, we occasionally
  write \(\Pr_\mu(E)\) even when \(\mu\) ranges over measures that may not be
  probability measures.
}
The fairness criteria we consider involve conditional independence relations. To
make sense of conditional independence relations more generally, for Borel
measurable \(f\) we define \(\EE_\mu[f \mid \C F]\) to be the Radon-Nikodym
derivative of the measure \(E \mapsto \EE_\mu[f \cdot \B 1_E]\) with respect to
the measure \(\mu\) restricted to the sub--\(\sigma\)-algebra of Borel sets \(\C
F\). (See \S~34 in \citet{billingsley2008probability}.) Similarly, we define
\(\EE_\mu[f \mid g]\) to be \(\EE_\mu[f \mid \sigma(g)]\), where \(\sigma(g)\)
denotes the sub--\(\sigma\)-algebra of the Borel sets generated by \(g\). In
cases where the condition can occur with non-zero probability, we can instead
make use of the elementary definition of discrete conditional probability.

\begin{lemma}
\label{lem:cond_prob}
  Let \(g\) be a Borel function on \(\C K\), and suppose \(\Pr_\mu(g = c) \neq
  0\) for some constant \(c \in \B R\). Then, we have that \(\mu\)-a.s., for any
  Borel function \(f\),
  \begin{equation*}
    \EE_\mu[f \mid g] \cdot \B 1_{g = c} = \frac{\EE_\mu[f \cdot \B 1_{g = c}]}
    {\Pr_\mu(g = c)} \cdot \B 1_{g = c}.
  \end{equation*}
\end{lemma}

See \citet{rao2005conditional} for proof.

With these notational conventions in place, we turn to establishing the
background conditions of Prop.~\ref{prop:k_shy}.

\begin{lemma}
\label{lem:banach}
  The set of totally bounded measures on a measure space \((V, \C M)\) form a
  complete topological vector space under the total variation norm, and hence a
  Banach space.
\end{lemma}

See, e.g., \citet{steele2019space} for proof. It follows from this that \(\B K\)
is a Banach space.

\begin{remark}
\label{rmk:Borel}
  Since \(\B K\) is a Banach space, it possesses a topology, and consequently a
  collection of Borel subsets. These Borel sets are to be distinguished from the
  Borel subsets of the underlying state space \(\C K\), which the elements of
  \(\B K\) measure. The requirement that the subset \(E\) of the convex set
  \(C\) be universally measurable in Proposition~\ref{prop:k_shy} is in
  reference to the \emph{Borel subsets of \(\B K\)}; the requirement that \(\mu
  \in \B K\) be a Borel measure is in reference to the \emph{Borel subsets of
  \(\C K\)}.
\end{remark}

Recall the definition of absolute continuity.

\begin{definition}
  Let \(\mu\) and \(\nu\) be measures on a measure space \((V, \C M)\). We say
  that a measure \(\nu\) is \emph{absolutely continuous with respect to
  \(\mu\)}---also written \(\nu \Lt \mu\)---if, whenever \(\mu[E] = 0\),
  \(\nu[E] = 0\).
\end{definition}

Absolute continuity is a closed property in the topology induced by the total
variation norm.

\begin{lemma}
\label{lem:abscont}
  Consider the space of totally bounded measures on a measure space \((V, \C
  M)\) and fix \(\mu\). The set of \(\nu\) such that \(\nu \Lt \mu\) is closed.
\end{lemma}

\begin{proof}
  Let \(\{\nu_i\}_{i \in \B N}\) be a convergent sequence of measures absolutely
  continuous with respect to \(\mu\). Let the limit of the \(\nu_i\) be \(\nu\).
  We seek to show that \(\nu \Lt \mu\). Let \(E \in \C M\) be an arbitrary set
  such that \(\mu[E] = 0\). Then, we have that
  \begin{align*}
    \nu[E]
      &= \lim_{n \to \infty} \nu_i[E] \\
      &= \lim_{n \to \infty} 0 \\
      &= 0,
  \end{align*}
  since \(\nu_i \Lt \mu\) for all \(i\). Since \(E\) was arbitrary, the result
  follows.
\end{proof}

Recall the definition of a pushforward measure.

\begin{definition}
  Let \(f : (V, \C M) \to (V', \C M')\) be a measurable function. Let \(\mu\) be
  a measure on \(V\). We define the \emph{pushforward measure} \(\mu \circ
  f^{-1}\) on \(V'\) by the map \(E' \mapsto \mu[f^{-1}(E')]\) for \(E' \in \C
  M'\).
\end{definition}

Within \(\B K\), in the case of counterfactual equalized odds and conditional
principal fairness, we define the subspace \(\bb K\) to be the set of totally
bounded measures \(\mu\) on \(\C K\) such that the pushforward measure \(\mu
\circ u^{-1}\) is absolutely continuous with respect to the Lebesgue measure
\(\lambda\) on \(\B R\) for all \(u \in \C U\). By the Radon-Nikodym theorem,
these pushforward measures arise from densities, i.e., for any \(\mu \in \bb
K\), there exists a unique \(f_\mu \in L^1(\B R)\) such that for any measurable
subset \(E\) of \(\B R\), we have
\begin{equation*}
  \mu \circ u^{-1}[E] = \int_{E} f_\mu \, \dx \lambda.
\end{equation*}
In the case of path-specific fairness, we require the joint distributions of the
counterfactual utilities to have a joint density. That is, we define the
subspace \(\bb K\) to be the set of totally bounded measures \(\mu\) on \(\C K\)
such that the pushforward measure \(\mu \circ (u^{\C A})^{-1}\) is absolutely
continuous with respect to Lebesgue measure on \(\B R^{\C A}\) for all \(u \in
\C U\). Here, we recall that
\begin{equation*}
  u^{\C A} : (a, (x_{a'})_{a' \in \C A}) \mapsto (u(x_{a'}))_{a' \in \C A}.
\end{equation*}
As before, there exists a corresponding density \(f_\mu \in L^1(\B R^{\C A})\).

We therefore see that \(\bb K\) extends in a natural way the notion of a \(\C
U\)- or \(\C U^{\C A}\)-fine distribution, and so, by a slight abuse of
notation, refer to \(\bb K\) as the set of \emph{\(\C U\)-fine measures on \(\C
K\)}.

Indeed, since \(\Pr_\mu(u(X) \in E, A = a) \leq \Pr_\mu(u(X) \in E)\), it also
follows that, for \(a \in \C A\) such that \(\Pr_\mu(A = a) > 0\), the
conditional distributions of \(u(X) \mid A = a\) are also absolutely continuous
with respect to Lebesgue measure, and so also have densities. For notational
convenience, we set \(f_{\mu, a}\) to be the function satisfying
\begin{equation*}
  \Pr_\mu(u(X) \in E, A = a) = \int_E f_{\mu, a} \, \dx \lambda,
\end{equation*}
so that \(f_{\mu} = \sum_{a \in \C A} f_{\mu, a}\).

Since absolute continuity is a closed condition, it follows that \(\bb K\) is a
closed subspace of \(\B K\). This leads to the following useful corollary of
Lemma~\ref{lem:abscont}.

\begin{corollary}
\label{cor:banach}
  The collection of \(\C U\)-fine measures on \(\C K\) is a Banach space.
\end{corollary}

\begin{proof}
  It is straightforward to see that \(\bb K\) is a subspace of \(\B K\). Since
  \(\bb K\) is a closed subset of \(\B K\) by Lemma~\ref{lem:abscont}, it is
  complete, and therefore a Banach space.
\end{proof}

We note the following useful fact about elements of \(\bb K\).

\begin{lemma}
\label{lem:mapping}
  Consider the mapping \(\mu \mapsto f_\mu\) from \(\bb K\) to \(L^1(\B R)\)
  given by associating a measure \(\mu\) with the Radon-Nikodym derivative of
  the pushforward measure \(\mu \circ u^{-1}\). This mapping is continuous.
  Likewise, the mapping \(\mu \mapsto f_{\mu,a}\) is continuous for all \(a \in
  \C A\), and, in the case of path-specific fairness, the mapping of \(\mu\) to
  the Radon-Nikodym derivative of \(\mu \circ (u^{\C A})^{-1}\) is continuous.
\end{lemma}

\begin{proof}
  We show only the first case. The others follow by virtually identical
  arguments.

  Let \(\epsilon > 0\) be arbitrary. Choose \(\mu \in \bb K\), and suppose that
  \(|\mu - \mu'|[\C K] < \epsilon\). Then, let
  \begin{align*}
    E^{\up}
      &= \{x \in \B R : f_\mu(x) > f_{\mu'}(x)\} \\
    E^{\low}
      &= \{x \in \B R : f_\mu(x) < f_{\mu'}(x)\}.
  \end{align*}
  Then \(E^{\up}\) and \(E^{\low}\) are disjoint, so we have that
  \begin{align*}
    \|f_\mu - f_{\mu'}\|_{L^1(\B R)}
      &= \left| \int_{E^{\up}} f_{\mu} - f_{\mu'} \, \dx \lambda \right| +
      \left| \int_{E^{\low}} f_{\mu} - f_{\mu'} \, \dx \lambda \right| \\
      &= |(\mu - \mu')[u^{-1}(E^{\up})]| + |(\mu - \mu')[u^{-1}(E^{\low})]| \\
      &< \epsilon,
  \end{align*}
  where the second equality follows by the definition of pushforward measures
  and the inequality follows from Lemma~\ref{lem:tot_var}. Since \(\epsilon\)
  was arbitrary, the claim follows.
\end{proof}

Finally, we define \(\bb Q\). We let \(\bb Q\) be the subset of \(\bb K\)
consisting of all \(\C U\)-fine probability measures, i.e., measures \(\mu \in
\B K\) such that:
\begin{enumerate}
  \item The measure \(\mu\) is \(\C U\)-fine;
  \item For all Borel sets \(E \subseteq \C K\), \(\mu[E] \geq 0\);
  \item The measure of the whole space is unity, i.e., \(\mu[\C K] = 1\).
\end{enumerate}

We conclude the background and notation by observing that threshold policies are
defined wholly by their thresholds for distributions in \(\bb K\) and \(\bb Q\).
Importantly, this observation does not hold when there are atoms on the utility
scale---which measures in \(\bb K\) lack---which can in turn lead to
counterexamples to Theorem~\ref{thm:dist}; see
Appendix~\ref{sec:counterexample}.

\begin{lemma}
\label{lem:simplethresh}
  Let \(\tau_0(x)\) and \(\tau_1(x)\) be two multiple threshold policies. If
  \(\tau_0(x)\) and \(\tau_1(x)\) have the same thresholds, then for any \(\mu
  \in \bb K\), \(\tau_0(X) = \tau_1(X)\) \(\mu\)-a.s. Similarly, for \(\mu \in
  \bb Q\), if
  \begin{equation*}
    \EE_\mu[\tau_0(X) \mid A = a] = \EE_\mu[\tau_1(X) \mid A = a]
  \end{equation*}
  for all \(a \in \C A\) such that \(\Pr_\mu(A = a) > 0\), then \(\tau_0(X) =
  \tau_1(X)\) \(\mu\)-a.s.

  Moreover, for \(\mu \in \bb K\) in the case of path-specific fairness, if
  \(\tau_0(x)\) and \(\tau_1(x)\) have the same thresholds, then
  \(\tau_0(X_{\Pi, A, a}) = \tau_1(X_{\Pi, A, a})\) \(\mu\)-a.s.\ for any \(a
  \in \C A\). Similarly, for \(\mu \in \bb Q\) in the case of path-specific
  fairness, if
  \begin{equation*}
    \EE_\mu[\tau_0(X_{\Pi, A, a})] = \EE_\mu[\tau_1(X_{\Pi, A, a})]
  \end{equation*}
  then \(\tau_0(X_{\Pi, A, a}) = \tau_1(X_{\Pi, A, a})\) \(\mu\)-a.s.\ as well.
\end{lemma}

\begin{proof}
  First, we show that threshold policies with the same thresholds are equal,
  then we show that threshold policies that distribute positive decisions across
  groups in the same way are equal.

  Let \(\{t_a\}_{a \in \C A}\) denote the shared set of thresholds. It follows
  that if \(\tau_0(x) \neq \tau_1(x)\), then \(u(x) = t_{\alpha(x)}\). Now,
  \begin{equation*}
    \Pr(u(X) = t_a, A = a) = \int_{t_a}^{t_a} f_{\mu, a} \, \dx \lambda = 0,
  \end{equation*}
  so \(\Pr_\mu(\tau_0(X) \neq \tau_1(X)) = 0\).
  Next, suppose
  \begin{equation*}
    \EE_\mu[\tau_0(X) \mid A = a] = \EE_\mu[\tau_1(X) \mid A = a].
  \end{equation*}
  If the thresholds of the two policies agree for all \(a \in \C A\) such that
  \(\Pr_\mu(A = a) > 0\), then we are done by the previous paragraph. Therefore,
  suppose \(t_a^0 \neq t_a^1\) for some suitable \(a \in \C A\), where \(t_a^i\)
  represents the threshold for group \(a \in \C A\) under the policy
  \(\tau_i(x)\). Without loss of generality, suppose \(t_a^0 < t_a^1\). Then, it
  follows that
  \begin{align*}
    \int_{t_a^{0}}^{t_a^1} f_{\mu,a} \, \dx \lambda
      &= \EE_\mu[\tau_0(X) \mid A = a] - \EE_\mu[\tau_1(X) \mid A = a] \\
      &= 0.
  \end{align*}
  Since \(\mu \in \bb Q\), \(\mu = |\mu|\), whence
  \begin{equation*}
    \Pr_{|\mu|}(t_0^a \leq u(X) \leq t_a^1 \mid A = a) = 0.
  \end{equation*}
  Since this is true for all \(a \in \C A\) such that \(\Pr_\mu(A = a) > 0\),
  \(\tau_0(X) = \tau_1(X)\) \(\mu\)-a.s.

  The proof in the case of path-specific fairness is almost identical.
\end{proof}

\subsubsection{Convexity, complete metrizability, and universal measurability}

The set of regular \(\C U\)-fine probability measures \(\bb Q\) is the set to
which we wish to apply Prop.~\ref{prop:k_shy}. To do so, we must show that \(\bb
Q\) is a convex and completely metrizable subset of \(\bb K\).

\begin{lemma}
\label{lem:convex}
  The set of regular probability measures \(\bb Q\) is convex and completely
  metrizable.
\end{lemma}

\begin{proof}
  The proof proceeds in two pieces. First, we show that the \(\C U\)-fine
  probability distributions are convex, as can be verified by direct
  calculation. Then, we show that \(\bb Q\) is closed and therefore complete in
  the original metric of \(\bb K\).

  We begin by verifying convexity. Let \(\mu, \mu' \in \bb Q\) and let \(E
  \subseteq \C K\) be an arbitrary Borel subset of \(\C K\). Then, choose
  \(\theta \in [0, 1]\), and note that
  \begin{align*}
    (\theta \cdot \mu + [1 - \theta] \cdot \mu') [E]
      &= \theta \cdot \mu[E] + [1 - \theta] \cdot \mu'[E] \\
      &\geq \theta \cdot 0 + [1 - \theta] \cdot 0 \\
      &= 0,
  \end{align*}
  and, likewise, that
  \begin{align*}
    (\theta \cdot \mu + [1 - \theta] \cdot \mu') [\C K]
      &= \theta \cdot \mu[\C K] + [1 - \theta] \cdot \mu'[\C K] \\
      &= \theta \cdot 1 + [1 - \theta] \cdot 1 \\
      &= 1.
  \end{align*}
  It remains only to show that \(\bb Q\) is completely metrizable. To prove
  this, it suffices to show that it is closed, since closed subsets of complete
  spaces are complete, and \(\bb K\) is a Banach space by Cor.~\ref{cor:banach},
  and therefore complete.

  Suppose \(\{\mu_i\}_{i \in \B N}\) is a convergent sequence of probability
  measures in \(\bb K\) with limit \(\mu\). Then
  \begin{equation*}
    \mu[E] = \lim_{i \to \infty} \mu_i[E] \geq \lim_{i \to \infty} 0 = 0
  \end{equation*}
  and
  \begin{equation*}
    \mu[\C K] = \lim_{i \to \infty} \mu_i[\C K] = \lim_{i \to \infty} 1 = 1.
  \end{equation*}
  Therefore \(\bb Q\) is closed, and therefore complete, and hence is a convex,
  completely metrizable subset of \(\bb K\).
\end{proof}

Next we prove that the set \(\bb E\) of regular \(\C U\)-fine densities over
which there exists a policy satisfying the relevant counterfactual fairness
definition that is not strongly Pareto dominated is universally measurable.

Recall the definition of universal measurability.

\begin{definition}
  Let \(V\) be a complete topological space. Then \(E \subseteq V\) is
  \emph{universally measurable} if \(V\) is measurable by the completion of
  every finite Borel measure on \(V\), i.e., if for every finite Borel measure
  \(\mu\), there exist Borel sets \(E'\) and \(S\) such that \(E\ \triangle\ E'
  \subseteq S\) and \(\mu[S] = 0\).
\end{definition}

We note that if a set is Borel, it is by definition universally measurable.
Moreover, if a set is open or closed, it is by definition Borel.

To show that \(\bb E\) is closed, we show that any convergent sequence in \(\bb
E\) has a limit in \(\bb E\). The technical complication of the argument stems
from the following fact that satisfying the fairness conditions, e.g.,
Eq.~\eqref{eq:counterfactual_fairness}, involves conditional expectations, about
which very little can be said in the absence of a density, and which are
difficult to compare when taken across distinct measures.

To handle these difficulties, we begin with a technical lemma,
Lemma~\ref{lem:ce_est}, which gives a coarse bound on how different the
conditional expectations of the same variable can be with respect to a
sub--\(\sigma\)-algebra \(\C F\) over two different distributions, \(\mu\) and
\(\mu'\), before applying the results to the proof of Lemma~\ref{lem:e_closed}.

\begin{definition}
\label{defn:standard_version}
  Let \(\mu\) be a measure on a measure space \((V, \C M)\), and let \(f\) be
  \(\mu\)-measurable. Consider the equivalence class of \(\C M\)-measurable
  functions \(C = \{g : g = f \text{ \(\mu\)-a.e.}\}\).\footnote{%
    Some authors define \(L^p(\mu)\) spaces to consist of such equivalence
    classes, rather than the definition we use here.
  } We say that any \(g \in C\) is a \emph{version} of \(f\), and that \(g \in
  C\) is a \emph{standard version} if \(g(v) \leq C\) for some constant \(C\)
  and all \(v \in V\).
\end{definition}

\begin{remark}
  It is straightforward to see that for \(f \in L^\infty(\mu)\), a standard
  version always exists with \(C = \|f\|_\infty\).
\end{remark}

\begin{remark}
  Note that in general, the conditional expectation \(\EE_{\mu'}[f \mid \C F]\)
  is defined only \(\mu'\)-a.e. If \(\mu\) is not assumed to be absolutely
  continuous with respect to \(\mu'\), it follows that
  \begin{equation}
  \label{eq:version_rmk}
    \| \EE_\mu[f \mid \C F] - \EE_{\mu'}[f \mid \C F] \|_{L^1(\mu)}
  \end{equation}
  is not entirely well-defined, in that its value depends on what version of
  \(\EE_{\mu'}[f \mid \C F]\) one chooses. For appropriate \(f\), however, one
  can nevertheless bound Eq.~\eqref{eq:version_rmk} for any standard version of
  \(\EE_{\mu'}[f \mid \C F]\).
\end{remark}

\begin{lemma}
\label{lem:ce_est}
  Let \(\mu\), \(\mu'\) be totally bounded measures on a measure space \((V, \C
  M)\). Let \(f \in L^\infty(\mu) \cap L^\infty(\mu')\). Let \(\C F\) be a
  sub--\(\sigma\)-algebra of \(\C M\). Let
  \begin{equation*}
    C = \max(\|f\|_{L^\infty(\mu)}, \|f\|_{L^\infty(\mu')}).
  \end{equation*}
  Then, if \(g\) is a standard version of \(\EE_{\mu'}[f \mid \C F]\), we have
  that
  \begin{equation}
  \label{eq:ce_bound}
    \int_V |\EE_\mu[f \mid \C F] - g| \, \dx \mu \leq 4 C \cdot |\mu - \mu'|[V].
  \end{equation}
\end{lemma}

\begin{proof}
  First, we note that both \(\EE_\mu[f \mid \C F]\) and \(g\) are \(\C
  F\)-measurable. Therefore, the sets
  \begin{equation*}
    E^{\up} = \{v \in V : \EE_\mu[f \mid \C F](v) > g(v)\}
  \end{equation*}
  and
  \begin{equation*}
    E^{\low} = \{v \in V : \EE_\mu[f \mid \C F](v) < g(v)\}
  \end{equation*}
  are in \(\C F\). Now, note that
  \begin{equation*}
    \int_V |\EE_\mu[f \mid \C F] - g| \, \dx \mu = \int_{E^{\up}} \EE_\mu[f \mid
    \C F] - g \, \dx \mu + \int_{E^{\low}} g - \EE_\mu[f \mid \C F] \, \dx \mu.
  \end{equation*}

  First consider \(E^{\up}\). Then, we have that
  \begin{align*}
    \int_{E^{\up}} \EE_\mu[f \mid \C F] - g \, \dx \mu
      &= \int_{E^{\up}} \EE_\mu[f \mid \C F] - g \, \dx \mu + \int_{E^{\up}} g -
      g \, \dx \mu' \\
      &\leq \left| \int_{E^{\up}} \EE_\mu[f \mid \C F] \, \dx \mu -
      \int_{E^{\up}} g \, \dx \mu' \right| + \int_{E^{up}} g \, \dx {|\mu -
      \mu'|} \\
      &\leq \left| \int_{E^{\up}} f \, \dx \mu - \int_{E^{\up}} f \, \dx \mu'
      \right| + \int_{E^{up}} C \, \dx {|\mu - \mu'|},
  \end{align*}
  where in the final inequality, we have used the fact that, since \(g\) is a
  standard version of \(\EE_{\mu'}[f \mid \C F]\),
  \begin{equation*}
    g(v) \leq \|\EE_{\mu'}[f \mid \C F]\|_{L^\infty(\mu')} \leq C
  \end{equation*}
  for all \(v \in V\), and the fact that, by the definition of conditional
  expectation,
  \begin{equation*}
    \int_E \EE_\nu[h \mid \C F] \, \dx \nu = \int_E h \, \dx \nu
  \end{equation*}
  for any \(E \in \C F\).

  Since \(f\) is everywhere bounded by \(C\), applying Lemma~\ref{lem:tot_var}
  yields that this final expression is less than or equal to \(2 C \cdot |\mu -
  \mu'|[V]\). An identical argument shows that
  \begin{equation*}
    \int_{E^{\low}} g - \EE_\mu[f \mid \C F] \, \dx \mu \leq 2 C \cdot |\mu -
    \mu'|[V],
  \end{equation*}
  whence the result follows.
\end{proof}

\begin{lemma}
\label{lem:e_closed}
  Let \(\bb E \subseteq \bb Q\) denote the set of joint densities on \(\C K\)
  such that there exists a policy satisfying the relevant fairness definition
  that is not strongly Pareto dominated. Then, \(\bb E\) is closed, and
  therefore universally measurable.
\end{lemma}

\begin{proof}
  For notational simplicity, we consider the case of counterfactual equalized
  odds. The proofs in the other two cases are virtually identical.

  Suppose \(\mu_i \to \mu\) in \(\bb K\), where \(\{\mu_i\}_{i \in \B N}
  \subseteq \bb E\). Then, by Lemma~\ref{lem:mapping}, \(f_{\mu_i, a} \to
  f_{\mu, a}\) in \(L^1(\B R)\). Moreover, by Lemma~\ref{lem:threshold}, there
  exists a sequence of threshold policies \(\{\tau_i(x)\}_{i \in \B N}\) such
  that both
  \begin{equation*}
    \EE_{\mu_i}[\tau(X)] = \min(b, \Pr_{\mu_i}(u(X) > 0))
  \end{equation*}
  and
  \begin{equation*}
    \EE_{\mu_i}[\tau_i(X) \mid A, Y(1)] = \EE_{\mu_i}[\tau_i(X) \mid Y(1)].
  \end{equation*}

  Let \(\{q_{a,i}\}_{a \in \C A}\) be defined by
  \begin{equation*}
    q_{a,i} = \EE_{\mu_i}[\tau_i(X) \mid A = a]
  \end{equation*}
  if \(\Pr_{\mu_i}(A = a) > 0\), and \(q_{a,i} = 0\) otherwise.

  Since \([0,1]^{\C A}\) is compact, there exists a convergent subsequence
  \(\{\{q_{a, n_i}\}_{a \in \C A}\}_{i \in \B N}\). Let it converge to the
  collection of quantiles \(\{q_a\}_{a \in \C A}\) defining, by
  Lemma~\ref{lem:quantile}, a multiple threshold policy \(\tau(x)\) over
  \(\mu\).

  Because \(\mu_i \to \mu\) and \(\{q_{a, n_i}\}_{a \in \C A} \to \{q_a\}_{a \in
  \C A}\), we have that
  \begin{equation*}
    \EE_\mu[\tau_{a, n_i}(X) \mid A = a] \to \EE_\mu[\tau(X) \mid A = a]
  \end{equation*}
  for all \(a \in \C A\) such that \(\Pr_\mu(A = a) > 0\). Therefore, by
  Lemma~\ref{lem:mapping}, \(\tau_{n_i}(X) \to \tau(X)\) in \(L^1(\mu)\).

  Choose \(\epsilon > 0\) arbitrarily. Then, choose \(N\) so large that for
  \(i\) greater than \(N\),
  \begin{align*}
    |\mu - \mu_{n_i}|[\C K] &< \tfrac {\epsilon} {10}, &\|\tau(X) -
    \tau_{n_i}(X)\|_{L^1(\mu)} &\leq \tfrac \epsilon {10}.
  \end{align*}
  Then, observe that \(\tau(x), \tau_i(x) \leq 1\), and recall that
  \begin{equation}
  \label{eq:cea}
    \EE_{\mu_{n_i}}[\tau_{n_i}(X) \mid A, Y(1)] = \EE_{\mu_{n_i}}[\tau_{n_i}(X)
    \mid Y(1)].
  \end{equation}
  Therefore, let \(g_i(x)\) be a standard version of
  \(\EE_{\mu_{n_i}}[\tau_{n_i}(X) \mid Y(1)]\) over \(\mu_{n_i}\). By
  Eq.~\eqref{eq:cea}, \(g_i(x)\) is also a standard version of
  \(\EE_{\mu_{n_i}}[\tau_{n_i}(X) \mid A, Y(1)]\) over \(\mu_{n_i}\). Then, by
  Lemma~\ref{lem:ce_est}, we have that
  \begin{align*}
    \| \EE_\mu [\tau(X) \mid A, Y(1)] &- \EE_{\mu_{n_i}}[\tau_{n_i}(X) \mid
    Y(1)] \|_{L^1(\mu)} \\
      &\leq \| \EE_\mu[\tau(X) \mid A, Y(1)] - \EE_\mu[\tau_{n_i}(X) \mid A,
      Y(1)] \|_{L^1(\mu)} \\
      &\hspace{0.5cm} + \| \EE_\mu[\tau_{n_i}(X) \mid A, Y(1)] - g_i(X)
        \|_{L^1(\mu)} \\
      &\hspace{0.5cm} + \| g_i(X) - \EE_\mu[\tau_{n_i}(X) \mid Y(1)]
        \|_{L^1(\mu)} \|_{L^1(\mu)} \\
      &\hspace{0.5cm} + \| \EE_\mu[\tau_{n_i}(X) \mid Y(1) - \EE_\mu[\tau(X)
        \mid Y(1)] \|_{L^1(\mu)} \\
      &< \frac \epsilon {10} + \frac {4 \epsilon} {10} + \frac {4 \epsilon} {10}
        + \frac \epsilon {10}.
  \end{align*}
  Since \(\epsilon > 0\) was arbitrary, it follows that, \(\mu\)-a.e.,
  \begin{equation*}
    \EE_\mu[\tau(X) \mid A, Y(1)] = \EE_\mu[\tau(X) \mid Y(1)].
  \end{equation*}

  Recall the standard fact that for independent random variables \(X\) and
  \(U\),
  \begin{equation*}
    \EE[f(X,U) \mid X] = \int f(X, u) \diff F_U(u),
  \end{equation*}
  where \(F_U\) is the distribution of \(U\).\footnote{%
    For a proof of this fact see, e.g., \citet{drhab2019conditional}.
  }
  Further recall that \(D = \B 1_{U_D \leq \tau(X)}\), where \(U_D \indep X,
  Y(1)\). It follows that
  \begin{equation*}
    \Pr_\mu(D = 1 \mid X, Y(1)) = \int_0^1 \B 1_{u_d < \tau(X)} \, \dx
    \lambda(u_d) = \tau(X).
  \end{equation*}
  Hence, by the law of iterated expectations,
  \begin{align*}
    \Pr_\mu(D = 1 \mid A, Y(1))
      &= \EE_\mu[\Pr_\mu(D = 1 \mid X, Y(1)) \mid A, Y(1)] \\
      &= \EE_\mu[\tau(X) \mid A, Y(1)] \\
      &= \EE_\mu[\tau(X) \mid Y(1)] \\
      &= \EE_\mu[\Pr_\mu(D = 1 \mid X, Y(1)) \mid Y(1)] \\
      &= \Pr_\mu(D = 1 \mid Y(1)).
  \end{align*}
  Therefore \(D \indep A \mid Y(1)\) over \(\mu\), i.e., counterfactual
  equalized odds holds for the decision policy \(\tau(x)\) over the distribution
  \(\mu\). Consequently \(\mu \in \bb E\), and so \(\bb E\) is closed and
  therefore universally measurable.
\end{proof}

\subsection{Shy Sets and Probes}
\label{sec:shyness-prelims2}

We require a number of additional technical lemmata for the proof of
Theorem~\ref{thm:dist}. The probe must be constructed carefully, so that, on the
utility scale, an arbitrary element of \(\bb Q\) is absolutely continuous with
respect to a typical perturbation. In addition, it is useful to show that a
number of properties are generic to simplify certain aspects of the proof of
Theorem~\ref{thm:dist}. For instance, Lemma~\ref{lem:condition} is used in
Theorem~\ref{thm:dist} to show that a certain conditional expectation is
generically well-defined, avoiding the need to separately treat certain corner
cases.

Cor.~\ref{cor:maximal} concerns the construction of the probe used in the proof
of Theorem~\ref{thm:dist}. Lemmata~\ref{lem:uncountable_sum}~to~\ref{lem:simple}
use Cor.~\ref{cor:maximal} to provide additional simplifications to the proof of
Theorem~\ref{thm:dist}.

\subsubsection{Maximal support}

First, to construct the probe used in the proof of Theorem~\ref{thm:dist}, we
require elements \(\mu \in \bb Q\) such that the densities \(f_\mu\) have
``maximal'' support. To produce such distributions, we use the following
measure-theoretic construction.

\begin{definition}
  Let \(\{E_\alpha\}_{\gamma \in \Gamma}\) be an arbitrary collection of
  \(\mu\)-measurable sets for some positive measure \(\mu\) on a measure space
  \((M, \C M)\). We say that \(E\) is the \emph{measure-theoretic union} of
  \(\{E_\gamma\}_{\gamma \in \Gamma}\) if \(\mu[E_\gamma \setminus E] = 0\) for
  all \(\gamma \in \Gamma\) and \(E = \bigcup_{i=1}^\infty E_{\gamma_i}\) for
  some countable subcollection \(\{\gamma_i\} \subseteq \B N\).
\end{definition}

While measure-theoretic unions themselves are known
(cf.~\citet{silva2008invitation}, \citet{rudin1991functional}), for
completeness, we include a proof of their existence, which, to the best of our
knowledge, is not found in the literature.

\begin{lemma}
\label{lem:mtu}
  Let \(\mu\) be a finite positive measure on a measure space \((V, \C M)\).
  Then an arbitrary collection \(\{E_\gamma\}_{\gamma \in \Gamma}\) of
  \(\mu\)-measurable sets has a measure-theoretic union.
\end{lemma}

\begin{proof}
  For each countable subcollection \(\Gamma' \subseteq \Gamma\), consider the
  ``error term''
  \begin{equation*}
    r(\Gamma') = \sup_{\gamma \in \Gamma} \mu \left[ E_{\gamma} \setminus
    \bigcup_{\gamma' \in \Gamma'} E_{\gamma'} \right]
  \end{equation*}
  We claim that the infimum of \(r(\Gamma')\) over all countable subcollections
  \(\Gamma' \subseteq \Gamma\) must be zero.

  For, toward a contradiction, suppose it were greater than or equal to
  \(\epsilon > 0\). Choose any set \(E_{\gamma_1}\) such that
  \(\mu[E_{\gamma_1}] \geq \epsilon\). Such a set must exist, since otherwise
  \(r(\emptyset) < \epsilon\). Choose \(E_{\gamma_2}\) such that
  \(\mu[E_{\gamma_2} \setminus E_{\gamma_1}] > \epsilon\). Again, some such set
  must exist, since otherwise \(r(\{\gamma_1\}) < \epsilon\). Continuing in this
  way, we construct a countable collection \(\{E_{\gamma_i}\}_{i \in \B N}\).

  Therefore, we see that
  \begin{align*}
    \mu[V] \geq \mu \left[ \bigcup_{i=1}^n E_{\gamma_i} \right] = \sum_{i=1}^n
    \mu \left[ E_{\gamma_i} \setminus \bigcup_{j=1}^i E_{\gamma_j} \right].
  \end{align*}
  By construction, every term in the final sum is greater than or equal to
  \(\epsilon\), contradicting the fact that \(\mu[V] < \infty\).

  Therefore, there exist countable collections \(\{\Gamma_n\}_{n \in \B N}\)
  such that \(r(\Gamma_n) < \frac 1 n\). It follows immediately that for all
  \(n\)
  \begin{equation*}
    r \left( \bigcup_{n \in \B N} \Gamma_n \right) \leq r(\Gamma_k)
  \end{equation*}
  for any fixed \(k \in \B N\). Consequently,
  \begin{equation*}
    r \left( \bigcup_{n \in \B N} \Gamma_n \right) = 0,
  \end{equation*}
  and \(\bigcup_{n \in \B N} \Gamma_n\) is countable.
\end{proof}

The construction of the ``maximal'' elements used to construct the probe in the
proof of Theorem~\ref{thm:dist} follows as a corollary of Lemma~\ref{lem:mtu}

\begin{corollary}
\label{cor:maximal}
  There are measures \(\mu_{\max, a} \in \bb Q\) such that for every \(a \in \C
  A\) and any \(\mu \in \bb K\),
  \begin{equation*}
    \lambda[\supp(f_{\mu, a}) \setminus \supp(f_{\mu_{\max}, a})] = 0.
  \end{equation*}
\end{corollary}

\begin{proof}
  Consider the collection \(\{\supp(f_{\mu, a})\}_{\mu \in \bb K}\). By
  Lemma~\ref{lem:mtu}, there exists a countable collection of measures
  \(\{\mu_i\}_{i \in \B N}\) such that for any \(\mu \in \bb K\),
  \begin{equation*}
    \lambda \left[ \supp(f_{\mu,a}) \setminus \bigcup_{i = 1}^{\infty}
    \supp(f_{\mu_i, a}) \right] = 0,
  \end{equation*}
  where, without loss of generality, we may assume that
  \(\lambda[\supp(f_{\mu_i, a})] > 0\) for all \(i \in \B N\). Such a sequence
  must exist, since, by the first hypothesis of Theorem~\ref{thm:dist}, for
  every \(a \in \C A\), there exists \(\mu \in \bb Q\) such that \(\Pr_\mu(A =
  a) > 0\). Therefore, we can define the probability measure \(\mu_{\max, a}\),
  where
  \begin{equation*}
    \mu_{\max, a} = \sum_{i=1}^n 2^{-i} \cdot \frac {\left| \mu_i \rest_{A = a}
    \right|} {\left |\mu_i \rest_{A = a} \right| [\C K]}.
  \end{equation*}
  It follows immediately by construction that
  \begin{equation*}
    \supp(f_{\mu_{\max}, a}) = \bigcup_{i=1}^\infty \supp(f_{\mu_i,a}),
  \end{equation*}
  and that \(\mu_{\max,a} \in \bb Q\).
\end{proof}

For notational simplicity, we refer to \(\supp(f_{\mu_{\max, a}})\) as \(S_a\)
throughout.

In the case of conditional principal fairness and path-specific fairness, we
need a mild refinement of the previous result that accounts for \(\omega\).

\begin{corollary}
\label{cor:maximal_ext}
  There are measures \(\mu_{\max, a, w} \in \bb Q\) defined for every \(w \in \C
  W = \img(\omega)\) and any \(a \in \C A\) such that for some \(\nu \in \bb
  K\), \(\Pr_\nu(W = w, A = a) > 0\). These measures have the property that for
  any \(\mu \in \bb K\),
  \begin{equation*}
    \lambda[\supp(f_{\mu', a, w}) \setminus \supp(f_{\mu_{\max}, a, w})] = 0,
  \end{equation*}
  where \(f_{\mu', a, w}\) is the density of the pushforward measure \((\mu'
  \rest_{W = w, A = a}) \circ u^{-1}\).
\end{corollary}

Recalling that \(|\img(\omega)| < \infty\), the proof is the same as
Cor.~\ref{cor:maximal}, and we analogously refer to \(\supp(f_{\mu_{\max, a,
w}})\) as \(S_{a,w}\). Here, we have assumed without loss of generality---as we
continue to assume in the sequel---that for all \(w \in \C W\), there is some
\(\mu \in \bb K\) such that \(\Pr_\mu(W = w) > 0\).

\begin{remark}
\label{rmk:dist_hyp}
  Because their support is maximal, the hypotheses of Theorem~\ref{thm:dist}, in
  addition to implying that \(\mu_{\max,a}\) is well-defined for all \(a \in \C
  A\), also imply that \(\Pr_{\mu_{\max,a}}(u(X) > 0) > 0\). In the case of
  conditional principal fairness, they further imply that \(\Pr_{\mu_{\max,a}}(W
  = w) > 0\) for all \(w \in \C W\) and \(a \in \C A\). Likewise, in the case of
  path-specific fairness, they further imply that \(\Pr_{\mu_{\max,a}}(W = w_i)
  > 0\) for \(i = 0, 1\) and some \(a \in \C A\).
\end{remark}

\subsubsection{Shy sets and probes}

In the following lemmata, we demonstrate that a number of useful properties are
generic in \(\bb Q\). We also demonstrate a short technical lemma,
Lemma~\ref{lem:simple}, which allows us to use these generic properties to
simplify the proof of Theorem~\ref{thm:dist}.

We begin with the following lemma, which is useful in verifying that certain
subspaces of \(\bb K\) form probes.

\begin{lemma}
\label{lem:probe}
  Let \(\bb W\) be a non-trivial finite dimensional subspace of \(\bb K\) such
  that \(\nu[\C K] = 0\) for all \(\nu \in \bb W\). Then, there exists \(\mu \in
  \bb K\) such that \(\lambda_{\bb W}[\bb Q - \mu] > 0\).
\end{lemma}

\begin{proof}
  Set
  \begin{equation*}
    \mu = \sum_{i=1}^n \frac {|\nu_i|}{|\nu_i|[\C K]},
  \end{equation*}
  where \(\nu_1, \ldots, \nu_n\) form a basis of \(\bb W\). Then, if \(|\beta_i|
  \leq \tfrac 1 {|\nu_i|[\C K]}\), it follows that
  \begin{equation*}
    \mu + \sum_{i = 1}^n \beta_i \cdot \nu_i \in \bb Q.
  \end{equation*}
  Since
  \begin{equation*}
    \lambda_n \left[ \prod_{i=1}^n \left[ - \frac 1 {|\nu_i|[\C K]}, \frac 1
    {|\nu_i|[\C K]} \right] \right] > 0,
  \end{equation*}
  it follows that \(\lambda_{\bb W}[\bb Q - \mu] > 0\).
\end{proof}

Next we show that, given a \(\nu \in \bb Q\), a generic element of \(\bb Q\)
``sees'' events to which \(\nu\) assigns non-zero probability. While
Lemma~\ref{lem:support} alone in principle suffices for the proof of
Theorem~\ref{thm:dist}, we include Lemma~\ref{lem:condition} both for conceptual
clarity and to introduce at a high level the style of argument used in the
subsequent lemmata and in the proof of Theorem~\ref{thm:dist} to show that a set
is shy relative to \(\bb Q\).

\begin{lemma}
\label{lem:condition}
  For a Borel set \(E \subseteq \C K\), suppose there exists \(\nu \in \bb Q\)
  such that \(\nu[E] > 0\). Then the set of \(\mu \in \bb Q\) such that \(\mu[E]
  > 0\) is prevalent.
\end{lemma}

\begin{proof}
  First, we note that the set of \(\mu \in \bb Q\) such that \(\mu[E] = 0\) is
  closed and therefore universally measurable. For, if \(\{\mu_i\}_{i \in \B N}
  \subseteq \bb Q\) is a convergent sequence with limit \(\mu\), then
  \begin{align*}
    \mu[E]
      &= \lim_{n \to \infty} \mu_i[E] \\
      &= \lim_{n \to \infty} 0 \\
      &= 0.
  \end{align*}
  Now, if \(\mu[E] > 0\) for all \(\mu \in \bb Q\), there is nothing to prove.
  Therefore, suppose that there exists \(\nu' \in \bb Q\) such that \(\nu'[E] =
  0\).

  Next, consider the measure \(\tilde \nu = \nu' - \nu\). Then, let \(\bb W =
  \Span(\tilde \nu)\). Since \(\tilde \nu \neq 0\) and
  \begin{equation*}
    \tilde \nu[\C K] = \nu'[\C K] - \nu[\C K] = 0,
  \end{equation*}
  it follows by Lemma~\ref{lem:probe} that \(\lambda_{\bb W}[\bb Q - \mu] > 0\)
  for some \(\mu\).

  Now, for arbitrary \(\mu \in \bb Q\), note that if \((\mu + \beta \cdot \tilde
  \nu)[E] = 0\), then
  \begin{equation*}
    \mu[E] - \beta \cdot \nu[E] = 0
  \end{equation*}
  i.e.,
  \begin{equation*}
    \beta = \frac {\mu[E]} {\nu[E]}.
  \end{equation*}
  A singleton has null Lebesgue measure, and so the set of \(\nu \in \bb W\)
  such that \((\mu + \nu)[E] = 0\) is \(\lambda_{\bb W}\)-null. Therefore, by
  Prop.~\ref{prop:k_shy}, the set of \(\mu \in \bb Q\) such that \(\mu[E] = 0\)
  is shy relative to \(\bb Q\), as desired.
\end{proof}

While Lemma~\ref{lem:condition} shows that a typical element of \(\bb Q\)
``sees'' individual events, in the proof of Theorem~\ref{thm:dist}, we require a
stronger condition, namely, that a typical element of \(\bb Q\) ``sees'' certain
uncountable collections of events. To demonstrate this more complex property, we
require the following technical result, which is closely related to the real
analysis folk theorem that any convergent uncountable ``sum'' can contain only
countably many non-zero terms. (See, e.g., \citet{benji2020sum}.)

\begin{lemma}
\label{lem:uncountable_sum}
  Suppose \(\mu\) is a totally bounded measure on \((V, \C M)\), \(f\) and \(g\)
  are \(\mu\)-measurable real-valued functions, and \(g \neq 0\) \(\mu\)-a.e.
  Then the set
  \begin{equation*}
    Z_\beta = \{v \in V : f(v) + \beta \cdot g(v) = 0\}
  \end{equation*}
  has non-zero \(\mu\) measure for at most countably many \(\beta \in \B R\).
\end{lemma}

\begin{proof}
  First, we show that for any countable collection \(\{\beta_i\}_{i \in \B N}
  \subseteq \B R\), the sum \(\sum_{i=1}^\infty \mu[Z_{\beta_i}]\) converges.
  Then, we show how this implies that \(\mu[Z_{\beta}] = 0\) for all but
  countably many \(\beta \in \B R\).

  First, we note that for distinct \(\beta, \beta' \in \B R\),
  \begin{equation*}
    Z_{\beta} \cap Z_{\beta'} \subseteq \{v \in V : (\beta - \beta') \cdot g(v)
    = 0\}.
  \end{equation*}
  Now, by hypothesis,
  \begin{equation*}
    \mu[\{v \in V : g(v) = 0\}] = 0,
  \end{equation*}
  and since \(\beta - \beta' \neq 0\), it follows that
  \begin{equation*}
    \mu[\{v \in V : (\beta - \beta') \cdot g(v) = 0\}] = 0
  \end{equation*}
  as well. Consequently, it follows that if \(\{Z_{\beta_i}\}_{i \in \B N}\) is
  a countable collection of distinct elements of \(\B R\), then
  \begin{align*}
    \sum_{i=1}^\infty \mu[Z_{\beta_i}]
      &= \mu \left[\bigcup_{i=1}^\infty Z_{\beta_i} \right] \\
      &\leq \mu[V] \\
      &< \infty.
  \end{align*}

  To see that this implies that \(\mu[Z_\beta] > 0\) for only countably many
  \(\beta \in \B R\), let \(G_\epsilon \subseteq \B R\) consist of those
  \(\beta\) such that \(\mu[Z_\beta] \geq \epsilon\). Then \(G_\epsilon\) must
  be finite for all \(\epsilon > 0\), since otherwise we could form a collection
  \(\{\beta_i\}_{i \in \B N} \subseteq G_\epsilon\), in which case
  \begin{equation*}
    \sum_{i=1}^\infty \mu[Z_{\beta_i}] \geq \sum_{i=1}^\infty \epsilon = \infty,
  \end{equation*}
  contrary to what was just shown. Therefore,
  \begin{equation*}
    \{\beta \in \B R : \mu[Z_\beta] > 0\} = \bigcup_{i=1}^\infty G_{1/i}
  \end{equation*}
  is countable.
\end{proof}

We now apply Lemma~\ref{lem:uncountable_sum} to the proof of the following
lemma, which states, informally, that, under a generic element of \(\bb Q\),
\(u(X)\) is supported everywhere it is supported under some particular fixed
element of \(\bb Q\). For instance, Lemma~\ref{lem:uncountable_sum} can be used
to show that for a generic element of \(\bb Q\), the density of \(u(X) \mid A =
a\) is positive \(\lambda \rest_{S_a}\)-a.e.

\begin{lemma}
\label{lem:support}
  Let \(\nu \in \bb Q\) and suppose \(\nu\) is supported on \(E\), i.e.,
  \(\nu[\C K \setminus E] = 0\). Then the set of \(\mu \in \bb Q\) such that
  \(\nu \circ u^{-1} \Lt (\mu \rest_E) \circ u^{-1}\) is prevalent relative to
  \(\bb Q\).
\end{lemma}

Lemma~\ref{lem:support} states, informally, that for generic \(\mu \in \bb Q\),
\(f_{\mu \rest_E}\) is supported everywhere \(f_\nu\) is supported.

\begin{proof}
  We begin by showing that the set of \(\mu \in \bb Q\) such that \(\nu \circ
  u^{-1} \Lt (\mu \rest_E) \circ u^{-1}\) is Borel, and therefore universally
  measurable. Then, we construct a probe \(\bb W\) and use it to show that this
  collection is finitely shy.

  To begin, let \(U_q\) denote the set of \(\mu \in \bb Q\) such that
  \begin{equation*}
    \nu \circ u^{-1}[\{|f_{\mu \rest_E}| = 0\}] < q.
  \end{equation*}
  We note that \(U_q\) is open. For, if \(\mu \in U_q\), then there exists some
  \(r > 0\) such that
  \begin{equation*}
    \nu \circ u^{-1}[\{|f_{\mu \rest_E}| < r\}] < q.
  \end{equation*}
  Let
  \begin{equation*}
    \epsilon = q - \nu \circ u^{-1}[\{|f_{\mu \rest_E}| < r\}].
  \end{equation*}
  Now, since \(\nu \circ u^{-1} \Lt \lambda\), there exists a \(\delta\) such
  that if \(\lambda[E'] < \delta\), then \(\nu \circ u^{-1}[E'] < \epsilon\).
  Choose \(\mu'\) arbitrarily so that \(|\mu - \mu'|[\C K] < \delta \cdot r\).
  Then, by Markov's inequality, we have that
  \begin{equation*}
    \lambda[\{|f_{\mu \rest_E} - f_{\mu' \rest_E}| > r\}] < \delta,
  \end{equation*}
  i.e.,
  \begin{equation*}
    \nu \circ u^{-1}[\{f_{\mu \rest_E} - f_{\mu' \rest_E}| > r\}] < \epsilon.
  \end{equation*}
  Now, we note that by the triangle inequality, wherever
  \(|f_{\mu' \rest_E}| = 0\), either \(|f_{\mu \rest_E}| < r\) or \(|f_{\mu
  \rest_E} - f_{\mu' \rest_E}| > r\). Therefore
  \begin{align*}
    \lambda[\{|f_{\mu' \rest_E}| = 0\}]
      &\leq \nu \circ u^{-1}[\{|f_{\mu \rest_E} - f_{\mu' \rest_E}| > r\}] + \mu
      \circ u^{-1} [\{|f_{\mu \rest_E}| < r] \\
      &< \epsilon + \mu \circ u^{-1}[\{|f_{\mu \rest_E}| < r] \\
      &< q.
  \end{align*}
  We conclude that \(\mu' \in U_q\), and so \(U_q\) is open.

  Note that \(\nu \circ u^{-1} \Lt (\mu \rest_E) \circ u^{-1}\) if and only if
  \begin{equation*}
    \lambda[\supp(f_\nu) \setminus \supp(f_{\mu \rest_E})] = 0.
  \end{equation*}
  By the definition of the support of a function, \(\lambda \rest_{\supp(f_\mu)}
  \Lt \mu \circ u^{-1}\). Therefore, it follows that
  \begin{equation*}
    \lambda[\supp(f_\mu) \setminus \supp(f_{\nu \rest_E})] = 0
  \end{equation*}
  if and only if
  \begin{equation*}
    \mu \circ u^{-1}[\supp(f_\mu) \setminus \supp(f_{\nu \rest_E})] = 0.
  \end{equation*}
  Then, it follows immediately that the set of \(\nu \in \bb Q\) such that \(\mu
  \circ u^{-1} \Lt (\nu \rest_E) \circ u^{-1}\) is \(\bigcap_{i=1}^n U_{1/i}\),
  which is, by construction, Borel, and therefore universally measurable.

  Now, since
  \begin{equation*}
    \Pr_\nu(u(X) < t) = \int_{-\infty}^t f_{\nu} \, \dx \lambda
  \end{equation*}
  is a continuous function of \(t\), by the intermediate value theorem, there
  exists \(t\) such that \(\Pr_\nu(u(X) \in S_0) = \Pr_\nu(u(X) \in S_1)\),
  where \(S_0 = \supp(f_\nu) \cap (-\infty, t)\) and \(S_1 = \supp(f_\nu) \cap
  [t, \infty)\). Then, we let
  \begin{equation*}
    \tilde \nu[E'] = \int_{E'} \B 1_{u^{- 1} (S_0)} - \B 1_{u^{-1}(S_1)} \, \dx
    \nu.
  \end{equation*}

  Take \(\bb W = \Span(\tilde \nu)\). Since \(\tilde \nu \neq 0\) and \(\tilde
  \nu[\C K] = 0\), we have by Lemma~\ref{lem:probe} that \(\lambda_{\bb W}[\bb Q
  - \mu] > 0\) for some \(\mu\).

  By the definition of a density, \(f_{\tilde \nu}\) is positive \((\tilde \nu
  \circ u^{-1})\)-a.e. Consequently, by the definition of \(\tilde \nu\),
  \(f_{\tilde \nu}\) is non-zero \((\mu \circ u^{-1})\)-a.e. Therefore, by
  Lemma~\ref{lem:uncountable_sum}, there exist only countably many \(\beta \in
  \B R\) such that the density of \((\mu + \beta \cdot \tilde \nu) \circ
  u^{-1}\) equals zero on a set of positive \(\mu \circ u^{-1}\)-measure. Since
  countable sets have \(\lambda\)-measure zero and \(\nu\) is arbitrary, the set
  of \(\mu \in \bb Q\) such that \(\nu \circ u^{-1} \Lt (\mu \rest_E) \circ
  u^{-1}\) is prevalent relative to \(\bb Q\) by Prop.~\ref{prop:k_shy}.
\end{proof}

The following definition and technical lemma are needed to extend
Theorem~\ref{thm:dist} to the cases of conditional principal fairness and
path-specific fairness, which involve additional conditioning on \(W =
\omega(X)\). In particular, one corner case we wish to avoid in the proof of
Theorem~\ref{thm:dist} is when the decision policy is non-trivial (i.e., some
individuals receive a positive decision and others do not) but from the
perspective of each \(\omega\)-stratum, the policy is trivial (i.e., everyone in
the stratum receives a positive or negative decision).
Definition~\ref{defn:overlap} formalizes this pathology, and
Lemma~\ref{lem:overlap} shows that this issue---under a mild hypothesis---does
not arise for a generic element of \(\bb Q\).

\begin{definition}
\label{defn:overlap}
  We say that \(\mu \in \bb Q\) \emph{overlaps utilities} when, for any
  budget-exhausting multiple threshold policy \(\tau(x)\), if
  \begin{equation*}
    0 < \EE_\mu[\tau(X)] < 1,
  \end{equation*}
  then there exists \(w \in \C W\) such that
  \begin{equation*}
    0 < \EE_\mu[\tau(X) \mid W = w] < 1.
  \end{equation*}

  If there exists a budget-exhausting multiple threshold policy \(\tau(x)\) such
  that
  \begin{equation*}
    0 < \EE_\mu[\tau(X)] < 1,
  \end{equation*}
  but for all \(w \in \C W\),
  \begin{equation*}
    \EE_\mu[\tau(X) \mid W = w] \in \{0, 1\},
  \end{equation*}
  then we say that \(\tau(x)\) \emph{splits utilities} over \(\mu\).
\end{definition}

Informally, having overlapped utilities prevents a budget-exhausting threshold
policy from having thresholds that fall on the utility scale exactly between the
strata induced by \(\omega\)---i.e., a threshold policy that splits utilities.
This is almost a generic condition in \(\bb Q\), as we shown in
Lemma~\ref{lem:overlap}.

\begin{lemma}
\label{lem:overlap}
  Let \(0 < b < 1\). Suppose that for all \(w \in \C W\) there exists \(\mu \in
  \bb Q\) such that \(\Pr_\mu(u(X) > 0, W = w) > 0\). Then almost every \(\mu
  \in \bb Q\) overlaps utilities.
\end{lemma}

\begin{proof}
  Our goal is to show that the set \(\bb E'\) of measures \(\mu \in \bb Q\) such
  that there exists a splitting policy \(\tau(x)\) is shy. To simplify the
  proof, we divide an conquer, showing that the set \(\bb E_{\Gamma}\) of
  measures \(\mu \in \bb Q\) such that there exists a splitting policy where the
  thresholds fall below \(w \in \Gamma \subseteq \C W\) and above \(w \notin
  \Gamma\) is Borel, before constructing a probe that shows that it is shy.
  Then, we argue that \(\bb E' = \bigcup_{\Gamma \subseteq \C W} \bb
  E_{\Gamma}\), which shows that \(\bb E'\) is shy.

  We begin by considering the linear map \(\Phi : \bb K \to \B R \times \B R^{\C
  W}\) given by
  \begin{equation*}
    \Phi(\mu) = \left( \Pr_\mu(u(X) = 0), \left( \Pr_\mu(W = w) \right)_{w \in
    \C W} \right).
  \end{equation*}
  For any \(\Gamma \subseteq \C W\), the sets
  \begin{align*}
    F^{\up}_{\Gamma}
      &= \{x \in \B R \times \B R^{\C W} : x_0 \geq b, b = \sum_{w \in \Gamma}
      x_w\}, \\
    F^{\low}_{\Gamma}
      &= \{x \in \B R \times \B R^{\C W} : x_0 \leq b, x_0 = \sum_{w \in \Gamma}
      x_w\},
  \end{align*}
  are closed by construction. Therefore, since \(\Phi\) is continuous,
  \begin{equation}
  \label{eq:shy_set}
    \bb E_{\Gamma} = \bb Q \cap \Phi^{-1} \left( \bigcup_{\Gamma \subseteq \C W}
    F^{\up}_{\Gamma} \cup F^{\low}_{\Gamma} \right)
  \end{equation}
  is closed, and therefore universally measurable.

  Note that by our hypothesis and Cor.~\ref{cor:maximal_ext}, for all \(w \in \C
  W\) there exists some \(a_w \in \C A\) such that
  \begin{equation*}
    \Pr_{\mu_{\max, a_w, w}}(u(X) > 0).
  \end{equation*}
  We use this to show that \(\bb E_{\Gamma}\) is shy. Pick \(w^* \in \C W\)
  arbitrarily, and consider the measures \(\nu_w\) for \(w \neq w^*\) defined by
  \begin{equation*}
    \nu_w = \frac {\mu_{\max, a_w, w} \rest_{u(X) > 0}} {\Pr_{\mu_{\max, a_w,
    w}}(u(X) > 0)} - \frac {\mu_{\max, a_{w^*}, w^*} \rest_{u(X) > 0}}
      {\Pr_{\mu_{\max, a_{w^*}, w^*}}(u(X) > 0)}.
  \end{equation*}
  We note that \(\nu_w[\C K] = 0\) by construction. Therefore, if \(\bb W_w =
  \Span(\nu_w)\), then \(\lambda_{\bb W_w}[\bb Q - \mu_w] > 0\) for some
  \(\mu_w\) by Lemma~\ref{lem:probe}.

  Moreover, we have that \(\Pr_\nu(u(X) > 0) = 0\) for all \(\nu \in \bb W_w\),
  i.e.,
  \begin{equation*}
  \label{eq:static}
    \Pr_\mu(u(X) > 0) = \Pr_{\mu + \nu}(u(X) > 0).
  \end{equation*}
  Now, since \(0 < b < 1\) and \(\omega\) partitions \(\C X\), it follows that
  \begin{equation*}
    \bb E_{\C W} = \bb E_{\emptyset} = \emptyset.
  \end{equation*}
  Since \(\lambda_{\bb W}[\emptyset] = 0\) for any subspace \(\bb W\), we can
  assume without loss of generality that \(\Gamma \neq \C W, \emptyset\).

  In that case, there exists \(w_{\Gamma} \in \C W\) such that if \(w^* \in
  \Gamma\), then \(w_{\Gamma} \notin \Gamma\), and vice versa. Without loss of
  generality, assume \(w_{\Gamma} \in \Gamma\) and \(w^* \notin \Gamma\). It
  then follows that for arbitrary \(\mu \in \bb Q\),
  \begin{equation*}
    \Phi(\mu + \beta \cdot \nu_{w_{\Gamma}}) = \Phi(\mu) + \beta \cdot \bb
    e_{w_{\Gamma}} - \beta \cdot \bb e_{w^*},
  \end{equation*}
  where \(\bb e_w\) is the basis vector corresponding to \(w \in \C W\). From
  this, it follows immediately by Eq.~\eqref{eq:static} that
  \begin{equation*}
    \mu + \beta \cdot \nu_{w_{\Gamma}} \in \bb E_{\Gamma}
  \end{equation*}
  only if
  \begin{equation*}
    \beta = \min(b, \Pr_\mu(u(X) > 0)) - \sum_{w \in \Gamma} \Pr_\mu(W = w).
  \end{equation*}
  This is a measure zero subset of \(\B R\), and so it follows that
  \begin{equation*}
    \lambda_{\bb W_{w_{\Gamma}}}[\bb E_{\Gamma} - \mu] = 0
  \end{equation*}
  for all \(\mu \in \bb K\). Therefore, by Prop.~\ref{prop:k_shy}, \(\bb
  E_{\Gamma}\) is shy in \(\bb Q\). Taking the union over \(\Gamma \subseteq \C
  W\), it follows by Prop.~\ref{prop:shy_axioms_rel} that \(\bigcup_{\Gamma
  \subseteq \C W} \bb E_{\Gamma}\) is shy.

  Now, we must show that \(\bb E' = \bigcup_{\Gamma \subseteq \C W} \bb
  E_{\Gamma}\). By construction, \(\bb E_{\Gamma} \subseteq \bb E'\), since the
  policy \(\tau(x) = \B 1_{\omega(x) \in \Gamma}\) is budget-exhausting and
  separates utilities. To see the reverse inclusion, suppose \(\mu \in \bb E'\),
  i.e., that there exists a budget-exhausting multiple threshold policy
  \(\tau(x)\) that splits utilities over \(\mu\). Then, let
  \begin{equation*}
    \Gamma_\mu = \{w \in \C W : \EE_\mu[\tau(X) \mid W = w] = 1\}.
  \end{equation*}
  Since \(\tau(x)\) is budget-exhausting, it follows immediately that \(\mu \in
  \bb E_{\Gamma_{\mu}}\). Therefore, \(\bb E' = \bigcup_{\Gamma \subseteq \C W}
  \bb E_{\Gamma}\), and so \(\bb E'\) is shy, as desired.
\end{proof}

We conclude our discussion of shyness and shy sets with the following general
lemma, which simplifies relative prevalence proofs by showing that one can,
without loss of generality, restrict one's attention to the elements of the shy
set itself in applying Prop.~\ref{prop:k_shy}.

\begin{lemma}
\label{lem:simple}
  Suppose \(E\) is a universally measurable subset of a convex, completely
  metrizable set \(C\) in a topological vector space \(V\). Suppose that for
  some finite-dimensional subspace \(V'\), \(\lambda_{V'}[C + v_0] > 0\) for
  some \(v_0 \in V\). If, in addition, for all \(v \in E\),
  \begin{equation}
  \label{eq:hyp}
    \lambda_{V'}[\{v' \in V' : v + v' \in E\}] = 0,
  \end{equation}
  then it follows that \(E\) is shy relative to \(C\).
\end{lemma}

\begin{proof}
  Let \(v\) be arbitrary. Then, either \((v + V') \cap E\) is empty or not.

  First, suppose it is empty. Since \(\lambda_{V'}[\emptyset] = 0\) by
  definition, it follows immediately that in this case \(\lambda_{V'}[E - v] =
  0\).

  Next, suppose the intersection is not empty, and let \(v + v^* \in E\) for
  some fixed \(v^* \in V'\). It follows that
  \begin{align*}
    \lambda_{V'}[E - v]
      &= \lambda_{V'}[\{v' \in V' : v + v' \in E \}] \\
      &= \lambda_{V'}[\{v' \in V' : (v + v^*) + v' \in E \}] \\
      &= 0,
  \end{align*}
  where the first equality follows by definition; the second equality follows by
  the translation invariance of \(\lambda_{V'}\), and the fact that \(v^* + V' =
  V'\); and the final inequality follows from Eq.~\eqref{eq:hyp}.

  Therefore \(\lambda_{V'}[E - v] = 0\) for arbitrary \(v\), and so \(E\) is
  shy.
\end{proof}

\subsection{Proof of Theorem~\ref{thm:dist}}
\label{sec:shyness-proof}

Using the lemmata above, we can prove Theorem~\ref{thm:dist}. We briefly
summarize what has been established so far:
\begin{itemize}
  \item \textbf{Lemma~\ref{lem:banach}:} The set \(\bb K\) of \(\C U\)-fine
    distributions on \(\C K\) is a Banach space;
  \item \textbf{Lemma~\ref{lem:convex}:} The subset \(\bb Q\) of \(\C U\)-fine
    probability measures on \(\C K\) is a convex, completely metrizable subset
    of \(\bb K\);
  \item \textbf{Lemma~\ref{lem:e_closed}:} The subset \(\bb E\) of \(\bb Q\) is
    a universally measurable subset of \(\bb K\), where \(\bb E\) is the set
    consisting of \(\C U\)-fine probability measures over which there exists a
    policy satisfying counterfactual equalized odds (resp., conditional
    principal fairness, or path-specific fairness) that is not strongly Pareto
    dominated.
\end{itemize}

Therefore, to apply Prop.~\ref{prop:k_shy}, it follows that what remains is to
construct a probe \(\bb W\) and show that \(\lambda_{\bb W}[\bb Q + \mu_0] > 0\)
for some \(\mu_0 \in \bb K\) but \(\lambda_{\bb W}[\bb E + \mu] = 0\) for all
\(\mu \in \bb K\).

\begin{proof}
  We divide the proof into three pieces. First, we illustrate how to construct
  the probe \(\bb W\) from a particular collection of distributions
  \(\{\nu_a^{\up}, \nu_a^{\low}\}_{a \in \C A}\). Second, we show that
  \(\lambda_{\bb W}[\bb E + \mu] = 0\) for all \(\mu \in \bb K\). For notational
  and expository simplicity, we focus in these first two sections on the case of
  counterfactual equalized odds. Therefore, in the third section, we show how to
  generalize the argument to conditional principal fairness and path-specific
  fairness.

  \medskip\noindent\emph{Construction of the probe.}\qquad
  We will construct our probe to address two different cases. We recall that, by
  Cor.~\ref{cor:exhaust}, any policy that is not strongly Pareto dominated must
  be a budget-exhausting multiple threshold policy with non-negative thresholds.
  In the first case, we consider when the candidate budget-exhausting multiple
  threshold policy is \(\B 1_{u(x) > 0}\). By perturbing the underlying
  distribution by \(\nu \in \bb W^{\low}\), we will be able to break the balance
  requirements implied by Eq.~\eqref{eq:counterfactual_equalized_odds}. In the
  second case, we treat the possibility that the candidate budget-exhausting
  multiple threshold policy has a non-trivial positive threshold for at least
  one group. By perturbing the underlying distribution by \(\nu \in \bb
  W^{\up}\) for an alternative set of perturbations \(\bb W^{\up}\), we will
  again be able to break the balance requirements.

  More specifically, to construct our probe \(\bb W = \bb W^{\up} + \bb
  W^{\low}\), we want \(\bb W^{\up}\) and \(\bb W^{\low}\) to have a number of
  properties. In particular, for all \(\nu \in \bb W\), perturbation by \(\nu\)
  should not affect whether the underlying distribution is a probability
  distribution, and should not affect how much of the budget is available to
  budget-exhausting policies. Specifically, for all \(\nu \in \bb W\),
  \begin{equation}
  \label{eq:preserve_prob}
    \int_{\C K} 1 \, \dx \nu = 0,
  \end{equation}
  and
  \begin{equation}
  \label{eq:preserve_budget}
    \int_{\C K} \B 1_{u(X) > 0} \, \dx \nu = 0.
  \end{equation}
  In fact, the amount of budget available to budget-exhausting policies will not
  change within group, i.e., for all \(a \in \C A\) and \(\nu \in \bb W\),
  \begin{equation}
  \label{eq:preserve_budget_ext}
    \int_{\C K} \B 1_{u(X) > 0, A = a} \, \dx \nu = 0.
  \end{equation}
  Additionally, for some distinguished \(y_0, y_1 \in \C Y\), non-zero
  perturbations in \(\nu^{\low} \in \bb W^{\low}\) should move mass between
  \(y_0\) and \(y_1\). That is, they should have the property that if
  \(\Pr_{|\nu^{\low}|}(A = a) > 0\), then
  \begin{equation}
  \label{eq:see_zero}
    \int_{\C K} \B 1_{u(X) < 0, Y = y_i, A = a} \, \dx \nu^{\low} \neq 0.
  \end{equation}
  Finally, perturbations in \(\bb W^{\up}\) should have the property that for
  any non-trivial \(t > 0\), some mass is moved either above or below \(t > 0\).
  More precisely, for any \(\mu \in \bb Q\) and any \(t\) such that
  \begin{equation*}
    0 < \Pr_\mu(u(X) > t \mid A = a) < 1,
  \end{equation*}
  if \(\nu^{\up} \in \bb W^{\up}\) is such that \(\Pr_{|\nu^{\up}|}(A = a) >
  0\), then
  \begin{equation}
  \label{eq:see_threshold}
    \int_{\C K} \B 1_{u(X) > t, A = a} \, \dx \nu^{\up} \neq 0.
  \end{equation}

  To carry out the construction, choose distinct \(y_0, y_1 \in \C Y\). Then,
  since
  \begin{equation*}
    \mu_{\max,a} \circ u^{-1}[S_a \cap [0, r_a)] - \mu_{\max,a} \circ u^{-1}[S_a
    \cap [r_a, \infty)]
  \end{equation*}
  is a continuous function of \(r_a\), it follows by the intermediate value
  theorem that we can partition \(S_a\) into three pieces,
  \begin{align*}
    S_a^{\low}
      &= S_a \cap (-\infty, 0), \\
    S^{\up}_{a,0}
      &= S_a \cap [0, r_a), \\
    S^{\up}_{a,1}
      &= S_a \cap [r_a, \infty),
  \end{align*}
  so that
  \begin{equation*}
    \Pr_{\mu_{\max,a}} \left( u(X) \in S^{\up}_{a,0} \right) =
    \Pr_{\mu_{\max,a}} \left( u(X) \in S^{\up}_{a,1} \right).
  \end{equation*}

  Recall that \(\C K = \C X \times \C Y\). Let \(\pi_{\C X} : \C K \to \C X\)
  denote projection onto \(\C X\), and \(\gamma_y : \C X \to \C K\) be the
  injection \(x \mapsto (x, y)\). We define
  \begin{align*}
    \nu_a^{\up}[E]
      &= \mu_{\max,a} \circ (\gamma_{y_1} \circ \pi_{\C X})^{-1} \left[ E \cap
        u^{-1} \left( S^{\up}_{a,1} \right) \right], \\
      &\hspace{1cm}- \mu_{\max,a} \circ (\gamma_{y_1} \circ \pi_{\C X})^{-1}
        \left[ E \cap u^{-1} \left( S^{\up}_{a,0} \right) \right], \\
    \nu_a^{\low}[E]
      &= \mu_{\max,a} \circ (\gamma_{y_1} \circ \pi_{\C X})^{-1} \left[ E \cap
        u^{-1} \left( S^{\low}_a \right) \right] \\
      &\hspace{1cm}- \mu_{\max,a} \circ (\gamma_{y_0} \circ \pi_{\C X})^{-1}
        \left[ E \cap u^{-1} \left( S^{\low}_a \right) \right].
  \end{align*}
  By construction, \(\nu_a^{\up}\) concentrates on
  \begin{equation*}
    \{y_1\} \times u^{-1}(S_a \cap [0, \infty)),
  \end{equation*}
  while \(\nu_a^{\low}\) concentrates on
  \begin{equation*}
    \{y_0, y_1\} \times u^{-1}(S_a \cap (-\infty, 0)).
  \end{equation*}
  Moreover, if we set
  \begin{align*}
    \bb W^{\up}
      &= \Span(\nu_a^{\up})_{a \in \C A}, \\
    \bb W^{\low}
      &= \Span(\nu_a^{\low})_{a \in \C A},
  \end{align*}
  then it is easy to see that
  Eqs.~\eqref{eq:preserve_prob}~to~\eqref{eq:see_zero} will hold. The only
  non-trivial case is Eq.~\eqref{eq:see_threshold}. However, by
  Cor.~\ref{cor:maximal}, the support of \(f_{\mu_{\max,a}}\) is maximal. That
  is, for \(\mu \in \bb Q\), if
  \begin{equation*}
    0 < \Pr_{\mu}(u(X) > t \mid A = a, u(X) > 0) < 1,
  \end{equation*}
  then it follows that \(0 < t < \sup S_a\). Either \(t \leq r_a\) or \(t >
  r_a\). First, assume \(t \leq r_a\); then, it follows by the construction of
  \(\nu_a^{\up}\) that
  \begin{align*}
    \nu_a^{\up} \circ u^{-1}[(t, \infty)]
      &= \int_{r_a}^\infty f_{\max,a} \, \dx \lambda - \int_t^{r_a} f_{\max,a}
      \, \dx \lambda \\
      &> \int_{r_a}^\infty f_{\max,a} \, \dx \lambda - \int_0^{r_a} f_{\max,a}
      \, \dx \lambda \\
      &= 0.
  \end{align*}
  Similarly, if \(t > r_a\),
  \begin{align*}
    \nu_a^{\up} \circ u^{-1}[(t, \infty)]
      &= \int_t^\infty f_{\max,a} \, \dx \lambda \\
      &> \int_{\sup S_a}^\infty f_{\max,a} \, \dx \lambda \\
      &= 0.
  \end{align*}
  Therefore Eq.~\eqref{eq:see_threshold} holds.

  Since \(\bb W\) is non-trivial\footnote{%
    In general, some or all of the \(\nu^{\low}\) may be zero depending on the
    \(\lambda\)-measure of \(S_a^{\low}\). However, as noted in
    Remark~\ref{rmk:dist_hyp}, the \(\nu_{a,i}^{\up}\) cannot be zero, since
    \(\Pr_{\mu_{\max,a}}(u(X) > 0) > 0\) for all \(a \in \C A\). Therefore \(\bb
    W \neq \{0\}\).
  }
  and \(\nu[\C K] = 0\) for all \(\nu \in \bb W\), it follows by
  Lemma~\ref{lem:probe} that \(\lambda_{\bb W}[\bb Q - \mu] > 0\) for some \(\mu
  \in \bb K\).

  \medskip\noindent\emph{Shyness.}\qquad
  Recall that, by Prop.~\ref{prop:shy_axioms_rel}, a set \(E\) is shy if and
  only if, for an arbitrary shy set \(E'\), \(E \setminus E'\) is shy. By
  Lemma~\ref{lem:condition}, a generic element of \(\mu \in \bb Q\) satisfies
  \(\Pr_{\mu}(u(X) > 0, Y(1) = y_i, A = a) > 0\) for \(i = 0, 1\), and \(a \in
  \C A\). Likewise, by Lemma~\ref{lem:support}, a generic \(\mu \in \bb Q\)
  satisfies \(\nu_a^{\up} \circ u^{-1} \Lt (\mu \rest_{\C X \times \{y_1\}})
  \circ u^{-1}\). Therefore, to simplify our task and recalling
  Remark~\ref{rmk:dist_hyp}, we may instead demonstrate the shyness of the set
  of \(\mu \in \bb Q\) such that:
  \begin{itemize}
    \item There exists a budget-exhausting multiple threshold policy \(\tau(x)\)
      with non-negative thresholds satisfying counterfactual equalized odds over
      \(\mu\);
    \item For \(i = 0, 1\),
      \begin{equation}
      \label{eq:imperturbable}
        \Pr_{\mu}(u(X) > 0, A = a, Y(1) = y_i) > 0;
      \end{equation}
    \item For all \(a \in \C A\),
      \begin{equation}
      \label{eq:nice_abscont}
        \nu_{a}^{\up} \circ u^{-1} \Lt (\mu \rest_{\alpha^{-1}(a) \times
        \{y_1\}}) \circ u^{-1}.
      \end{equation}
  \end{itemize}
  By a slight abuse of notation, we continue to refer to this set as \(\bb E\).
  We note that, by the construction of \(\bb W\), Eq.~\eqref{eq:imperturbable}
  is not affected by perturbation by \(\nu \in \bb W\), and
  Eq.~\eqref{eq:nice_abscont} is not affected by perturbation by \(\nu^{\low}
  \in \bb W\).

  In particular, by Lemma~\ref{lem:simple}, it suffices to show that
  \(\lambda_{\bb W}[\bb E - \mu] = 0\) for \(\mu \in \bb E\).

  Therefore, let \(\mu \in \bb E\) be arbitrary. Let the budget-exhausting
  multiple threshold policy satisfying counterfactual equalized odds over it be
  \(\tau(x)\), so that
  \begin{equation*}
    \EE_\mu[\tau(X)] = \min(b, \Pr_\mu(u(X) > 0)),
  \end{equation*}
  with thresholds \(\{t_a\}_{a \in \C A}\). We split into two cases based on
  whether \(\tau(X) = \B 1_{u(X) > 0}\) \(\mu\)-a.s.\ or not.

  In both cases, we make use of the following two useful observations.

  First, note that as \(\bb E \subseteq \bb Q\), if \(\mu + \nu\) is not a
  probability measure, then \(\mu + \nu \notin \bb E\). Therefore, without loss
  of generality, we assume throughout that \(\mu + \nu\) is a probability
  measure.

  Second, suppose \(\tau'(x)\) is a policy satisfying counterfactual equalized
  odds over some \(\nu \in \bb Q\). Then, if \(0 < \EE_\mu[\tau'(X)] < 1\), it
  follows that for all \(a \in \C A\),
  \begin{equation}
  \label{eq:all_nontrivial}
    0 < \EE_\mu[\tau'(X) \mid A = a] < 1.
  \end{equation}
  For, suppose not. Then, without loss of generality, there must be \(a_0, a_1
  \in \C A\) such that
  \begin{equation*}
    \EE_\mu[\tau'(X) \mid A = a_0] = 0
  \end{equation*}
  and
  \begin{equation*}
    \EE_\mu[\tau'(X) \mid A = a_1] > 0.
  \end{equation*}
  But then, by the law of iterated expectation, there must be some \(\C Y'
  \subseteq \C Y\) such that \(\mu[\C X \times \C Y'] > 0\) and so,
  \begin{equation*}
    \B 1_{\C X \times \C Y'} \cdot \EE_\mu [\tau'(X) \mid A = a_1, Y(1)]
      > 0
      = \B 1_{\C X \times \C Y'} \cdot \EE_\mu[\tau'(X) \mid A = a_0, Y(1)],
  \end{equation*}
  contradicting the fact that \(\tau'(x)\) satisfies counterfactual equalized
  odds over \(\mu\). Therefore, in what follows, we can assume that
  Eq.~\eqref{eq:all_nontrivial} holds.

  Our goal is to show that \(\lambda_{\bb W}[\bb E - \mu] = 0\).

  \begin{case}[\(\tau(X) = \B 1_{u(X) > 0}\)]
    We argue as follows. First, we show that \(\B 1_{u(X) > 0}\) is the unique
    budget-exhausting multiple threshold policy with non-negative thresholds
    over \(\mu + \nu\) for all \(\nu \in \bb W\). Then, we show that the set of
    \(\nu \in \bb W\) such that \(\B 1_{u(x) > 0}\) satisfies counterfactual
    equalized odds over \(\mu + \nu\) is a \(\lambda_{\bb W}\)-null set.

    We begin by observing that \(\bb W^{\low} \neq \{0\}\). For, if that were
    the case, then Eq.~\eqref{eq:all_nontrivial} would not hold for \(\tau(x)\).

    Next, we note that by Eq.~\eqref{eq:preserve_budget}, for any \(\nu \in \bb
    W\),
    \begin{equation*}
      \Pr_{\mu + \nu}(u(X) > 0) = \Pr_\mu(u(X) > 0)
    \end{equation*}
    and so
    \begin{equation*}
      \EE_{\mu + \nu}[\B 1_{u(X) > 0}] = \min(b, \Pr_{\mu + \nu}(u(X) > 0)).
    \end{equation*}
     If \(\tau'(x)\) is a feasible multiple threshold policy with non-negative
     thresholds and \(\tau'(X) \neq \B 1_{u(X) > 0}\) \((\mu + \nu)\)-a.s.,
     then, as a consequence,
    \begin{equation*}
      \EE_{\mu + \nu}[\tau'(X)] < \Pr_{\mu + \nu}(u(X) > 0) \leq b.
    \end{equation*}
    Therefore, it follows that \(\B 1_{u(X) > 0}\) is the unique
    budget-exhausting multiple threshold policy over \(\mu + \nu\) with
    non-negative thresholds.

    Now, note that if counterfactual equalized odds holds with decision policy
    \(\tau(x) = \B 1_{u(x) > 0}\), then, by
    Eq.~\eqref{eq:counterfactual_fairness} and Lemma~\ref{lem:cond_prob}, we
    must have that
    \begin{equation*}
      \Pr_{\mu + \nu}(u(X) > 0 \mid A = a, Y(1) = y_1) = \Pr_{\mu + \nu}(u(X) >
      0 \mid A = a', Y(1) = y_1)
    \end{equation*}
    for \(a, a' \in \C A\).\footnote{%
      To ensure that both quantities are well-defined, here and throughout the
      remainder of the proof we use the fact that by
      Eqs.~\eqref{eq:preserve_budget_ext}~and~\eqref{eq:imperturbable},
      \(\Pr_{\mu + \nu}(u(X) > 0, A = a, Y(1) = y_1) > 0\).
    }

    Now, we will show that a typical element of \(\bb W\) breaks this balance
    requirement. Choose \(a^*\) such that \(\nu^{\low}_{a*} \neq 0\). Recall
    that \(\nu\) is fixed, and let \(\nu' = \nu - \beta_{a^*}^{\low} \cdot
    \nu^{\low}_{a^*}\). Let
    \begin{equation*}
      p_a = \Pr_{\mu + \nu'}(u(X) > 0 \mid A = a', Y(1) = y_1).
    \end{equation*}
    Note that it cannot be the case that \(p_a = 0\) for all \(a \in \C A\),
    since, by Eq.~\eqref{eq:imperturbable},
    \begin{equation*}
      \Pr_{\mu + \nu'}(u(X) > 0 \mid Y(1) = y_1) > 0.
    \end{equation*}
    Therefore, by the foregoing discussion, either \(p_{a^*} > 0\) or \(p_{a^*}
    = 0\) and we can choose \(a' \in \C A\) such that \(p_{a'} > 0\). Since the
    \(\nu_a^{\low}\), \(\nu_{a,i}^{\up}\) are all mutually singular, it follows
    that counterfactual equalized odds can only hold over \(\mu + \nu\) if
    \begin{equation*}
    \label{eq:pa_prime}
      p_{a'} = \Pr_{\mu + \nu}(u(X) > 0 \mid A = a^*, Y(1) = y_1).
    \end{equation*}
    Now, we observe that by Lemma~\ref{lem:cond_prob}, that
    \begin{equation*}
      \Pr_{\mu + \nu}(u(X) > 0 \mid A = a^*, Y(1) = y_1) = \frac {\eta}
      {\pi + \beta_{a^*}^{\low} \cdot \rho}
    \end{equation*}
    where
    \begin{align*}
      \eta
        &= \Pr_\mu(u(X) > 0, A = a^*, Y(1) = y_1) \\
      \pi
        &= \Pr_\mu(A = a^*, Y(1) = y_1), \\
      \rho
        &= \int_{\C K} \B 1_{A = a^*, Y(1) = y_1} \, \dx \nu_{a^*}^{\low}.
    \end{align*}
    since
    \begin{align*}
      0
        &= \int_{\C K} \B 1_{u(X) > 0, A = a^*, Y(1) = y_1} \, \dx
          \nu_{a^*}^{\low}, \\
      0
        &\neq \int_{\C K} \B 1_{A = a^*, Y(1) = y_1} \, \dx \nu_{a^*}^{\low}.
    \end{align*}
    Here, the equality follows by the fact that \(\nu^{\low}\) is supported on
    \(S^{\low}_a \times \{y_0, y_1\}\) and the inequality from
    Eq.~\eqref{eq:see_zero}.

    Therefore, if, in the first case, \(p_{a'} > 0\), then counterfactual
    equalized odds only holds if
    \begin{equation*}
      \beta_{a^*}^{\low} = \frac {e - p_{a'} \cdot \pi} {p_{a'}
      \cdot \rho},
    \end{equation*}
    since, as noted above, \(\rho \neq 0\) by Eq.~\eqref{eq:see_zero}. In the
    second case, if \(p_{a'} = 0\), then counterfactual equalized odds can only
    hold if
    \begin{equation*}
      e = p_{a^*} \cdot \pi = 0.
    \end{equation*}
    Since we chose \(a'\) so that \(p_{a^*} > 0\) if \(p_{a'} = 0\) and \(\pi >
    0\) by Eq.~\eqref{eq:imperturbable}, this is impossible.

    In either case, we see that the set of \(\beta_{a^*}^{\low} \in \B R\) such
    that there a budget-exhausting threshold policy with positive thresholds
    satisfying counterfactual equalized odds over \(\mu + \nu' +
    \beta_{a^*}^{\low} \cdot \nu_{a^*}^{\low}\) has \(\lambda\)-measure zero.
    That is
    \begin{equation*}
      \lambda_{\Span(\nu_{a^*}^{\low})}[\bb E - \mu - \nu'] = 0.
    \end{equation*}
    Since \(\nu'\) was arbitrary, it follows by Fubini's theorem that
    \(\lambda_{\bb W}[\bb E - \mu] = 0\).
  \end{case}

  \begin{case}[\(\tau(X) \neq \B 1_{u(X) > 0}\)]
    Our proof strategy is similar to the previous case. First, we show that, for
    a given fixed \(\nu^{\low} \in \bb W^{\low}\), there is a unique candidate
    policy \(\tilde \tau(x)\) for being a budget-exhausting multiple threshold
    policy with non-negative thresholds and satisfying counterfactual equalized
    odds over \(\mu + \nu^{\low} + \nu^{\up}\) for any \(\nu^{\up} \in \bb
    W^{\up}\). Then, we show that the set of \(\nu^{\up}\) such that \(\tilde
    \tau(X)\) satisfies counterfactual equalized odds has \(\lambda_{\bb
    W^{\up}}\) measure zero. Finally, we argue that this in turn implies that
    the set of \(\nu \in \bb W\) such that there exists a Pareto efficient
    policy satisfying counterfactual equalized odds over \(\mu + \nu\) has
    \(\lambda_{\bb W}\)-measure zero.

    We seek to show that \(\lambda_{\bb W^{\up}}[\bb E - (\mu + \nu^{\low})] =
    0\). To begin, we note that since \(\nu_{a,i}^{\up}\) concentrates on
    \(\{y_1\} \times \C X\) for all \(a \in \C A\), it follows that
    \begin{equation*}
      \EE_{\mu + \nu^{\low}}[d(X) \mid A = a, Y(1) = y_0] = \EE_{\mu +
      \nu^{\low} + \nu^{\up}}[d(X) \mid A = a, Y(1) = y_0]
    \end{equation*}
    for any \(\nu^{\up} \in \bb W^{\up}\).

    Now, suppose there exists some \(\nu^{\up} \in \bb W^{\up}\) such that there
    exists a budget-exhausting multiple threshold policy \(\tilde \tau(x)\) with
    non-negative thresholds such that counterfactual equalized odds is satisfied
    over \(\mu + \nu^{\low} + \nu^{\up}\). (If not, then we are done and
    \(\lambda_{\bb W^{\up}}[\bb E - (\mu + \nu^{\low})] = 0\), as the measure of
    the empty set is zero.) Let
    \begin{equation*}
      p = \EE_{\mu + \nu^{\low}}[\tilde \tau(X) \mid A = a, Y(1) = y_0].
    \end{equation*}
    Suppose that \(\tilde \tau'(x)\) is an alternative budget-exhausting
    multiple threshold policy with non-negative thresholds such that
    counterfactual equalized odds is satisfied. We seek to show that \(\tau'(X)
    = \tau(X)\) \((\mu + \nu^{\low} + \nu^{\up})\)-a.e.\ for any \(\nu^{\up} \in
    \bb W^{\up}\). Toward a contradiction, suppose that for some \(a_0 \in \C
    A\),
    \begin{equation*}
      \EE_{\mu + \nu^{\low}}[\tilde \tau'(X) \mid A = a_0, Y(1) = y_0] < p.
    \end{equation*}
    Since, by Eq.~\eqref{eq:imperturbable}, \(\Pr_{\mu + \nu^{\low}}(A = a_0,
    Y(1) = y_0) > 0\), it follows that
    \begin{equation*}
      \EE_{\mu + \nu^{\low}}[\tilde \tau'(X) \mid A = a_0] < \EE_{\mu +
      \nu^{\low}}[\tilde \tau(X) \mid A = a_0].
    \end{equation*}
    Therefore, since \(\tilde \tau(x)'\) is budget exhausting, there must be
    some \(a_1\) such that
    \begin{equation*}
      \EE_{\mu + \nu^{\low}}[\tilde \tau'(X) \mid A = a_1] > \EE_{\mu +
      \nu^{\low}}[\tilde \tau(X) \mid A = a_1].
    \end{equation*}
    From this, it follows \(\tilde \tau'(x)\) can be represented by a threshold
    greater than or equal to that of \(\tilde \tau(x)\) on \(\alpha^{-1}(a_1)\),
    and hence
    \begin{align*}
      \EE_{\mu + \nu^{\low}}[\tilde \tau'(X) \mid A = a_1, Y(1) = y_0]
        &\geq \EE_{\mu + \nu^{\low}}[\tilde \tau(X) \mid A = a_0, Y(1) = y_0] \\
        &= p \\
        &> \EE_{\mu + \nu^{\low}}[\tilde \tau'(X) \mid A = a_0, Y(1) = y_0],
    \end{align*}
    contradicting the fact that \(\tilde \tau'(x)\) satisfies counterfactual
    equalized odds.

    By the fact that \(\nu^{\low}\) is supported on \(u^{-1}((-\infty, 0])\),
    the preceding discussion, and Lemma~\ref{lem:simplethresh}, it follows that
    \begin{equation*}
      \tilde \tau(X) = \tilde \tau'(X) \quad (\mu \rest_{\C X \times
      \{y_0\}})\text{-a.e.}
    \end{equation*}
    By Eq.~\eqref{eq:nice_abscont}, it follows that \(\tilde \tau(X) = \tilde
    \tau'(X)\) \(\nu^{\up}_{a,i}\)-a.e.\ for \(i = 0, 1\). As a consequence,
    \begin{equation*}
      \tilde \tau(X) = \tilde \tau'(X) \quad (\mu + \nu^{\low} +
      \nu^{up})\text{-a.e.}
    \end{equation*}
    for all \(\nu^{\up} \in \bb W^{\up}\). Therefore \(\tilde \tau(X)\) is,
    indeed, unique, as desired.

    Now, we note that since \(\tau(X) \neq \B 1_{u(X) > 0}\), it follows that
    \(\EE[\tau(X)] < \Pr_\mu(u(X) > 0)\). It follows that \(\EE_\mu[\tau(X)] =
    b\), since \(\tau(x)\) is budget exhausting. Therefore, by
    Eq.~\eqref{eq:preserve_budget}, it follows that for any budget-exhausting
    policy \(\tilde \tau(X)\), \(\EE[\tilde \tau(X)] = b\), and so \(\tilde
    \tau(X) \neq \B 1_{u(X) > 0}\) over \(\mu + \nu\).

    Therefore, fix \(\nu^{\low}\) and \(\tilde \tau(X)\). By
    Eq.~\eqref{eq:all_nontrivial}, there is some \(a^*\) such that
    \begin{equation*}
      0 < \Pr_{\mu + \nu^{\low}}(u(X) > \tilde t_{a^*} \mid A = a^*) < 1.
    \end{equation*}
    Then, it follows by Eq.~\eqref{eq:see_threshold} that
    \begin{equation*}
      \int_{\C K} \B 1_{u(X) > \tilde t_{a^*}} \, \dx \nu^{\up}_{a^*} \neq
      0.
    \end{equation*}
    Fix \(\nu' = \nu - \beta_{a^*}^{\up} \cdot \nu_{a^*}^{\up}\). Then, for some
    \(a \neq a^*\), set
    \begin{equation*}
      p^* = \EE_{\mu + \nu'}[\tilde \tau(X) \mid A = a, Y(1) = y_1].
    \end{equation*}
    Since the \(\nu_a^{\low}\), \(\nu_a^{\up}\) are all mutually singular, it
    follows that counterfactual equalized odds can only hold over \(\mu + \nu\)
    if
    \begin{equation*}
      p^* = \Pr_{\mu + \nu}(u(X) > \tilde t_{a^*} \mid A = a^*, Y(1) = y_1).
    \end{equation*}
    Now, we observe that by Lemma~\ref{lem:cond_prob}, that
    \begin{align}
    \label{eq:case_2_exp}
    \begin{split}
      \Pr_{\mu + \nu}(u(X) > \tilde t_{a^*} &\mid A = a^*, Y(1) = y_1) \\
        &= \frac {\eta + \beta_a^{\up} \cdot \gamma} \pi
    \end{split}
    \end{align}
    where
    \begin{align*}
      \eta
        &= \Pr_{\mu + \nu^{\low}}(u(X) > \tilde t_{a^*} \mid A = a^*, Y(1) =
          y_1), \\
      \pi
        &= \Pr_{\mu + \nu^{\low}}(A = a^*, Y(1) = y_1), \\
      \gamma
        &= \int_{\C K} \B 1_{u(X) > \tilde t_{a^*}, A = a^*, Y(1) = y_1} \, \dx
          \nu_a^{\up},
    \end{align*}
    and we note that
    \begin{equation*}
      0 = \int_{\C K} \B 1_{A = a^*, Y(1) = y_1} \, \dx \nu_a^{\low}.
    \end{equation*}
    Eq.~\eqref{eq:case_2_exp} can be rearranged to
    \begin{equation*}
      (p^* \cdot \pi - \eta) - \beta \cdot \gamma = 0.
    \end{equation*}
    This can only hold if
    \begin{equation*}
      \beta = \frac {p^* \cdot \pi - \eta} \gamma,
    \end{equation*}
    since by Eq.~\eqref{eq:see_threshold}, \(\gamma \neq 0\). Since any
    countable subset of \(\B R\) is a \(\lambda\)-null set,
    \begin{equation*}
      \lambda_{\Span(\nu_{a^*}^{\up})}[\bb E - \mu - \nu'] = 0.
    \end{equation*}
    Since \(\nu'\) was arbitrary, it follows by Fubini's theorem that
    \(\lambda_{\bb W^{\up}}[\bb E - \mu - \nu^{\low}] = 0\) in this case as
    well. Lastly, since \(\nu^{\low}\) was also arbitrary, applying Fubini's
    theorem a final time gives that \(\lambda_{\bb W}[\bb E - \mu] = 0\).
  \end{case}

  \medskip\noindent\emph{Conditional principal fairness and path-specific fairness.}\qquad
  The extension of these results to conditional principal fairness and
  path-specific fairness is straightforward. All that is required is a minor
  modification of the probe.

  In the case of conditional principal fairness, we set
  \begin{align*}
    \nu_{a,w}^{\up}[E]
      &= \mu_{\max,a,w} \circ (\gamma_{(y_1,y_1)} \circ \pi_{\C X})^{-1}[E \cap
        u^{-1}(S^{\up}_{a,1})], \\
      &\hspace{1cm}- \mu_{\max,a} \circ (\gamma_{(y_1,y_1)} \circ \pi_{\C
        X})^{-1}[E \cap u^{-1}(S^{\up}_{a,w})], \\
    \nu_{a,w}^{\low}[E]
      &= \mu_{\max,a,w} \circ (\gamma_{(y_1,y_1)} \circ \pi_{\C X})^{-1}[E \cap
        u^{-1}(S^{\low}_a)] \\
      &\hspace{1cm}- \mu_{\max,a} \circ (\gamma_{(y_0,y_0)} \circ \pi_{\C
        X})^{-1}[E \cap u^{-1}(S^{\low}_{a,w})],
  \end{align*}
  where \(\gamma_{(y,y')} : \C X \to \C K\) is the injection \(x \mapsto (x, y,
  y')\).
  Our probe is then given by
  \begin{align*}
    \bb W^{\up}
    &= \Span(\nu_{a,w}^{\up}), \\
    \bb W^{\low}
    &= \Span(\nu_{a,w}^{\low}),
  \end{align*}
  almost as before.

  The proof otherwise proceeds virtually identically, except for two points.
  First, recalling Remark~\ref{rmk:dist_hyp}, we use the fact that a generic
  element of \(\bb Q\) satisfies \(\Pr_\mu(A = a, W = w) > 0\) in place of
  \(\Pr_\mu(A = a) > 0\) throughout. Second, we use the fact that \(\omega\)
  overlaps utility in place of Eq.~\eqref{eq:all_nontrivial}. In particular, If
  \(\omega\) does not overlap utilities for a generic \(\mu \in \bb Q\), then,
  by Lemma~\ref{lem:overlap}, there exists \(w \in \C W\) such that
  \(\Pr_\mu(u(X) > 0, W = w) = 0\) for all \(\mu \in \bb Q\). If this occurs, we
  can show that no budget-exhausting multiple threshold policy with positive
  thresholds satisfies conditional principal fairness, exactly as we did to show
  Eq.~\eqref{eq:all_nontrivial}.

  In the case of path-specific fairness, we instead define
  \begin{align*}
    S_{a,w}^{\low}
    &= S_{a,w} \cap (-\infty, r_{a,w}), \\
    S_{a,w}^{\up}
    &= S_{a,w} \cap [r_{a,w}, \infty),
  \end{align*}
  where \(r_{a,w}\) is chosen so that
  \begin{equation*}
    \Pr_{\mu_{\max,a,w}}(u(X) \in S_{a,w}^{\low}) = \Pr_{\mu_{\max,a,w}}(u(X)
    \in S_{a,w}^{\up}).
  \end{equation*}
  Let \(\pi_X\) denote the projection from \(\C K = \C A \times \C X^{\C A}\)
  given by
  \begin{equation*}
    \left( a, (x_{a'})_{a' \in \C A} \right) \mapsto x_a.
  \end{equation*}
  Let \(\pi_{a'}\) denote the projection from the \(a'\)-th component. (That is,
  given \(\mu \in \bb K\), the distribution of \(X_{\Pi, A, a'}\) over \(\mu\)
  is given by \(\mu \circ \pi^{-1}_{a'}\) and the distribution of \(X\) is given
  by \(\mu \circ \pi^{-1}_X\).)
  Then, we let \(\tilde \mu_{\max,a,w}\) be the measure on \(\C X\)
  given by
  \begin{align*}
    \tilde \mu_{\max,a,w}[E]
    &= \mu_{\max,a,w}[E \cap (u \circ \pi_a)^{-1}(s_{a,w}^{\up})] \\
    &\hspace{1cm} - \mu_{\max,a,w}[E \cap (u \circ \pi_a)^{-1}(S_{a,w}^{\low})].
  \end{align*}
  Finally, let \(\phi : \C A \to \C A\) be a
  permutation of the groups with no fixed points, i.e., so that \(a' \neq
  \phi(a')\) for all \(a' \in \C A\). Then, we define
  \begin{equation*}
    \nu_{a'} = \delta_{a'} \times \tilde \mu_{\max,\phi(a'),w_1} \times \prod_{a
    \neq \phi(a')}
    \mu_{\max,a,w_1} \circ \pi_a^{-1},
  \end{equation*}
  where \(\delta_a\) is the measure on \(\C A\) given by \(\delta_a[\{a'\}] =
  \B 1_{a = a'}\). Then, simply let
  \begin{equation*}
    \bb W = \Span(\nu_a')_{a'\in \C A}.
  \end{equation*}

  Since \(\tilde \mu_{\max,a,w}[\C X] = 0\) for all \(a \in \C A\), it follows
  that \(\nu_{a,w} \circ \pi_X^{-1} = 0\), i.e.,
  \begin{equation*}
    \Pr_\mu(X \in E) = \Pr_{\mu + \nu}(X \in E)
  \end{equation*}
  for any \(\nu \in \bb W\) and \(\mu \in \bb Q\). Therefore
  Eqs.~\eqref{eq:preserve_prob}~and~\eqref{eq:preserve_budget} hold. Moreover,
  the \(\nu_a\) satisfy the following strengthening of
  Eq.~\eqref{eq:see_threshold}. Perturbations in \(\bb W\) have the property
  that for any non-trivial \(t\)---not necessarily positive---some of the mass
  of \(u(X_{\Pi, A, a})\) is moved either above or below \(t\). More precisely,
  for any \(\mu \in \bb Q\) and any \(t\) such that
  \begin{equation*}
    0 < \Pr_\mu(u(X) > t \mid A = a) < 1,
  \end{equation*}
  if \(\nu \in \bb W\) is such that \(\Pr_{|\nu|}(A = \phi^{-1}(a)) > 0\), then
  \begin{equation}
    \int_{\C K} \B 1_{u(X_{\Pi, A, a}) > t} \, \dx \nu_a \neq 0.
  \end{equation}
  This stronger property means that we need not separately treat the case where
  \(\tau(X) = \B 1_{u(X) > 0}\) \(\mu\)-a.e.

  Other than this difference the proof proceeds in the same way, except for two
  points. First, we again make use of the fact that \(\omega\) can be assumed to
  overlap utilities in place of Eq.~\eqref{eq:all_nontrivial}, as in the case of
  conditional principal fairness. Second, \(w_0\) and \(w_1\) take the place of
  \(y_0\) and \(y_1\). In particular, to establish the uniqueness of \(\tilde
  \tau(x)\) given \(\mu\) and \(\nu^{\low}\) in the second case, instead of
  conditioning on \(y_0\), we instead condition on \(w_0\), where, following the
  discussion in Remark~\ref{rmk:dist_hyp} and Lemma~\ref{lem:condition}, this
  conditioning is well-defined for a generic element of \(\bb Q\).
\end{proof}

We have focused on causal definitions of fairness, but the thrust of our
analysis applies to non-causal conceptions of fairness as well. Below we show
that policies constrained to satisfy (non-counterfactual) equalized
odds~\citep{hardt2016equality} are generically strongly Pareto dominated, a
result that follows immediately from our proof above.

\begin{definition}
\label{defn:eo}
  \emph{Equalized odds} holds for a decision policy \(d(x)\) when
  \begin{equation}
  \label{eq:eo}
    d(X) \indep A \mid Y.
  \end{equation}
\end{definition}

We note that \(Y\) in Eq.~\eqref{eq:eo} does \emph{not} depend on our choice of
\(d(x)\), but rather represents the realized value of \(Y\), e.g., under some
\emph{status quo} decision making rule.

\begin{corollary}
\label{cor:eo}
  Suppose \(\C U\) is a set of utilities consistent modulo \(\alpha\). Further
  suppose that for all \(a \in \C A\) there exist a \(\C U\)-fine distribution
  of \(X\) and a utility \(u \in \C U\) such that \(\Pr(u(X) > 0, A = a) > 0\),
  where \(A = \alpha(X)\). Then, for almost every \(\C U\)-fine distribution of
  \(X\) and \(Y\) on \(\C X \times \C Y\), any decision policy \(d(x)\)
  satisfying equalized odds is strongly Pareto dominated.
\end{corollary}

\begin{proof}
  Consider the following maps. Distributions of \(X\) and \(Y(1)\), i.e.,
  probability measures on \(\C X \times \C Y\), can be embedded in the space of
  joint distributions on \(X\), \(Y(0)\), and \(Y(1)\) via pushing forward by
  the map \(\iota\), where \(\iota : (x, y) \mapsto (x, y, y)\). Likewise, given
  a fixed decision policy \(D = d(X)\), joint distributions of \(X\), \(Y(0)\),
  and \(Y(1)\) can be projected onto the space of joint distributions of \(X\)
  and \(Y\) by pushing forward by the map \(\pi_d : (x, y_0, y_1) \mapsto (x,
  y_{d(x)})\). Lastly, we see that the composition of \(\iota\) and
  \(\pi_d\)---regardless of our choice of \(d(x)\)---is the identity, as shown
  in the diagram below.
  \begin{equation*}
    \begin{tikzpicture}[xscale = 3, yscale = 2]
      \node at (0,1) (A) {\(\C X \times \C Y\)};
      \node at (1,1) (B) {\(\C X \times \C Y \times \C Y\)};
      \node at (1,0) (C) {\(\C X \times \C Y\)};
      \draw[->] (A) to node[above] {\(\iota\)} (B);
      \draw[->] (B) to node[left] {\(\pi_d\)} (C);
      \draw[->] (A) to node[below left] {\(\id\)} (C);
    \end{tikzpicture}
  \end{equation*}
  We note also that counterfactual equalized odds holds for \(\mu\) exactly when
  equalized odds holds for \(\mu \circ (\pi_d \circ \iota)^{-1}\). The result
  follows immediately from this and Theorem~\ref{thm:dist}.
\end{proof}

\subsection{Proof of Theorem~\ref{thm:fpr}}
\label{app:fpr}

The proof of Theorem~\ref{thm:fpr} is simpler than the proof of
Theorem~\ref{thm:dist}, but uses some of the same machinery. As before, let \(\C
K = \C A \times [0,1]\) denote the state space, and \(\B K\) denote the set of
measures on \([0, 1] \times \C A\). Let \(\bb K\) denote the measures \(\mu \in
\B K\) that are absolutely continuous with respect to \(\lambda \times \delta\),
where \(\lambda\) is Lebesgue measure and \(\delta\) is the counting measure on
\(\C A\)---i.e., measures such that the restriction to \([0,1] \times \{a\}\)
has a density for all \(a \in \C A\). Applying Cor.~\ref{cor:banach} with \(\C U
= \{\pi\}\), where \(\pi : (u, a) \mapsto u\), shows that \(\bb K\) is a Banach
space. As before, we let \(\bb Q \subseteq \bb K\) denote the probability
simplex, i.e., the set of all \(\mu \in \bb K\) such that \(\mu[\C K] = 1\) and
\(\mu[E] \geq 0\) for all Borel sets \(E\).

Let \(R = r(X)\) denote risk.\footnote{%
     Here, since the measures are on the risk scale, rather than on \(X\), we
     write \(R\) for notational simplicity.
}
By Lemma~\ref{lem:maximize}, we have that a policy is utility maximizing if and
only if \(\B 1_{R > t} \leq d(X) \leq \B 1_{R \geq t}\) a.s. By Since
\(\Pr_\mu(R = t) = 0\) for any absolutely continuous measure \(\mu\), we see
that, in fact, a policy is utility maximizing if and only if \(\B 1_{R > t} =
d(X)\) \(\mu\)-a.s.

Consider the sets
\[
  \bb E_{\T {FP}} = \left\{\mu \in \C K : (\forall a \in \C A) \, \frac {\B
  E_\mu[(1 - R) \cdot \B 1_{A = a, R > t}]} {\B E_\mu[(1 - R) \cdot \B 1_{A =
a}]} = \frac {\B E_\mu[(1 - R) \cdot \B 1_{R > t}]} {\B E_\mu[1 - R]} \right\},
\]
and
\[
  \bb E_{\T {DP}} = \{\mu \in \C K : (\forall a \in \C A) \, \Pr_\mu(R > t \mid
  A = a) = \Pr_\mu(R > t)\}.
\]
We note that since the mapping \(\mu \mapsto \Pr_\mu(E)\) is a continuous
function on \(\C K\), the set of measures such that \(\Pr_\mu(A = a, R > t) =
0\) or \(\Pr_\mu(R > t) = 0\) is closed and shy. It follows that \(\bb E_{\T
{FP}}\) and \(\bb E_{\T {DP}}\) are Borel. We note that, by definition, \(\bb
E_{\T {DP}}\) is the set of risk distributions such that there exists a
utility-maximizing policy satisfying demographic parity (when the condition is
well defined). Likewise, an application of the law of iterated expectations
yields that \(\bb E_{\T {FP}}\) is the set of distributions satisfying equalized
false positive rates (when the condition is well-defined).

With this context in place, we can move on to the proof of
Theorem~\ref{thm:fpr}.

\begin{proof}[Proof of Theorem~\ref{thm:fpr}]
  First we construct the probe. Choose distinct \(a_0, a_1 \in \C A\)
  arbitrarily, and let \(\tilde \nu\) be defined by
  \[
    \tilde \nu[E] = \frac {\lambda[E_{a_0} \cap [0,t)]} t - \frac
    {\lambda[E_{a_1} \cap [0,t)]} t,
  \]
  where \(E_a = E \cap \{A = a\}\). Then, by Lemma~\ref{lem:probe}, there exists
  \(\mu_0\) such that \(\lambda_{\bb W}[\C K + \nu_0] > 0\), where \(\bb K = \D
  {Span}(\tilde \nu)\).

  We see that
  \begin{align*}
    \Pr_{\tilde \nu}(R < t, A = a_0)
      &= 1,
    &\Pr_{\tilde \nu}(R > t, A = a_0)
      &= 0, \\
    \Pr_{\tilde \nu}(R < t, A = a_1)
      &= -1,
    &\Pr_{\tilde \nu}(R > t, A = a_1)
      &= 0.
  \end{align*}

  It follows that
  \begin{align*}
    \Pr_{\mu + \beta \cdot \tilde \nu}(R > t \mid A = a_0)
        &= \frac {e_0} {p_0 + \beta},
    & \Pr_{\mu + \beta \cdot \tilde \nu}(R > t, \mid A = a_1)
        &= \frac {e_1} {p_1 - \beta},
  \end{align*}
  where
  \begin{align*}
    e_0
      &= \Pr_\mu(R > t, A = a_0),
    & e_1
      &= \Pr_\mu(R > t, A = a_1), \\
    p_0
      &= \Pr_\mu(A = a_0),
    & p_1
      &= \Pr_\mu(A = a_1).
  \end{align*}
  Note that by Lemmata~\ref{lem:support}~and~\ref{lem:simple}, we can assume
  that \(e_0, e_1 > 0\). It follows, rearranging terms, that
  \[
    \Pr_{\mu + \beta \cdot \tilde \nu}(R > t \mid A = a_0) = \Pr_{\mu + \beta
    \cdot \tilde \nu}(R > t \mid A = a_1)
  \]
  if and only if
  \[
    \beta = \frac {e_0 \cdot p_1 - e_1 \cdot p_0} {e_0 + e_1},
  \]
  which is a measure-zero subset of \(\beta \in \B R\). Therefore \(\lambda_{\bb
  W}[\bb E_{\T {DP}} + \mu] = 0\).

  In the same way, we observe that
  \begin{align*}
      \B E_{\tilde \nu}[R \cdot \B 1_{A = a_0, R < t}]
        &= \frac t 2,
      & \B E_{\tilde \nu}[R \cdot \B 1_{A = a_0, R > t}]
        &= 0 \\
      \B E_{\tilde \nu}[R \cdot \B 1_{A = a_1, R < t}]
        &= -\frac t 2,
      & \B E_{\tilde \nu}[R \cdot \B 1_{A = a_1, R > t}]
        &= 0,
  \end{align*}
  and so
  \begin{align*}
    \frac {\B E_{\mu + \beta \cdot \tilde \nu}[(1 - R) \cdot \B 1_{A = a_0, R >
    t}]} {\B E_{\mu + \beta \cdot \tilde \nu}[(1 - R) \cdot \B 1_{A = a_0}]}
        &= \frac {e_0'} {p_0' + \beta \cdot \frac t 2},
    & \frac {\B E_{\mu + \beta \cdot \tilde \nu}[(1 - R) \cdot \B 1_{A = a_1, R
    > t}]} {\B E_{\mu + \beta \cdot \tilde \nu}[(1 - R) \cdot \B 1_{A = a_1}]}
        &= \frac {e_1'} {p_1' - \beta \cdot \frac t 2},
  \end{align*}
  where
  \begin{align*}
    e_0'
      &= \B E_\mu[(1 - R) \cdot \B 1_{A = a_0, R > t}],
    & e_1'
      &= \B E_\mu[(1 - R) \cdot \B 1_{A = a_a, R > t}], \\
    p_0'
      &= \B E_\mu[(1 - R) \cdot \B 1_{A = a_0}],
    & p_1'
      &= \B E_\mu[(1 - R) \cdot \B 1_{A = a_0}].
  \end{align*}
  As before, \(\mu + \beta \cdot \tilde \nu \in \bb E_{\T {FP}}\) if and only if
  \[
    \beta = \frac 2 t \cdot \frac {e_0 \cdot p_1 - e_1 \cdot p_0} {e_0 + e_1},
  \]
  which is again a measure-zero subset of \(\beta \in \B R\). Therefore
  \(\lambda_{\bb W}[\bb E_{\T {FP}} + \mu] = 0\).

  Therefore it follows that both \(\bb E_{\T {FP}}\) and \(\bb E_{\T {DP}}\) are
  shy.
\end{proof}

\subsection{Proof of Corollary~\ref{cor:committee}}
\label{app:committee}

The proof of Corollary~\ref{cor:committee} is a straightforward application of
Theorem~\ref{thm:dist}. We begin by giving the complete theorem statement.

\begin{corollary}
\label{cor:committee-full}
  Consider a utility of the form given in Eq.~\eqref{eq:agg_util}, where \(v\)
  is monotonically increasing in both coordinates and \(m(x) \geq 0\). Suppose
  that for all \(a \in \C A\) there exist an \(\{m\}\)-fine distribution of \(X\)
  such that \(\Pr(m(X) > 0, A = a) > 0\), where \(A = \alpha(X)\). Then,
  \begin{itemize}
    \item For almost every \(\{m\}\)-fine distribution of \(X\) and \(Y(1)\),
      no utility-maximizing decision policy satisfies counterfactual equalized
      odds.
    \item If \(|\img(\omega)| < \infty\) and there exists an \(\{m\}\)-fine
      distribution of \(X\) such that \(\Pr(A = a, W = w) > 0\) for all \(a \in
      \C A\) and \(w \in \img(\omega)\), where \(W = \omega(X)\), then, for
      almost every \(\{m\}\)-fine joint distribution of \(X\), \(Y(0)\), and
      \(Y(1)\), no utility-maximizing decision policy satisfies conditional
      principal fairness.
    \item If \(|\img(\omega)| < \infty\) and there exists a \(\{m\}\)-fine
      distribution of \(X\) such that \(\Pr(A = a, W = w_i) > 0\) for all \(a
      \in \C A\) and some distinct \(w_0, w_1 \in \img(\omega)\), then, for
      almost every \(\{m\}^{\C A}\)-fine joint distributions of \(A\) and the
      counterfactuals \(X_{\Pi, A, a'}\), no utility-maximizing decision policy
      satisfies path-specific fairness.
  \end{itemize}
\end{corollary}

Recall that for notational simplicity, we refer to the distinguished utility as
\(u^*\), where
\[
    u^*(d) = v \left(\B E[m(X) \cdot d(X)], \B E[\B 1_{\alpha(X) = a_1} \cdot
    d(X)] \right).
\]

\begin{proof}[Proof of Corollary~\ref{cor:committee}]
    Consider the subset \(S\) of \(\B R^2\) consisting of all pairs
    \[
        (\B E[m(X) \cdot d(X)], \B E[\B 1_{\alpha(X) = a_1} \cdot d(X)]),
    \]
    where \(d\) ranges over feasible policies. We note that, for \(\theta \in
    [0, 1]\),
    \[
        \theta \cdot \B E[m(X) \cdot d_0(X)] + [1 - \theta] \cdot \B E[m(X)
        \cdot d_1(X)] = \B E[m(X) \cdot (\theta \cdot d_0(X) + [1 - \theta]
        \cdot d_1(X))],
    \]
    and similarly for \(\B E[\B 1_{\alpha(X) = a_1} \cdot d(X)]\). Likewise,
    \[
        \theta \cdot \B E[d_0(X)] + [1 - \theta] \cdot \B E[d_1(X)] = \B
        E[\theta \cdot d_0(X) + (1 - \theta) \cdot d_1(X)],
    \]
    and so convex combinations of feasible policies are feasible. It follows
    that \(S\) is convex.

    Now, consider a point \((x_0, y_0)\) at which \(v(x,y)\) is maximized on
    \(S\). Since \(v\) is monotonically increasing in both coordinates, we must
    have that
    \[
        S \cap \left( (x_0, y_0) + \B R^2_{\geq 0} \right) = \{(x_0, y_0)\},
    \]
    and so, by the separating hyperplane theorem, there exists \((h_0, h_1) \in
    \B R^2\) and \(t \in \B R\) such that
    \begin{alignat*}{2}
        (h_0, h_1)^\top (x_0, y_0)
            &= t, \\
        (h_0, h_1)^\top \bb c
            &> t,
            &&\quad \left( \bb c \in (x_0, y_0) + \B R^2_{\geq 0}, \bb c \neq
            (x_0, y_0) \right) \\
        (h_0, h_1)^\top \bb s
            &< t.
            &&\quad \left( \bb s \in S, \bb s \neq (x_0, y_0) \right)
    \end{alignat*}
    We note that it follows that both \(h_0\) and \(h_1\) are positive, since,
    otherwise, without loss of generality, if \(h_0 = 0\), then
    \[
        (h_0, h_1)^\top (x_0 + \epsilon, y_0) = h_1 \cdot y_0 = (h_0, h_1)^\top
        (x_0, y_0) = 0,
    \]
    contrary to assumption, since
    \[
        (x_0 + \epsilon, y_0) \in (x_0, y_0) + \B R^2_{\geq 0}, (x_0 + \epsilon,
        y_0) \neq (x_0, y_0).
    \]
    Let \(\lambda_* = h_1 / h_0\), and consider the collection of utilities
    \[
        \C U = \{m(x) + \lambda \cdot \B 1_{\alpha(x) = a_1}\}_{\lambda > 0}.
    \]
    Since \(m(x) \geq 0\), \(\C U\) is consistent modulo \(\alpha\).

    We need the further claim that any policy that is utility maximizing for
    some \(u(x) \in \C U\) is Pareto efficient. For, suppose that \(d_0(x)\)
    were utility-maximizing for \(u_0(x)\) but not Pareto efficient, i.e., there
    existed \(d_1(x)\) and \(u_1(x)\) such that \(u_0(d_1) = u_0(d_0)\) but
    \(u_1(d_1) > u_1(d_0)\).
    Then, we would have for any \(u \in \C U\) that
    \begin{align*}
        u(d_1) - u(d_0)
          &= u(d_1) - u(d_0) - (u_0(d_1) - u_0(d_0)) \\
          &= (\lambda - \lambda_0) \cdot \left( \B E[d_1(X) \cdot \B
          1_{\alpha(X) = a_1}] - \B E[d_0(X) \cdot \B 1_{\alpha(X) = a_1}]
        \right).
    \end{align*}
    First suppose that \(\lambda_1 > \lambda_0\). Then, it follows from the fact
    that \(u_1(d_1) > u_1(d_0)\) that
    \[
        \B E[d_1(X) \cdot \B 1_{\alpha(X) = a_1}] > \B E[d_0(X) \cdot \B
        1_{\alpha(X) = a_1}].
    \]
    Now, if \(0 < \lambda_2 < \lambda_1\) and \(u(x) = m(x) + \lambda_2 \cdot \B
    1_{\alpha(x) = a_1}\), then we have that \(u_2(d_1) < u_2(d_0)\). In the
    same way, if \(\lambda_1 < \lambda_0\), we could choose \(\lambda_2 >
    \lambda_1\) such that \(u_2(d_1) < u_2(d_0)\). Therefore \(d_1\) does not
    Pareto dominate \(d_0\), contrary to hypothesis. Therefore, any policy that
    is utility maximizing for some \(u(x) \in \C U\) is Pareto efficient.

    In particular, it follows that the policy that maximizes \(u(x) = m(x) +
    \lambda_* \cdot \B 1_{\alpha(x) = a_1}\) in expectation is Pareto efficient.
    By the construction of the separating hyperplane, this is also the policy
    that maximizes \(u^*\), and so the policy that maximizes \(u^*\) is Pareto
    efficient. Therefore, under the hypotheses of Theorem~\ref{thm:dist}, for
    almost every joint distribution, the utility maximizing policy does not
    satisfy counterfactual equalized odds, conditional principal fairness, or
    path-specific fairness.
\end{proof}

\subsection{General Measures on \texorpdfstring{\(\C K\)}{K}}
\label{sec:counterexample}

Theorem~\ref{thm:dist} is restricted to \(\C U\)-fine and \(\C U^{\C A}\)-fine
distributions on the state space. The reason for this restriction is that when
the distribution of \(X\) induces atoms on the utility scale, threshold policies
can possess additional---or even infinite---degrees of freedom
when the threshold falls exactly on an atom. In particular circumstances, these
degrees of freedom can be used to ensure causal fairness notions, such as
counterfactual equalized odds, hold in a locally robust way. In particular, the
generalization of Theorem~\ref{thm:dist} beyond \(\C U\)-fine measures to all
totally bounded measures on the state space is false, as illustrated by the
following proposition.

\begin{proposition}
\label{prop:counterexample}
  Consider the set \(\bb E' \subseteq \B K\) of distributions---not necessarily
  \(\C U\)-fine---on \(\C K = \C X \times \C Y\) over which there exists a
  Pareto efficient policy satisfying counterfactual equalized odds. There exist
  \(b\), \(\C X\), \(\C Y\), and \(\C U\) such that \(\bb E'\) is \emph{not}
  relatively shy.
\end{proposition}

\begin{proof}
  We adopt the notational conventions of Section~\ref{sec:shyness-prelims}. We
  note that by Prop.~\ref{prop:shy_axioms_rel}, a set can only be shy if it has
  empty interior. Therefore, we will construct an example in which an open ball
  of distributions on \(\C K\) in the total variation norm all allow for a
  Pareto efficient policy satisfying counterfactual equalized odds, i.e., are
  contained in \(\bb E'\).

  Let \(b = \tfrac 3 4\), \(\C Y = \{0, 1\}\), \(\C A = \{a_0, a_1\}\), and \(\C
  X = \{0, 1\} \times \{a_0, a_1\} \times \B R\). Let \(\alpha : \C X \to \C A\)
  be given by \(\alpha : (y, a, v) \mapsto a\) for arbitrary \((y, a, v) \in \C
  X\). Likewise, let \(u : \C X \to \B R\) be given by \(u : (y, a, v) \mapsto
  v\). Then, if \(\C U = \{u\}\), \(\C U\) is vacuously consistent modulo
  \(\alpha\). Consider the joint distribution \(\mu\) on \(\C K = \C X \times \C
  Y\) where for all \(y, y' \in \C Y\), \(a \in \C A\), and \(u \in \B R\),
  \begin{equation*}
    \Pr_\mu(X = (a, y, u), Y(1) = y') = \frac 1 4 \cdot \B 1_{y = y'} \cdot
    \Pr_\mu(u(X) = u),
  \end{equation*}
  where, over \(\mu\), \(u(X)\) is distributed as a \(\tfrac 1 2\)-mixture of
  \(\unif(1, 2)\) and \(\delta(1)\); that is, \(\Pr(u(X) = 1) = \tfrac 1 2\) and
  \(\Pr(a < u(X) < b) = \tfrac {b - a} 2\) for \(0 \leq a \leq b < 1\).

  We first observe that there exists a Pareto efficient threshold policy
  \(\tau(x)\) such that counterfactual equalized odds is satisfied with respect
  to the decision policy \(\tau(X)\). Namely, let
  \begin{equation*}
    \tau(a, y, u) =
      \begin{cases}
        1           & u > 1, \\
        \frac 1 2   & u = 1, \\
        0           & u < 1.
      \end{cases}
  \end{equation*}
  Then, it immediately follows that \(\EE[\tau(X)] = \tfrac 3 4 = b\). Since
  \(\tau(x)\) is a threshold policy and exhausts the budget, it is utility
  maximizing by Lemma~\ref{lem:maximize}. Moreover, if \(D = \B 1_{U_D \leq
  \tau(X)}\) for some \(U_D \sim \unif(0, 1)\) independent of \(X\) and
  \(Y(1)\), then \(D \indep A \mid Y(1)\). Since \(u(X) \indep A, Y(1)\), it
  follows that
  \begin{align*}
    \Pr_\mu(D = 1 &\mid A = a, Y(1) = y) \\
      &\hspace{0.5cm}= \Pr(U_D \leq \tau(X) \mid A = a, Y(1) = y) \\
      &\hspace{0.5cm}= \Pr(U_D \leq \tau(X)) \\
      &\hspace{0.5cm}= \EE_\mu[\tau(X)],
  \end{align*}
  Therefore Eq.~\eqref{eq:counterfactual_equalized_odds} is satisfied, i.e.,
  counterfactual equalized odds holds. Now, using \(\mu\), we construct an open
  ball of distributions over which we can construct similar threshold policies.
  In particular, suppose \(\mu'\) is any distribution such that \(|\mu -
  \mu'|[\C K] < \tfrac 1 {64}\). Then, we claim that there exists a
  budget-exhausting threshold policy satisfying counterfactual equalized odds
  over \(\mu'\). For, we note that
  \begin{align*}
    \Pr_{\mu'}(U > 1) < \Pr_\mu(U > 1) + \frac 1 {64} = \frac {33} {64}, \\
    \Pr_{\mu'}(U \geq 1) > \Pr_\mu(U \geq 1) - \frac 1 {64} = \frac {63} {64},
  \end{align*}
  and so any threshold policy \(\tau'(x)\) satisfying \(\EE[\tau'(X)] = b =
  \tfrac 3 4\) must have \(t = 1\) as its threshold.

  We will now construct a threshold policy \(\tau'(x)\) satisfying
  counterfactual equalized odds over \(\mu'\). Consider a threshold policy of
  the form
  \begin{equation*}
    \tau'(a, y, u) =
      \begin{cases}
        1       & u > 1, \\
        p_{a,y} & u = 1, \\
        0       & u < 1.
      \end{cases}
  \end{equation*}

  For notational simplicity, let
  \begin{align*}
    q_{a,y}
      &= \Pr_{\mu'}(A = a, Y = y, U > 1), \\
    r_{a,y}
      &= \Pr_{\mu'}(A = a, Y = y, U = 1), \\
    \pi_{a,y}
      &= \Pr_{\mu'}(A = a, Y = y).
  \end{align*}
  Then, we have that
  \begin{align*}
    \EE_{\mu'}[\tau'(X)]
      &= \sum_{a,y} q_{a,y} + p_{a,y} \cdot r_{a,y}, \\
    \EE_{\mu'}[\tau'(X) \mid A = a, Y = y]
      &= \frac {q_{a,y} + p_{a,y} \cdot r_{a,y}} {\pi_{a,y}}.
  \end{align*}
  Therefore, the policy will be budget exhausting if
  \begin{equation*}
    \sum_{a,y} q_{a,y} + p_{a,y} \cdot r_{a,y} = \tfrac 3 4,
  \end{equation*}
  and it will satisfy counterfactual equalized odds if
  \begin{align}
  \label{eq:system}
  \begin{split}
    \pi_{a_1, 0} \cdot (q_{a_0, 0} &+ p_{a_0, 0} \cdot r_{a_0, 0}) \\
      &= \pi_{a_0, 0} \cdot (q_{a_1, 0} + p_{a_1, 0} \cdot r_{a_1, 0}), \\
    \pi_{a_1, 1} \cdot (q_{a_0, 1} &+ p_{a_0, 1} \cdot r_{a_0, 1}) \\
      &= \pi_{a_0, 1} \cdot (q_{a_1, 1} + p_{a_1, 1} \cdot r_{a_1, 1}),
  \end{split}
  \end{align}
  since, as above,
  \begin{equation*}
    \Pr(D = 1 \mid A = a, Y(1) = y) = \EE[\tau'(X) \mid A = a, Y(1) = y].
  \end{equation*}

  Again, for notational simplicity, let
  \begin{equation*}
    S = \frac {\tfrac 3 4 - \Pr_{\mu'}(U > 1)} {\Pr_{\mu'}(U = 1)}.
  \end{equation*}
  Then, a straightforward algebraic manipulation shows that
  Eq.~\eqref{eq:system} is solved by setting \(p_{a_0, y}\) to be
  \begin{equation*}
    \frac {S \cdot \pi_{a_0, y} \cdot (r_{a_0, y} + r_{a_1, y}) + \pi_{a_0, y}
    \cdot q_{a_1, y} - \pi_{a_1, y} \cdot q_{a_0, y}} {r_{a_0, y} \cdot
    (\pi_{a_0, y} + \pi_{a_1, y})},
  \end{equation*}
  and \(p_{a_1, y}\) to be
  \begin{equation*}
    \frac {S \cdot \pi_{a_1, y} \cdot (r_{a_0, y} + r_{a_1, y}) + \pi_{a_1, y}
    \cdot q_{a_0, y} - \pi_{a_0, y} \cdot q_{a_1, y}} {r_{a_1, y} \cdot
    (\pi_{a_0, y} + \pi_{a_1, y})}.
  \end{equation*}
  In order for \(\tau'(x)\) to be a well-defined policy, we need to show that
  \(p_{a,y} \in [0,1]\) for all \(a \in \C A\) and \(y \in \C Y\). To that end,
  note that
  \begin{align*}
    q_{a,y}
      &= \Pr_{\mu'}(A = a, Y = y, U > 1), \\
    r_{a,y}
      &= \Pr_{\mu'}(A = a, Y = y, U = 1), \\
    \pi_{a,y}
      &= \Pr_{\mu'}(A = a, Y = y), \\
    r_{a_0, y} + r_{a_1, y}
      &= \Pr_{\mu'}(Y = y, U = 1), \\
    \pi_{a_0,y} + \pi_{a_1,y}
      &= \Pr_{\mu'}(Y = y), \\
    S
      &= \frac {\frac 3 4 - \Pr_{\mu'}(U > 1)} {\Pr_{\mu'}(U = 1)}. \\
  \end{align*}
  Now, we recall that \(|\Pr_{\mu'}(E) - \Pr_\mu(E)| < \frac 1 {64}\) for any
  event \(E\) by hypothesis. Therefore,
  \begin{alignat*}{3}
    \frac {7} {64} \leq\ 
      & \makebox[\myl]{\(q_{a,y}\)}
      & \ \leq \frac {9} {64}, \\
    \frac 7 {64} \leq\ 
      & \makebox[\myl]{\(r_{a,y}\)}
      & \ \leq \frac 9 {64}, \\
    \frac 7 {64} \leq\ 
      & \makebox[\myl]{\(\pi_{a,y}\)}
      & \ \leq \frac {17} {64}, \\
    \frac {15} {64} \leq\ 
      & \makebox[\myl]{\(r_{a_0, y} + r_{a_1, y}\)}
      & \ \leq \frac {17} {64}, \\
    \frac {31} {64} \leq\ 
      & \pi_{a_0,y} + \pi_{a_1,y}
      & \ \leq \frac {33} {64}, \\
    \frac {15} {31} \leq\ 
      & \makebox[\myl]{\(S\)}
      & \ \leq \frac {17} {33}.
   \end{alignat*}
  Using these bounds and the expressions for \(p_{a,y}\) derived above, we see
  that
  \begin{align*}
    \frac {629} {3069} < p_{a,y} < \frac {6497} {7161},
  \end{align*}
  and hence \(p_{a,y} \in [0, 1]\) for all \(a \in \C A\) and \(y \in \C Y\).

  Therefore, the policy \(\tau'(x)\) is well-defined, and, by construction, is
  budget-exhausting and therefore utility-maximizing by
  Lemma~\ref{lem:maximize}. It also satisfies counterfactual equalized odds by
  construction. Since \(\mu'\) was arbitrary, it follows that the set of
  distributions on \(\C K\) such that there exists a Pareto efficient policy
  satisfying counterfactual equalized odds contains an open ball, and hence is
  not shy.
\end{proof}

\section{Theorem~\ref{thm:path_specific} and Related Results}

We first prove a variant of Theorem~\ref{thm:path_specific} for general,
continuous covariates \(\C X\). Then, we extend and generalize
Theorem~\ref{thm:path_specific} using the theory of finite Markov chains,
offering a proof of the theorem different from the sketch included in the main
text.

\subsection{Extension to Continuous Covariates}

Here we follow the proof sketch in the main text for
Theorem~\ref{thm:path_specific}, which assumes a finite covariate-space \(\C
X\). In that case, we start with a point \(x^*\) with maximum decision
probability \(d(x^*)\), and then assume, toward a contradiction, that there
exists a point with strictly lower decision probability. The general case is
more involved since it is not immediately clear that the maximum value of
\(d(x)\) is achieved with positive probability in \(\C X\). We start with the
lemma below before proving the main result.

\begin{lemma}
\label{lem:d_pi}
  A decision policy \(d(x)\) satisfies path-specific fairness with \(W = X\) if
  and only if any \(a' \in \C A\),
  \begin{equation*}
    \EE[d(X_{\Pi, A, a'}) \mid X] = d(X).
  \end{equation*}
\end{lemma}

\begin{proof}
  First, suppose that \(d(x)\) satisfies path-specific fairness. To show the
  result, we use the standard fact that for independent random variables \(X\)
  and \(U\),
  \begin{equation}
  \label{eq:efxu}
    \EE[f(X,U) \mid X] = \int f(X, u) \diff F_U(u),
  \end{equation}
  where \(F_U\) is the distribution of \(U\). \citep[For a proof of this fact
  see, for example,][]{drhab2019conditional}

  Now, we have that
  \begin{align*}
    \EE[D_{\Pi, A, a'} \mid X_{\Pi, A, a'}]
    &= \EE[\mathbb{1}_{U_D \leq d(X_{\Pi, A, a'})} \mid X_{\Pi, A, a'}] \\
    &= \int_0^1 \mathbb{1}_{u \leq d(X_{\Pi, A, a'})} \diff u \\
    &= d(X_{\Pi, A, a'}),
  \end{align*}
  where the first equality follows from the definition of \(D_{\Pi, A, a'}\),
  and the second from Eq.~\eqref{eq:efxu}, since the exogenous variable \(U_D
  \sim \unif(0,1)\) is independent of the counterfactual covariates \(X_{\Pi, A,
  a'}\). An analogous argument shows that \(\EE[D \mid X] = d(X)\).

  Finally, conditioning on \(X\), we have
  \begin{align*}
    \EE[d(X_{\Pi, A, a'}) \mid X]
      &= \EE[\EE[D_{\Pi, A, a'} \mid X_{\Pi, A, a'}] \mid X] \\
      &= \EE[\EE[D_{\Pi, A, a'} \mid X_{\Pi, A, a'}, X] \mid X] \\
      &= \EE[D_{\Pi, A, a'} \mid X] \\
      &= \EE[D \mid X] \\
      &= d(X),
  \end{align*}
  where the second equality follows from the fact that \(D_{\Pi, A, a'} \indep X
  \mid X_{\Pi, A, a'}\), the third from the law of iterated expectations, and
  the fourth from the definition of path-specific fairness.

  Next, suppose that
  \begin{equation*}
    \EE[d(X_{\Pi, A, a'} \mid X] = d(X)
  \end{equation*}
  for all \(a' \in \C A\). Then, since \(W = X\) and \(X \indep U_D\), using
  Eq.~\eqref{eq:efxu}, we have that for all \(a' \in \C A\),
  \begin{align*}
    \EE[D_{\Pi, A, a'} \mid X]
      &= \EE[\EE[\B 1_{U_D \leq d(X_{\Pi, A, a'})} \mid X_{\Pi, A, a'}, X] \mid
        X] \\
      &= \EE[\EE[d(X_{\Pi, A, a'}) \mid X_{\Pi, A, a'}, X] \mid X] \\
      &= \EE[d(X_{\Pi, A, a'}) \mid X] \\
      &= d(X) \\
      &= \EE[d(X) \mid X] \\
      &= \EE[D \mid X].
  \end{align*}
  This is exactly Eq.~\eqref{eq:path_specific_fairness}, and so the result
  follows.
\end{proof}

We are now ready to prove a continuous variant of
Theorem~\ref{thm:path_specific}. The technical hypotheses of the theorem ensure
that the conditional probability measures \(\Pr(E \mid X)\) are ``sufficiently''
mutually non-singular distributions on \(\C X\) with respect to the distribution
of \(X\)---for example, the conditions ensure that the conditional distribution
of \(X_{\Pi, A, a} \mid X\) does not have atoms that \(X\) itself does not have,
and \emph{vice versa}. For notational and conceptual simplicity, we only
consider the case of trivial \(\zeta\), i.e., where \(\zeta(x) = \zeta(x')\) for
all \(x, x' \in \C X\).

\begin{proposition}
\label{prop:path_specific}
  Suppose that
  \begin{enumerate}
    \item For all \(a \in \C A\) and any event \(S\) satisfying \(\Pr(X \in S
      \mid A = a) > 0\), we have, a.s.,
      \begin{equation*}
        \Pr(X_{\Pi, A, a} \in S \lor A = a \mid X) > 0.
      \end{equation*}
    \item For all \(a \in \C A\) and \(\epsilon > 0\), there exists \(\delta >
      0\) such that for any event \(S\) satisfying \(\Pr(X \in S \mid A = a) <
      \delta\), we have, a.s.,
      \begin{equation*}
        \Pr(X_{\Pi, A, a} \in S, A \neq a \mid X) < \epsilon.
      \end{equation*}
  \end{enumerate}
  Then, for \(W = X\), any \(\Pi\)-fair policy \(d(x)\) is constant a.s. (i.e.,
  \(d(X) = p\) a.s.\ for some \(0 \leq p \leq 1\)).
\end{proposition}

\begin{proof}
  Let \(d_{\max} = \|d(x)\|_\infty\), the essential supremum of \(d\). To
  establish the theorem statement, we show that \(\Pr(d(X) = d_{\max} \mid A
  = a) = 1\) for all \(a \in \C A\). To do that, we begin by showing that there
  exists some \(a \in \C A\) such that \(\Pr(d(X) = d_{\max} \mid A = a) > 0\).

  Assume, toward a contradiction, that for all \(a \in \C A\),
  \begin{equation}
  \label{eq:contradiction}
    \Pr(d(X) = d_{\max} \mid A = a) = 0.
  \end{equation}
  Because \(\C A\) is finite, there must be some \(a_0 \in \C A\) such that
  \begin{equation}
  \label{eq:dmax}
    \Pr(d_{\max} - d(X) < \epsilon \mid A = a_0) > 0
  \end{equation}
  for all \(\epsilon > 0\).

  Choose \(a_1 \neq a_0\). We show that for values of \(x\) such that \(d(x)\)
  is close to \(d_{\max}\), the distribution of \(d(X_{\Pi, A, a_1}) \mid X =
  x\) must be concentrated near \(d_{\max}\) with high probability to satisfy
  the definition of path-specific fairness, in
  Eq.~\eqref{eq:path_specific_fairness}. But, under the assumption in
  Eq.~\eqref{eq:contradiction}, we also show that the concentration occurs with
  low probability, by the continuity hypothesis in the statement of the theorem,
  establishing the contradiction.

  Specifically, by Markov's inequality, for any \(\rho > 0\), a.s.,
  \begin{align*}
    \Pr(d_{\max} - d(X_{\Pi, A, a_1}) \geq \rho \mid X)
      &\leq \frac {\EE[d_{\max} - d(X_{\Pi, A, a_1}) \mid X]} {\rho} \\
      &= \frac {d_{\max} - d(X)} {\rho},
  \end{align*}
  where the final equality follows from Lemma~\ref{lem:d_pi}. Rearranging, it
  follows that for any \(\rho > 0\), a.s.,
  \begin{equation}
  \label{eq:ineq1}
    \Pr(d_{\max} - d(X_{\Pi, A, a_1}) < \rho \mid X) \geq 1 - \frac {d_{\max} -
    d(X)} {\rho}.
  \end{equation}

  Now let \(S = \{x \in \C X : d_{\max} - d(x) < \rho\}\). By the second
  hypothesis of the theorem, we can choose \(\delta\) sufficiently small that if
  \begin{equation*}
    \Pr(X \in S \mid A = a_1) < \delta
  \end{equation*}
  then, a.s.,
  \begin{equation*}
    \Pr(X_{\Pi, A, a_1} \in S, A \neq a_1 \mid X) < \tfrac 1 2.
  \end{equation*}
  In other words, we can chose \(\delta\) such that if
  \begin{equation*}
    \Pr(d_{\max} - d(X) < \rho \mid A = a_1) < \delta
  \end{equation*}
  then, a.s.,
  \begin{equation*}
    \Pr(d_{\max} - d(X_{\Pi, A, a_1}) < \rho, A \neq a_1 \mid X) < \tfrac 1 2
  \end{equation*}

  By Eq.~\eqref{eq:contradiction}, we can choose \(\epsilon > 0\) so small that
  \begin{equation*}
    \Pr(d_{\max} - d(X) < \epsilon \mid A = a_1) < \delta.
  \end{equation*}
  Then, we have that
  \begin{equation*}
    \Pr(d_{\max} - d(X_{\Pi, A, a_1}) < \epsilon, A \neq a_1 \mid X) < \tfrac 1
    2
  \end{equation*}
  a.s. Further, by the definition of the essential supremum and \(a_0\), and the
  fact that \(a_0 \neq a_1\), we have that
  \begin{equation*}
    \Pr(d_{\max} - d(X) < \tfrac \epsilon 2, A \neq a_1) > 0.
  \end{equation*}
  Therefore, with positive probability, we have that
  \begin{align*}
  \label{eq:ineq2}
    1 - \frac {d_{\max} - d(X)} {\epsilon}
      &> 1 - \frac { \frac \epsilon 2} \epsilon \\
      &= \frac 1 2 \\
      &> \Pr(d_{\max} - d(X_{\Pi, A, a_1}) < \epsilon, A \neq a_1
        \mid X).
  \end{align*}
  This contradicts Eq.~\eqref{eq:ineq1}, and so it cannot be the case that
  \(\Pr(d(X) = d_{\max} \mid A = a_0) = 0\), meaning \(\Pr(d(X) = d_{\max} \mid
  A = a_0) > 0\).

  Now, we show that \(\Pr(d(X) = d_{\max} \mid A = a_1) = 1\). Suppose, toward a
  contradiction, that
  \begin{equation*}
    \Pr(d(X) < d_{\max} \mid A = a_1) > 0.
  \end{equation*}
  Then, by the first hypothesis, a.s.,
  \begin{equation*}
    \Pr(d(X_{\Pi, A, a_1}) < d_{\max} \lor A = a_1 \mid X) > 0
  \end{equation*}
  As a consequence,
  \begin{align*}
    d_{\max}
      &= \EE[d(X) \mid d(X) = d_{\max}, A = a_0] \\
      &= \EE[ \EE[d(X_{\Pi, A, a_1}) \mid X] \mid d(X) = d_{\max}, A = a_0] \\
      &< \EE[ \EE[d_{\max} \mid X] \mid d(X) = d_{\max}, A = a_0] \\
      &= \EE[d_{\max} \mid d(X) = d_{\max}, A = a_0] \\
      &= d_{\max},
  \end{align*}
  where we can condition on the set \(\{d(X) = d_{\max}, A = a_0\}\) since
  \(\Pr(d(X) = d_{\max} \mid A = a_0) > 0\); and the second equality above
  follows from Lemma~\ref{lem:d_pi}. This establishes the contradiction, and so
  \(\Pr(d(X) = d_{\max} \mid A = a_1) = 1\).

  Finally, we extend this equality to all \(a \in \C A\). Since, \(\Pr(d(X) \neq
  d_{\max} \mid A = a_1) = 0\), we have, by the second hypothesis of the
  theorem, that, a.s.,
  \begin{equation*}
    \Pr(d(X_{\Pi, A, a_1}) \neq d_{\max}, A \neq a_1 \mid X) = 0.
  \end{equation*}
  Since, by definition, \(\Pr(X_{\Pi, A, a_1} = X \mid A = a_1) = 1\), and
  \(\Pr(d(X) = d_{\max} \mid A = a_1) = 1\), we can strengthen this to
  \begin{equation*}
    \Pr(d(X_{\Pi, A, a_1}) \neq d_{\max} \mid X) = 0.
  \end{equation*}

  Consequently, a.s.,
  \begin{align*}
    d(X) &= \EE[d(X_{\Pi, A, a}) \mid X]\\
    & = \EE[d_{\max} \mid X]\\
    & = d_{\max},
  \end{align*}
  where the first equality follows from Lemma~\ref{lem:d_pi}, establishing the
  result.
\end{proof}

\subsection{A Markov Chain Perspective}
\label{sec:markov}

The theory of Markov chains illuminates---and allows us to extend---the proof of
Theorem~\ref{thm:path_specific}. Suppose \(\C X = \{x_1, \ldots,
x_n\}\).\footnote{%
  Because of the technical difficulties associated with characterizing the
  long-run behavior of arbitrary infinite Markov chains, we restrict our
  attention in this section to Markov chains with finite state spaces.
} For any \(a' \in \C A\), let \(P_{a'} = [p^{a'}_{i,j}]\) where \(p^{a'}_{i,j}
= \Pr(X_{\Pi, A, a'} = x_j \mid X = x_i)\). Then \(P_{a'}\) is a stochastic
matrix.

To motivate the subsequent discussion, we first note that this perspective
conceptually simplifies some of our earlier results. Lemma~\ref{lem:d_pi} can be
recast as stating that when \(W = X\), a policy \(d\) is \(\Pi\)-fair if and
only if \(P_{a'} d = d\)---i.e., if and only if \(d\) is a 1-eigenvector of
\(P_{a'}\)---for all \(a' \in \C A\).

The 1-eigenvectors of Markov chains have a particularly simple structure, which
we derive here for completeness.

\begin{lemma}
\label{lem:markov}
  Let \(S_1, \ldots, S_m\) denote the recurrent classes of a finite Markov chain
  with transition matrix \(P\). If \(d\) is a 1-eigenvector of \(P\), then \(d\)
  takes a constant value \(p_k\) on each \(S_k\), \(k = 1, \ldots, m\), and
  \begin{equation}
  \label{eq:pi_fair_character}
    d_i = \sum_{k=1}^m \left[ \lim_{n \to \infty} \sum_{j \in S_k} P_{ij}^n
    \right] \cdot p_k.
  \end{equation}
\end{lemma}

\begin{remark}
  We note that \(\lim_{n \to \infty} \sum_{j \in S_k} P^n_{i,j}\) always exists
  and is the probability that the Markov chain, beginning at state \(i\), is
  eventually absorbed by the recurrent class \(S_k\).
\end{remark}

\begin{proof}
  Note that, possibly by reordering the states, we can arrange that the
  stochastic matrix \(P\) is in canonical form, i.e., that
  \begin{equation*}
    P =
      \begin{bmatrix}
        B   & \\
        R'  & Q
      \end{bmatrix},
  \end{equation*}
  where \(Q\) is a sub-stochastic matrix, \(R\) is non-negative, and
  \begin{equation*}
    B =
      \begin{bmatrix}
        P_1 &     &       & \\
          & P_2   &       & \\
          &     & \ddots  & \\
          &     &       & P_m
      \end{bmatrix}
  \end{equation*}
  is a block-diagonal matrix with the stochastic matrix \(P_i\) corresponding to
  the transition probabilities on the recurrent set \(S_i\) in the \(i\)-th
  position along the diagonal.

  Now, consider a \(1\)-eigenvector \(v = [v_1\ v_2]^\top\) of \(P\). We must
  have that \(P v = v\), i.e., \(B v_1 = v_1\) and \(R' v_1 + Q v_2 = v_2\).
  Therefore \(v_1\) is a 1-eigenvector of \(B\). Since \(B\) is block diagonal,
  and each diagonal element is a positive stochastic matrix, it follows by the
  Perron-Frobenius theorem that the 1-eigenvectors of \(B\) are given by
  \(\Span(\bb 1_{S_i})_{i=1, \ldots, m}\), where \(\bb 1_{S_i}\) is the vector
  which is 1 at index \(j\) if \(j \in S_i\) and is 0 otherwise.

  Now, for \(v_1 \in \Span(\bb 1_{S_i})_{i=1, \ldots, m}\), we must find \(v_2\)
  such that \(R' v_1 + Q v_2 = v_2\).

  Note that every finite Markov chain \(M\) can be canonically associated with
  an absorbing Markov chain \(M^{\abs}\) where the set of states of \(M^{\abs}\)
  is exactly the union of the transitive states of \(M\) and the recurrent sets
  of \(M\). (In essence, one tracks which state of \(M\) the Markov chain is in
  until it is absorbed by one of the recurrent sets, at which point the entire
  recurrent set is treated as a single absorbent state.) The transition matrix
  \(P^{\abs}\) associated with \(M^{\abs}\) is given by
  \begin{equation*}
    P^{\abs} =
      \begin{bmatrix}
        I & \\
        R & Q
      \end{bmatrix}
  \end{equation*}
  where \(R = R' [\bb 1_{S_1}\ \ldots\ \bb 1_{S_m}]\). In particular, it follows
  that \(v = [v_1\ v_2]^\top\) is a 1-eigenvector of \(P\) if and only if \([T
  v_1\ v_2]^\top\) is a 1-eigenvector of \(P^{\abs}\), where \(T : \bb 1_{S_i}
  \mapsto \bb e_i\).

  Now, if \(v\) is a 1-eigenvector of \(P^{\abs}\), then it is a 1-eigenvector
  of \((P^{\abs})^k\) for all \(k\). Since \(Q\) is sub-stochastic, the series
  \(\sum_{k=0}^\infty Q^k\) converges to \((I - Q)^{-1}\). Since
  \begin{equation*}
    (P^{\abs})^k =
      \begin{bmatrix}
        I                 & \\
        (I + Q + \cdots + Q^{k-1}) R  & Q^k
      \end{bmatrix},
  \end{equation*}
  it follows that
  \begin{equation*}
    \lim_{k \to \infty} (P^{\abs})^k =
      \begin{bmatrix}
        I         & \\
        (I - Q)^{-1} R   & 0
      \end{bmatrix}.
  \end{equation*}
  Therefore, if \(v = [v_1\ v_2]^\top\) is a 1-eigenvector of \(P^{\abs}\), we
  must have that \((I - Q)^{-1} R v_1 = v_2\). By Theorem~3.3.7 in
  \citet{kemeny1976finite}, the \((i,k)\) entry of \((I - Q)^{-1} R\) is exactly
  the probability that, conditional on \(X_0 = x_i\), the Markov chain is
  eventually absorbed by the recurrent set \(S_k\). This is, in turn, by the
  Chapman-Kolmogorov equations and the definition of \(S_k\), equal to \(\lim_{n
  \to \infty} \sum_{j \in S_k} P^n_{i, j}\), and therefore the result follows.
\end{proof}

We arrive at the following simple necessary condition on \(\Pi\)-fair policies.

\begin{corollary}
\label{cor:markov}
  Suppose \(\C X\) is finite, and define the stochastic matrix \(P = \tfrac 1
  {|\C A|} \sum_{a' \in \C A} P_{a'}\). If \(d(x)\) is a \(\Pi\)-fair policy
  then it is constant on the recurrent classes of \(P\).
\end{corollary}

\begin{proof}
  By Lemma~\ref{lem:d_pi}, \(d\) is \(\Pi\)-fair if and only if \(P_{a'} d = d\)
  for all \(a' \in \C A\). Therefore,
  \begin{equation}
    \frac 1 {|\C A|} \sum_{a' \in \C A} P_{a'} d = \frac 1 {|\C A|} \sum_{a' \in
    \C A} d = d,
  \end{equation}
  and so \(d\) is a 1-eigenvector of \(P\). Therefore it is constant on the
  recurrent classes of \(P\) by Lemma~\ref{lem:markov}.
\end{proof}

We note that Theorem~\ref{thm:path_specific} follows immediately from this.

\begin{proof}[Proof of Theorem~\ref{thm:path_specific}]
  Note that \(\tfrac 1 {|\C A|} \sum_{a \in \C A} P_a\) decomposes into a block
  diagonal stochastic matrix, where each block corresponds to a single stratum
  of \(\zeta\) and is irreducible. Consequently, each stratum forms a recurrent
  class, and the result follows.
\end{proof}

\section{Proofs of Propositions~\ref{prop:main_cee} and~\ref{prop:pred_parity}}

The proofs of Proposition~\ref{prop:main_cee} and
Proposition~\ref{prop:pred_parity} rely on certain shared theory about beta
distributions. We begin by reviewing this theory before moving onto the proofs
of the respective propositions.

\subsection{Beta distributions and stochastic dominance}
\label{app:beta_theory}

We begin by introducing incomplete beta functions and distributions.

\begin{definition}
  The \emph{incomplete beta function} \(I_t(\alpha, \beta)\) for \(t \in (0,1]\)
  is given by
  \begin{equation*}
    I_t(\alpha, \beta) = \int_0^t x^{\alpha - 1} (1 - x)^{\beta - 1}.
  \end{equation*}
  A random variable \(X\) is said to be distributed as \(\bbeta_t(\alpha,
  \beta)\) if \(X \sim Y \mid Y < t\) for \(Y \sim \bbeta(\alpha, \beta)\);
  equivalently, if \(X\) has PDF
  \begin{equation*}
    \B 1_{x \in (0, t)} \cdot \frac {x^{\alpha - 1} (1 - x)^{\beta - 1}}
    {I_t(\alpha, \beta)}.
  \end{equation*}
\end{definition}

An important property relating different beta distributions is stochastic
dominance.

\begin{definition}
  Let \(X\) and \(Y\) be random variables with CDFs \(F_X(t)\) and \(F_Y(t)\),
  respectively. We say that \(X\) \emph{stochastically dominates} \(Y\), written
  \(Y \leq_{\st} X\), if \(F_X(t) \leq F_Y(t)\) for all \(t \in \B R\). We say
  that \(X\) \emph{strictly stochastically dominates} \(Y\), i.e., \(Y <_{\st}
  X\), if \(F_X(t) = F_Y(t)\) implies that either \(F_X(t) = F_Y(t) = 0\) or
  \(F_X(t) = F_Y(t) = 1\).
\end{definition}

Stochastic dominance has the following useful property.

\begin{lemma}
\label{lem:mean}
  Suppose \(Y \leq_{\st} X\). Then, for any monotonically non-decreasing
  \(\varphi : \B R \to \B R\), we have that \(\B E[\varphi(Y)] \leq \B
  E[\varphi(X)]\).
\end{lemma}

For proof, see (1.A.7) in \cite{shaked2007stochastic}. We will need to improve
the inequality to a strict inequality. We begin with the simplest case.

\begin{lemma}
\label{lem:dv}
  Suppose \(Y <_{\st} X\), where \(X\) and \(Y\) are positive and \(F_X(t)\),
  \(F_Y(t)\) are continuous. Then, \(\B E[Y \cdot \B 1_{Y > t}] < \B E[X \cdot
  \B 1_{X > t}]\) for any \(t > 0\) such that there exists \(t' \geq
  t\) with \(F_X(t') < F_Y(t')\).
\end{lemma}

\begin{proof}
  Since since \(F_Y(x) \geq F_X(x)\) for all \(x\) and, in particular, \(F_Y(x)
  > F_X(x)\) on an open interval containing \(t'\), we have that
  \begin{align*}
    \B E[Y \cdot \B 1_{Y > t}]
        &= t \cdot [1 - F_Y(t)] + \int_t^\infty 1 - F_Y(x) \, \dx x \\
        &< t \cdot [1 - F_X(t)] + \int_t^\infty 1 - F_X(x) \, \dx x \\
        &= \B E[X \cdot \B 1_{X > t}]
  \end{align*}
  where we have applied the Darth Vader rule to calculate the
  expectations \citep{muldowney2012darth}.
\end{proof}

This leads to the following modest technical generalization.

\begin{lemma}
\label{lem:monotone}
  Suppose \(Y <_{\st} X\), where \(X\) and \(Y\) are positive and \(F_X(t)\) and
  \(F_Y(t)\) are continuous. Suppose that \(t > 0\) is such that there exists
  \(t' < t\) with \(F_X(t') < F_Y(t')\), and \(f : \B R_{>0} \to \B R_{>0}\) is
  monotonically decreasing and continuous. Then
  \[
    \B E[f(Y) \cdot \B 1_{Y < t}] > \B E[f(X) \cdot \B 1_{X < t}].
  \]
\end{lemma}

\begin{proof}
  Consider the transformed variables \(f(X)\) and \(f(Y)\). Since \(f\) is
  monotonically decreasing and continuous, it is invertible, and in particular
  \[
    \Pr(f(X) < x) = \Pr(X > f^{-1}(x)) = 1 - F_X(f^{-1}(x)).
  \]
  It follows that the CDFs of \(f(X)\) and \(f(Y)\) are \(1 - F_X(f^{-1}(x))\)
  and \(1 - F_Y(f^{-1}(x))\), respectively. Then, observe that since \(F_X(x)
  \leq F_Y(x)\) for all \(x \in \B R\),
  \[
    1 - F_X(x) \geq 1 - F_Y(x)
  \]
  for all \(x \in \B R\), i.e., \(f(X) \leq_{\st} f(Y)\). In particular, since
  \(F_X(x) = 0\) if and only if \(1 - F_X(x) = 1\) and \emph{vice versa}, it
  follows from the invertibility of \(f\) that \(f(X) <_{\st} f(Y)\). Therefore,
  after noting that \(X < t\) if and only if \(f(X) > f(t)\) and similarly for
  \(Y\), the result follows from Lemma~\ref{lem:dv}.
\end{proof}

It is relatively straightforward to characterize the stochastic dominance
relationships between various (incomplete) beta distributions according to
\(\alpha\) and \(\beta\); see \cite{arab2020convex} for full details. Here, we
require only the following result, which closely follows the proof of Theorem~1
there.

\begin{lemma}
\label{lem:st}
  If \(X \sim \bbeta_t(\alpha_0, \beta_0)\), \(Y \sim \bbeta_t(\alpha_1,
  \beta_1)\), where \(\alpha_0 \geq \alpha_1\) and \(\beta_0 \leq \beta_1\),
  then \(Y \leq_{\st} X\). If, in addition, either \(\alpha_0 > \alpha_1\) or
  \(\beta_0 < \beta_1\), then \(Y <_{\st} X\).
\end{lemma}

\begin{proof}
  Consider the CDFs \(F_X(s)\) and \(F_Y(s)\). We will use the difference \(G(s)
  = F_X(s) - F_Y(s)\) to demonstrate the result. The case where \(\alpha_0 =
  \alpha_1\) and \(\beta_0 = \beta_1\) is trivial, so we restrict our attention
  to the case where one of the inequalities is strict. For simplicity, we assume
  that \(\alpha_0 > \alpha_1\); the case where \(\beta_0 < \beta_1\) is
  virtually identical.

  In particular, observe that \(G(0) = G(t) = 0\), and that for \(s \in (0,
  t)\),
  \begin{align*}
    G'(s)
      &= \frac {s^{\alpha_0 - 1} (1 - s)^{\beta_0 - 1}} {I_t(\alpha_0,
      \beta_0)} - \frac {s^{\alpha_1 - 1} (1 - s)^{\beta_1 - 1}} {I_t(\alpha_1,
      \beta_1)} \\
      &= \frac {s^{\alpha_0 - 1} (1 - s)^{\beta_0 - 1}} {I_t(\alpha_1, \beta_1)}
      \cdot \left[ s^{\alpha_1 - \alpha_0} (1 - s)^{\beta_1 - \beta_0} - \frac
      {I_t(\alpha_1, \beta_1)} {I_t(\alpha_0, \beta_0)} \right].
  \end{align*}
  We consider the two multiplicands in the final expression. We note that the
  first is greater than zero for all \(s \in (0, t)\). Therefore, for \(s \in
  (0, t)\), \(G'(s) = 0\) if and only if
  \begin{equation*}
    s^{\alpha_1 - \alpha_0} (1 - s)^{\beta_1 - \beta_0} = \frac {I_t(\alpha_1,
    \beta_1)} {I_t(\alpha_0, \beta_0)}.
  \end{equation*}
  Now, since \(\alpha_0 > \alpha_1\) and \(\beta_0 \leq \beta_1\), it follows
  that the left-hand side of the previous expression is strictly decreasing. In
  particular, \(G'(s) = 0\) for at most one \(s \in (0, t)\). Since \(G(0) =
  G(t) = 0\) and \(G(t)\) is non-constant, it follows that \(G'(s) = 0\) for
  exactly one \(s \in (0, t)\) by Rolle's theorem. In particular, either \(G(s)
  > 0\) for all \(s \in (0,t)\) or \(G(s) < 0\) for all \(t \in (0,t)\).

  Consequently,
  \begin{equation*}
    s^{\alpha_1 - \alpha_0} (1 - s)^{\beta_1 - \beta_0} - \frac {I_t(\alpha_1,
    \beta_1)} {I_t(\alpha_0, \beta_0)}
  \end{equation*}
  changes sign for some \(s_0 \in (0, t)\). Since the minuend is strictly
  decreasing, it follows that \(G'(s) > 0\) for \(s \in (0, s_0)\). Therefore,
  in particular, \(G(s) > 0\) for all \(s \in (0, s_0)\), and hence for \(s \in
  (0, t)\). Therefore \(F_X(s) > F_Y(s)\) for \(s \in (0,t)\), and so \(Y
  <_{\st} X\).
\end{proof}

\subsection{Proof of Proposition~\ref{prop:main_cee}}
\label{app:main_cee}

Our proof is a relatively straightforward application of Sard's theorem.
Intuitively, counterfactual predictive parity imposes three constraints on
\(\alpha_0\), \(\beta_0\), \(\alpha_1\), \(\beta_1\), and the group-specific
thresholds \(t_0\) and \(t_1\). These constraints are sufficiently smooth that
the zero locus should take the form of a 3-manifold, by the inverse function
theorem. Projecting onto the first four coordinates (i.e., eliminating \(t_0\)
and \(t_1\)) gives rise to a set of measure zero by Sard's theorem.

\begin{proof}[Proof of Proposition~\ref{prop:main_cee}]
  We begin by noting that the risk distributions, conditional on \(Y(1) = i\),
  for \(i = 0, 1\), take a particularly simple form.
  \begin{align*}
    r(X) \mid A = a_i, Y(1) = 1
      &\sim \bbeta(\alpha_i + 1, \beta_i), \\
    r(X) \mid A = a_i, Y(1) = 0
      &\sim \bbeta(\alpha_i, \beta_i + 1).
  \end{align*}
  This follows upon noting that
  \begin{align*}
    \Pr(r(X) < t, Y(1) = 1 \mid A = a_i)
      &= \frac 1 {B(\alpha_i, \beta_i)} \int_0^1 x \cdot \B 1_{x < t} \cdot
      x^{\alpha_i - 1} (1 - x)^{\beta_i - 1} \, \dx x \\
      &= \frac {I_t(\alpha_i + 1, \beta_i)} {B(\alpha_i, \beta_i)}, \\
    \Pr(r(X) < t, Y(1) = 0 \mid A = a_i)
      &= \frac 1 {B(\alpha_i, \beta_i)} \int_0^1 (1 - x) \cdot \B 1_{x < t}
      \cdot x^{\alpha_i - 1} (1 - x)^{\beta_i - 1} \, \dx x \\
      &= \frac {I_t(\alpha_i, \beta_i + 1)} {B(\alpha_i, \beta_i)}.
  \end{align*}
  By Cor.~\ref{cor:exhaust}, if there exists a Pareto efficient policy
  satisfying counterfactual equalized odds, then it must correspond to some
  multiple threshold policy. In particular, by
  Eq.~\eqref{eq:counterfactual_equalized_odds} and the fact that the policy is
  budget exhausting, and using Prop.~\ref{prop:threshold} with \(\lambda = 0\),
  there must be \(t_0\), \(t_1\) such that
  \begin{align*}
    \frac {I_{t_0}(\alpha_0, \beta_0)} {B(\alpha_0, \beta_0)} + \frac
    {I_{t_1}(\alpha_1, \beta_1)} {B(\alpha_1, \beta_1)}
      &= 2 - b \\
    \frac {I_{t_0}(\alpha_0 + 1, \beta_0)} {B(\alpha_0 + 1, \beta_0)}
      &= \frac {I_{t_1}(\alpha_1 + 1, \beta_1)} {B(\alpha_1 + 1, \beta_1)} \\
    \frac {I_{t_0}(\alpha_0, \beta_0 + 1)} {B(\alpha_0, \beta_0 + 1)}
      &= \frac {I_{t_1}(\alpha_1, \beta_1 + 1)} {B(\alpha_1, \beta_1 + 1)}.
  \end{align*}
  Here, the first equality encodes budget exhaustion, the second the fact that
  \(\Pr(D = 1 \mid A = a_0, Y(1) = 1) = \Pr(D = 1 \mid A = a_1, Y(1) = 1)\), and
  the third the fact that \(\Pr(D = 1 \mid A = a_0, Y(1) = 0) = \Pr(D = 1 \mid A
  = a_1, Y(1) = 0)\).

  Let \(\bb f : \B R^6 \to \B R^3\) be given by
  \begin{align*}
    f_0(\alpha_0, \beta_0, t_0, \alpha_1, \beta_1, t_1)
      &= \frac {I_{t_0}(\alpha_0, \beta_0)} {B(\alpha_0, \beta_0)} + \frac
      {I_{t_1}(\alpha_1, \beta_1)} {B(\alpha_1, \beta_1)} - (2 - b), \\
    f_1(\alpha_0, \beta_0, t_0, \alpha_1, \beta_1, t_1)
      &= \frac {I_{t_0}(\alpha_0 + 1, \beta_0)} {B(\alpha_0 + 1, \beta_0)} -
      \frac {I_{t_1}(\alpha_1 + 1, \beta_1)} {B(\alpha_1 + 1, \beta_1)}, \\
    f_2(\alpha_0, \beta_0, t_0, \alpha_1, \beta_1, t_1)
      &= \frac {I_{t_0}(\alpha_0, \beta_0 + 1)} {B(\alpha_0, \beta_0 + 1)} -
      \frac {I_{t_1}(\alpha_1, \beta_1 + 1)} {B(\alpha_1, \beta_1 + 1)}.
  \end{align*}
  Then, given \(\alpha_0\), \(\beta_0\), \(\alpha_1\) and \(\beta_1\), there
  exists a Pareto optimal policy satisfying counterfactual equalized odds only
  if there exist \(t_0\) and \(t_1\) such that
  \[
    \bb f(\alpha_0, \beta_0, t_0, \alpha_1, \beta_1, t_1) = \bb 0.
  \]
  Let \(D \bb f\) denote the Jacobian of \(\bb f\). If we can show that \(\bb
  f\) is smooth and \(D \bb f\) has full rank, then the proof is complete. For,
  by Theorem~5.12 in \cite{lee2013introduction}, it follows that \(\bb
  f^{-1}(\bb 0) \subseteq \B R^6\) is a smooth 3-manifold. The restriction of
  the map
  \[
    \pi : (\alpha_0, \beta_0, t_0, \alpha_1, \beta_1, t_1) \mapsto (\alpha_0,
    \beta_0, \alpha_1, \beta_1)
  \]
  to \(\bb f^{-1}(\bb 0)\) is smooth, and so by Sard's theorem (see Theorem~6.10
  in \cite{lee2013introduction}), since \(\B R^{4}_{>0}\) is a trivially a
  smooth 4-manifold, the measure of the singular values of \(\pi \rest_{\bb
  f^{-1}(\bb 0)}\) is zero. However, since the maximum rank of \(D \pi\) on
  \(\bb f^{-1}(\bb 0)\) is three, every point of \(\bb f^{-1}(\bb 0)\) is
  singular, and consequently, the whole image of \(\pi\) is singular, i.e.,
  \(\pi(\bb f^{-1}(\bb 0))\) has measure zero. However, as argued above, the set
  of \((\alpha_0, \beta_0, \alpha_1, \beta_1)\) such that there exists a Pareto
  efficient distribution satisfying counterfactual equalized odds is a subset of
  \(\pi(\bb f^{-1}(\bb 0))\).

  Therefore, it remains only to show that \(\bb f\) is smooth, and that \(D \bb
  f\) has full rank for all \((\alpha_0, \beta_0, t_0, \alpha_1, \beta_1, t_1)
  \in \B R^2_{>0} \times [0,1] \times \B R^2_{>0} \times [0,1]\). This
  verification is a routine exercise in linear algebra and multivariable
  calculus, and is given below.

  \medskip\noindent\emph{Smoothness.}\qquad
  We note that since smooth functions are closed under
  composition, it suffices to show that \(I_t(\alpha, \beta)\) is a smooth
  function of \(t\), \(\alpha\), and \(\beta\). First, we consider partial
  derivatives with respect to \(\alpha\) and \(\beta\). If we could
  differentiate under the integral sign, then we would have that
  \begin{equation}
  \label{eq:duthis}
    \frac {\partial^{n+m}} {\partial \alpha^n \partial \beta^m} I_t(\alpha,
    \beta) = \int_0^t \log(x)^n \log(1 - x)^m x^{\alpha - 1} (1 - x)^{\beta - 1}
    \, \dx
    x.
  \end{equation}
  Recall the well-known condition for the Leibniz integral rule that if
  \(\tfrac {\partial} {\partial x} \phi(x,t) |_{x = x'}\) is dominated by some
  integrable \(g(t)\) for all \(x'\) in some neighborhood of \(x\),
  then\footnote{%
    See, e.g., Theorem~6.28 in \cite{klenke2020probability}.
  }
  \[
    \frac {\dx} {\dx x} \int_0^t \phi(x,t) \, \dx t = \int_0^t \frac {\partial}
    {\partial x} \phi(x, t) \, \dx t.
  \]
  Since the integrand \(\log(x)^n \log(1-x)^m x^{\alpha - 1} (1 - x)^{\beta -
  1}\) is strictly decreasing in both \(\alpha\) and \(\beta\) for \(x \in
  (0,1)\), it suffices merely to show that \(\log(x)^n \log(1-x)^m x^{\alpha-1}
  (1 - x)^{\beta-1}\) is integrable on \((0,1)\) for all \(\alpha > 0\) and
  \(\beta > 0\). Moreover, since
  \begin{itemize}
    \item The whole integrand is bounded on \((\epsilon, 1 - \epsilon)\),
    \item The factor \(\log(1-x)^m (1-x)^{\beta-1}\) is bounded on \((0,
      \epsilon)\),
    \item The factor \(\log(x)^n x^{\alpha-1}\) is bounded on \((1-\epsilon,
      1)\),
  \end{itemize}
  it suffices to show that \(\log(x)^n x^{\alpha-1}\) is integrable on
  \((0,\epsilon)\) and that \(\log(1-x)^m (1-x)^{\beta-1}\) is integrable on
  \((1-\epsilon, 1)\). Up to the change of variables \(x \mapsto 1 - x\), since
  \(n\), \(m\), \(\alpha\), and \(\beta\) are arbitrary, we see that it suffices
  to verify that \(\log(x)^n x^{\alpha-1}\) alone is integrable on \((0,1)\).
  Integrating by parts, we have that
  \[
    \int_0^t \log(x)^n x^{\alpha-1} \, \dx = \left[ \frac {\log(x)^n x^\alpha}
    \alpha \right]_0^1 - n \int_0^1 \log(x)^{n-1} x^{\alpha-1} \, \dx x.
  \]
  Since, by l'Hôpital's rule, \(\lim_{x \to 0} \log(x)^n x^\alpha = 0\), this
  expression equals
  \[
    - n \int_0^1 \log(x)^{n-1} x^{\alpha-1} \, \dx x.
  \]
  Since \(x^{\alpha-1}\) is integrable on \(0,1\) we see inductively that so is
  \(\log(x)^n x^{\alpha-1}\). Therefore, we can differentiate under the integral
  sign, and Eq.~\eqref{eq:duthis} holds.

  Now, taking derivatives with respect to \(t\) yields that
  \[
    \frac {\partial^{n+m+1}} {\partial t \partial \alpha^n \partial \beta^m}
    I_t(\alpha, \beta) = \log(t)^n \log(1 - t)^m t^{\alpha - 1} (1 - t)^{\beta -
    1},
  \]
  which is a polynomial in smooth functions of \(t\), and hence is smooth in
  \(t\). Therefore
  \[
    \frac {\partial^{n + m + k}} {\partial t^k \partial \alpha^n \partial
    \beta^m} I_t(\alpha, \beta)
  \]
  exists and is a polynomial in \(t\) for all \(k > 1\).

  By continuity, the orders of the partial derivatives can be switched
  arbitrarily (see, e.g., Theorem~9.40 in \cite{rudin1976principles}), and so it
  follows that \(\bb f\) is smooth.

  \medskip\noindent\emph{Full rank.}\qquad
  By the rank-nullity theorem, it suffices to show that
  the column rank of \(D \bb f\) is three. However, letting \(B_{\alpha,
  \beta} \sim \bbeta(\alpha, \beta)\) we have, by the results of the previous
  section, that the first three columns (i.e., partial derivatives with respect
  to \(\alpha_0\), \(\beta_0\), and \(t_0\), respectively) of \(D \bb f\) are
  given by
  \[
    \begin{bmatrix}
      \B E[\log(B_{\alpha_0, \beta_0}) \cdot \B 1_{B_{\alpha_0, \beta_0} < t_0}]
        & \B E[\log(B_{\beta_0, \alpha_0}) \cdot \B 1_{B_{\beta_0, \alpha_0} < 1
        - t_0}]
        & \frac {t_0^{\alpha_0 - 1} \cdot (1 - t_0)^{\beta_0 - 1}} {B(\alpha_0,
        \beta_0)} \\
      \B E[\log(B_{\alpha_0+1, \beta_0}) \cdot \B 1_{B_{\alpha_0+1, \beta_0} <
      t_0}]
        & \B E[\log(B_{\beta_0, \alpha_0+1}) \cdot \B 1_{B_{\beta_0, \alpha_0+1}
        < 1 - t_0}]
        & \frac{t_0^{\alpha_0} \cdot (1 - t_0)^{\beta_0 - 1}} {B(\alpha_0 + 1,
        \beta_0)} \\
      \B E[\log(B_{\alpha_0, \beta_0+1}) \cdot \B 1_{B_{\alpha_0, \beta_0+1} <
      t_0}]
        & \B E[\log(B_{\beta_0+1, \alpha_0}) \cdot \B 1_{B_{\beta_0+1, \alpha_0}
        < 1 - t_0}]
        & \frac {t_0^{\alpha_0 - 1} \cdot (1 - t_0)^{\beta_0}} {B(\alpha_0,
        \beta_0+1)}
    \end{bmatrix}.
  \]
  It is easy to see that the first two columns are necessarily independent,
  since all entries are negative but, by
  Lemmata~\ref{lem:monotone}~and~\ref{lem:st}, taking \(f(x) = - \log(x)\), we
  have that
  \[
    \B E[\log(B_{\alpha_0+1, \beta_0}) \cdot \B 1_{B_{\alpha_0+1, \beta_0} <
    t_0}] > \B
    E[\log(B_{\alpha_0, \beta_0+1}) \cdot \B 1_{B_{\alpha_0, \beta_0+1} < t_0}]
  \]
  while
  \[
    \B E[\log(B_{\beta_0, \alpha_0+1}) \cdot \B 1_{B_{\beta_0, \alpha_0+1} < 1 -
    t_0}] < \B E[\log(B_{\beta_0+1, \alpha_0}) \cdot \B 1_{B_{\beta_0+1,
    \alpha_0} < 1 - t_0}].
  \]
  Now observe that the final three columns of \(D \bb f\)---i.e., its partial
  derivatives with respect to \(\alpha_1\), \(\beta_1\), and \(t_1\),
  respectively---are given by
  \[
    \begin{bmatrix}
      \B E[\log(B_{\alpha_1, \beta_1}) \cdot \B 1_{B_{\alpha_1, \beta_1} < t_1}]
        & \B E[\log(B_{\beta_1, \alpha_1}) \cdot \B 1_{B_{\beta_1, \alpha_1} < 1
        - t_1}]
        & \frac {t_1^{\alpha_1 - 1} \cdot (1 - t_1)^{\beta_1 - 1}} {B(\alpha_1,
        \beta_1)} \\
      -\B E[\log(B_{\alpha_1+1, \beta_1}) \cdot \B 1_{B_{\alpha_1+1, \beta_1} <
      t_1}]
        & -\B E[\log(B_{\beta_1, \alpha_1+1}) \cdot \B 1_{B_{\beta_1,
        \alpha_1+1} < 1 - t_1}]
        & -\frac{t_1^{\alpha_1} \cdot (1 - t_1)^{\beta_1 - 1}} {B(\alpha_1 + 1,
        \beta_1)} \\
      -\B E[\log(B_{\alpha_1, \beta_1+1}) \cdot \B 1_{B_{\alpha_1, \beta_1+1} <
      t_1}]
        & -\B E[\log(B_{\beta_1+1, \alpha_1}) \cdot \B 1_{B_{\beta_1+1,
        \alpha_1} < 1 - t_1}]
        & -\frac {t_1^{\alpha_1 - 1} \cdot (1 - t_1)^{\beta_1}} {B(\alpha_1,
        \beta_1+1)}
    \end{bmatrix}.
  \]
  In particular, we notice that in the fourth column---i.e., partial derivatives
  with respect to \(\alpha_1\)---the element in the first row is negative, while
  the elements in the second and third row are positive. Therefore, the fourth
  column is necessarily independent of the first two columns---i.e., the partial
  derivatives with respect to \(\alpha_0\) and \(\beta_0\)---in which every
  element is negative. Therefore, we have proven that there are three linearly
  independent columns, and so the rank of \(D \bb f\) is full.
\end{proof}

\subsection{Proof of Proposition~\ref{prop:pred_parity}}
\label{app:pred_parity}

To prove the proposition, we must use our characterizations of the conditional
tail risks of the beta distribution proven in Appendix~\ref{app:beta_theory}
above. Note that in Proposition~\ref{prop:pred_parity}, for expositional
clarity, we parameterize beta distributions in terms of their mean \(\mu\) and
sample size \(v\); here, for mathematical simplicity, we parameterize them in
terms of successes, \(\alpha\), and failures, \(\beta\), where \(\mu = \tfrac
\alpha {\alpha + \beta}\) and \(v = \alpha + \beta\).

Using the theory above, we begin by proving a modest generalization of
Prop.~\ref{prop:pred_parity}.

\begin{lemma}
\label{lem:pred_parity_general}
  Suppose \(\C A = \{a_0, a_1\}\), and consider the family \(\C U\) of utility
  functions of the form
  \begin{equation*}
    u(x) = r(x) + \lambda \cdot \B 1_{\alpha(x) = a_1},
  \end{equation*}
  indexed by \(\lambda \geq 0\), where \(r(x) = \EE[Y(1) \mid X = x]\). Suppose
  the conditional distributions of \(r(X)\) given \(A\) are beta distributed,
  i.e.,
  \begin{equation*}
    \C D( r(X) \mid A = a ) = \bbeta(\alpha_a, \beta_a),
  \end{equation*}
  with \(\alpha_{a_1} < \alpha_{a_0}\) and \(\beta_{a_0} <
  \beta_{a_1}\). Then any policy satisfying counterfactual predictive parity is
  strongly Pareto dominated.
\end{lemma}

\begin{proof}
  Suppose there were a Pareto efficient policy satisfying counterfactual
  predictive parity. Let \(\lambda = 0\). Then, by Prop.~\ref{prop:threshold},
  we may without loss of generality assume that there exist thresholds
  \(t_{a_0}\), \(t_{a_1}\) such that a threshold policy
  \(\tau(x)\) witnessing Pareto efficiency is given by
  \begin{equation*}
    \tau(x) = \begin{cases}
      1 & r(x) > t_{\alpha(x)}, \\
      0 & r(x) < t_{\alpha(x)}.
    \end{cases}
  \end{equation*}
  (Note that by our distributional assumption, \(\Pr(u(x) = t) = 0\) for all \(t
  \in [0, 1]\).) Since \(\lambda \geq 0\), we must have that \(t_{a_0} \geq
  t_{a_1}\). Since \(b < 1\), \(0 < t_{a_0}\). Therefore,
  \begin{align*}
    \EE[Y(1) \mid A = a_0, D = 0]
      &= \EE[r(X) \mid A = a_0, u(X) < t_{a_0}] \\
      &\geq \EE[r(X) \mid A = a_0, u(X) < t_{a_1}] \\
      &> \EE[r(X) \mid A = a_1, u(X) < t_{a_1}] \\
      &= \EE[Y(1) \mid A = a_1, D = 0],
  \end{align*}
  where the first equality follows by the law of iterated expectation, the
  second from the fact that \(t_{a_1} \leq t_{a_0}\), the third from our
  distributional assumption and Lemmata~\ref{lem:mean}~and~\ref{lem:st}, and the
  final again from the law of iterated expectation. However, since
  counterfactual predictive parity is satisfied, \(\EE[Y(1) \mid A = a_0, D = 0]
  = \EE[Y(1) \mid A = a_1, D = 0]\), which is a contradiction. Therefore, no
  such threshold policy exists.
\end{proof}

After accounting for the difference in parameterization,
Prop.~\ref{prop:pred_parity} follows as a corollary.

\begin{proof}[Proof of Prop.~\ref{prop:pred_parity}]
  Since \(\mu_{a_0} > \mu_{a_1}\), \(\alpha_{a_0} = v \cdot \mu_{a_0} > v \cdot
  \mu_{a_1} = \alpha_{a_1}\) and \(\beta_{a_0} = v \cdot (1 - \mu_{a_0}) < v
  \cdot (1 - \mu_{a_1}) = \beta_{a_1}\). Therefore \(\beta_{a_0} < \beta_{a_1}\)
  and \(\alpha_{a_1} < \alpha_{a_0}\), and so, by
  Lemma~\ref{lem:pred_parity_general}, the proposition follows.
\end{proof}

\vskip 0.2in
\bibliography{bibli}

\end{document}